\documentclass{article}
\usepackage{epubtk}
\usepackage{graphicx}
\usepackage{amsmath}
\usepackage{amssymb}
\usepackage{textcomp}
\usepackage{xspace}

\showlistoftablesfalse 

\newcommand{\mum}{\textmu\hspace{-0.2pt}m\xspace}
\newcommand{\muHz}{\textmu\hspace{-0.2pt}Hz\xspace}
\newcommand{\muN}{\textmu\hspace{-0.2pt}N\xspace}
\newcommand{\Hz}{Hz\super{-1/2}\xspace}
\newcommand{\HzHz}{Hz~Hz\super{-1/2}\xspace}

\begin{document}

\title{Gravitational Wave Detection by Interferometry \\(Ground and Space)}

\author{%
\epubtkAuthorData{Matthew Pitkin}
        {Scottish Universities Physics Alliance (SUPA)\\
        School of Physics and Astronomy,\
        University of Glasgow \\
        Glasgow G12 8QQ, U.K.}
        {matthew.pitkin@glasgow.ac.uk}
        {}
\and
\epubtkAuthorData{Stuart Reid}
        {Scottish Universities Physics Alliance (SUPA)\\
        School of Physics and Astronomy,
        University of Glasgow \\
        Glasgow G12 8QQ, U.K.}
        {stuart.reid.2@glasgow.ac.uk}
    {}
\and
\epubtkAuthorData{Sheila Rowan}
        {Scottish Universities Physics Alliance (SUPA)\\
        School of Physics and Astronomy,
        University of Glasgow \\
        Glasgow G12 8QQ, U.K.}
        {sheila.rowan@glasgow.ac.uk}
        {}
\and
\epubtkAuthorData{Jim Hough}
        {Scottish Universities Physics Alliance (SUPA)\\
        School of Physics and Astronomy,
        University of Glasgow \\
        Glasgow G12 8QQ, U.K.}
        {james.hough@glasgow.ac.uk}
        {}
}

\date{}
\maketitle

\begin{abstract}
Significant progress has been made in recent years on the development of
gravitational-wave detectors. Sources such as coalescing compact binary
systems, neutron stars in low-mass X-ray binaries, stellar collapses and
pulsars are all possible candidates for detection. The most promising design
of gravitational-wave detector uses test masses a long distance apart and
freely suspended as pendulums on Earth or in drag-free spacecraft.  The
main theme of this review is a discussion of the mechanical and optical
principles used in the various long baseline systems in operation around the
world -- LIGO (USA), Virgo (Italy/France), TAMA300 and LCGT (Japan), and
GEO600 (Germany/U.K.) -- and in LISA, a proposed space-borne interferometer. A
review of recent science runs from the current generation of ground-based
detectors will be discussed, in addition to highlighting the astrophysical
results gained thus far. Looking to the future, the major upgrades to LIGO
(Advanced LIGO), Virgo (Advanced Virgo), LCGT and GEO600 (GEO-HF) will be
completed over the coming years, which will create a network of detectors
with the significantly improved sensitivity required to detect
gravitational waves. Beyond this, the concept and design of possible
future ``third generation'' gravitational-wave detectors, such as the
Einstein Telescope (ET), will be discussed.
\end{abstract}

\epubtkKeywords{gravitational waves, laser interferometry, gravitational wave
detectors, science runs, Interferometric gravitational wave detectors, Noise
sources, Data analysis}

%\newpage

%\epubtkUpdate
%    [Id=A,
%     ApprovedBy=subjecteditor,
%     AcceptDate={17 June 2011},
%     PublishDate={11 July 2011},
%     Type=major]{%
%For the update the author list has changed to be Matthew Pitkin,
%Stuart Reid, Sheila Rowan and Jim Hough. There have been minor updates
%to Sections \ref{section:introduction}, \ref{section:gravwaves} and
%\ref{section:Detection}; major updates to Sections~\ref{section:noise}
%and \ref{section:interferometry}; Section~\ref{section:construction}
%has been renamed and includes entirely new material on the operation
%of, and results from, the first generation of gravitational wave
%detectors and upgrades that are under way; and
%Section~\ref{section:space} also includes major updates about the
%status of LISA. The number of references has increased from 110 to 324.}

%%%%%%%%%%%%%%%%%%%%%%%%%%%%%%%%%%%%%%%%%%%%%%%%%%%%%%%%%%%%%%%%%%%%%%%%%%%%%%%%
%%%%%%%%%%%%%%%%%%%%%%%%%%%%%%%%%%%%%%%%%%%%%%%%%%%%%%%%%%%%%%%%%%%%%%%%%%%%%%%%
%%%%%%%%%%%%%%%%%%%%%%%%%%%%%%%%%%%%%%%%%%%%%%%%%%%%%%%%%%%%%%%%%%%%%%%%%%%%%%%%

\newpage

\section{Introduction}
\label{section:introduction}

Gravitational waves, one of the more exotic predictions of Einstein's General
Theory of Relativity may, after decades of controversy over their existence, be
detected within the next five years.

Sources such as interacting black holes, coalescing compact binary systems,
stellar collapses and pulsars are all possible candidates for detection;
observing signals from them will significantly boost our understanding of the
Universe. New unexpected sources will almost certainly be found and time will
tell what new information such discoveries will bring. Gravitational waves are
ripples in the curvature of space-time and manifest themselves as fluctuating
tidal forces on masses in the path of the wave. The first gravitational-wave
detectors were based on the effect of these forces on the fundamental resonant
mode of aluminium bars at room temperature. Initial instruments were constructed
by Joseph Weber~\cite{Weber1, Weber2} and subsequently developed by others.
Reviews of this early work are given in~\cite{Tyson, Douglass}. Following the
lack of confirmed detection of signals, aluminium bar systems operated at and
below the temperature of liquid helium were developed~\cite{Astone, Prodi,
Amaldi, Heng}, although work in this area is now subsiding, with only two
detectors, Auriga~\cite{AURIGA} and Nautilus~\cite{NAUTILUS}, continuing to
operate. Effort also continues to be pursued into cryogenic spherical bar
detectors, which are designed to have a wider bandwidth than the cylindrical
bars, with the two prototype detectors the Dutch MiniGRAIL~\cite{MiniGRAIL,
Gottardi:2007} and Brazilian M\'{a}rio Schenberg~\cite{Schenberg, Aguiar:2006}.
However, the most promising design of gravitational-wave detectors, offering the
possibility of very high sensitivities over a wide range of frequency, uses
widely-separated test masses freely suspended as pendulums on Earth or
in a drag-free craft in space; laser interferometry provides a means
of sensing the motion of the masses produced as they interact with a
gravitational wave.

Ground-based detectors of this type, based on the pioneering work of Forward
and colleagues (Hughes Aircraft)~\cite{Forward}, Weiss and colleagues (MIT)
\cite{Weiss}, Drever and colleagues (Glasgow/Caltech) \cite{Drever1,
Drever2} and Billing and colleagues (MPQ Garching)~\cite{Billing}, will be
used to observe sources whose radiation is emitted at frequencies above a few
Hz, and space-borne detectors, as originally envisaged by Peter Bender and Jim
Faller~\cite{BenderFaller1, BenderFaller2} at JILA, will be developed for
implementation at lower frequencies.

Gravitational-wave detectors of long baseline have been built in a number of
places around the world; in the USA (LIGO project led by a Caltech/MIT
consortium)~\cite{LIGOS5, LIGOweb}, in Italy (Virgo project, a joint
Italian/French venture)~\cite{Acernese:2007, VIRGOweb}, in Germany (GEO600
project built by a collaboration centred on the University of Glasgow, the
University of Hannover, the Max Planck Institute for Quantum Optics, the Max
Planck Institute for Gravitational Physics (Albert Einstein Institute), Golm and
Cardiff University)~\cite{Willke:2007, GEOweb} and in Japan (TAMA300
project)~\cite{TAMAStatus, TAMAweb}. A space-borne detector, called
LISA~\cite{LISA, NASAweb, ESAweb}, was until earlier this year (2011)
under study as a joint ESA/NASA mission as one L-class candidate
within the ESA Cosmic Visions program (a recent meeting detailing
these missions can be found here~\cite{ESACosmicVisions}). Funding
constraints within the US now mean that ESA must examine the
possibility of flying an L-class mission with European-only
funding. The official ESA statement on the next steps for LISA can be
found here~\cite{LISAESAstatement}. When completed, this detector
array would have the capability of detecting gravitational wave
signals from violent astrophysical events in the Universe, providing
unique information on testing aspects of general relativity and
opening up a new field of astronomy.

It is also possible to observe the tidal effects of a passing gravitational
wave by Doppler tracking of separated objects. For example, Doppler tracking of
spacecraft allows the Earth and an interplanetary spacecraft to be used as test
masses, where their relative positions can be monitored by comparing the nearly
monochromatic microwave signal sent from a ground station with the coherently
returned signal sent from the spacecraft~\cite{Estabrook:1975}. By comparing
these signals, a Doppler frequency time series $\Delta \nu / \nu_0$, where
$\nu_0$ is the central frequency from the ground station, can be generated.
Peculiar characteristics within the Doppler time series, caused by the passing
of gravitational waves, can be studied in the approximate frequency band of
10\super{-5} to 0.1~Hz. Several attempts have been made in recent decades to
collect such data (Ulysses, Mars Observer, Galileo, Mars Global Surveyor,
Cassini) with broadband frequency sensitivities reaching 10\super{-16}
(see~\cite{Armstrong:2006} for a thorough review of gravitational-wave
searches using Doppler tracking). There are currently no plans for
dedicated experiments using this technique; however, incorporating
Doppler tracking into another planetary mission would provide a
complimentary precursor mission before dedicated experiments such as
LISA are launched.

The technique of Doppler tracking to search for gravitational-wave signals can
also be performed using pulsar-timing experiments.  Millisecond
pulsars~\cite{Lorimer:2008} are known to be very precise clocks, which
allows the effects of a passing gravitational wave to be observed
through the modulation in the time of arrival of pulses from the
pulsar. Many noise sources exist and, for this reason, it is necessary
to monitor a large array of pulsars over a long observation time.
Further details on the techniques used and upper limits that have been
set with pulsar timing experiments can be found from groups such as
the European Pulsar Timing Array~\cite{Janssen:2008}, the North
American Nanohertz Observatory for Gravitational
Waves~\cite{Jenet:2006,Jenet:2009}, and the Parkes Pulsar Timing
Array~\cite{Hobbs:2008}.

All the above detection methods cover over 13 orders of magnitude in frequency
(see Figure~\ref{fig:fullspectrum}) equivalent to covering from radio waves to
X-rays in the electromagnetic spectrum.  This broadband coverage allows us to
probe a wide range of potential sources.

\epubtkImage{full_gw_spectrum.png}{%
\begin{figure}[htbp]
  \centerline{\includegraphics[width=\textwidth]{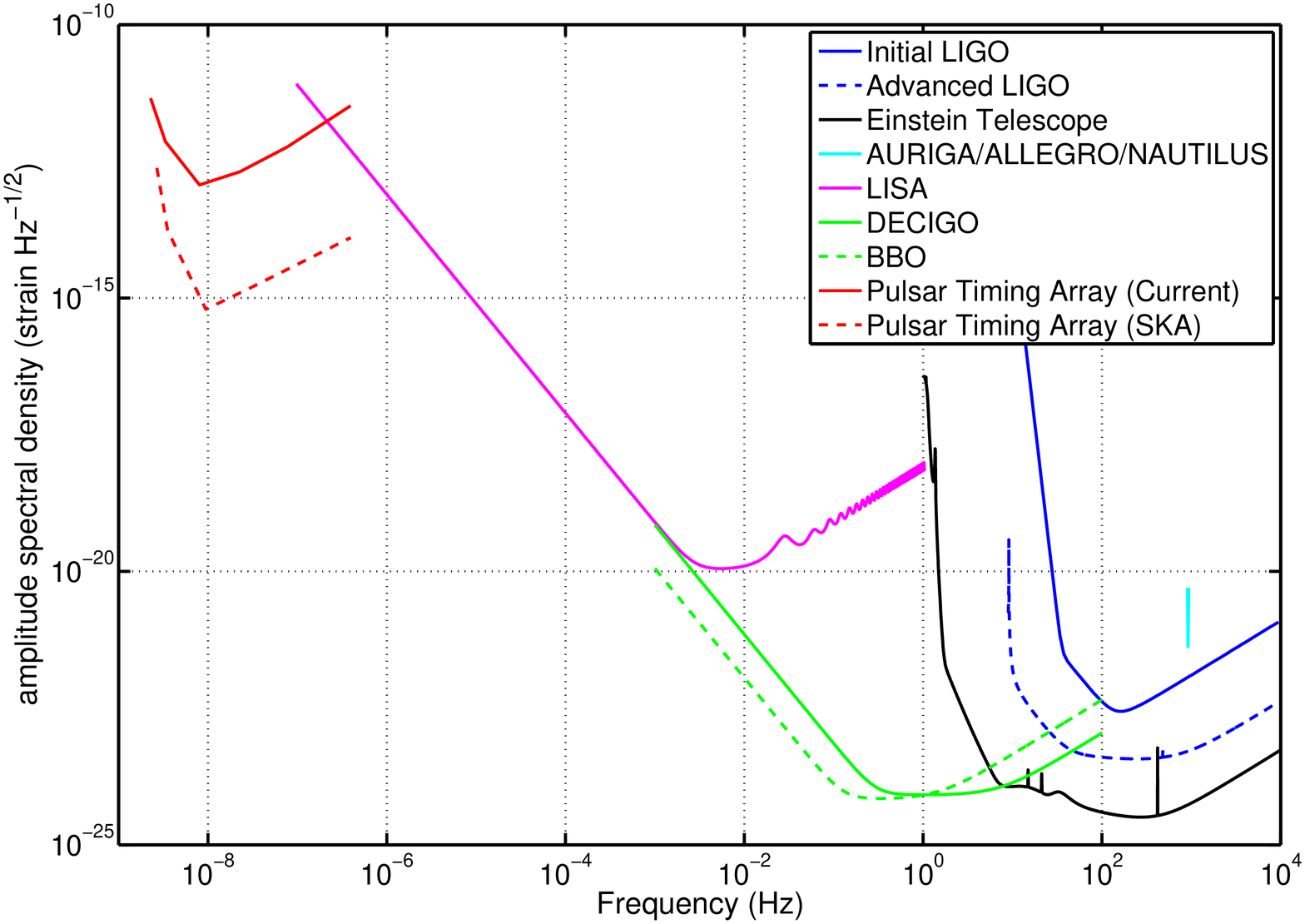}}
  \caption{The sensitivity of various gravitational-wave detection
    techniques across 13 orders of magnitude in frequency. At the low
    frequency end the sensitivity curves for pulsar timing arrays
    (based on current observations and future observations with the
    Square Kilometre Array~\cite{SKA}) are extrapolated from Figure~4
    in~\cite{Yardley:2010}. In the mid-range LISA, DECIGO and BBO are
    described in more detail in Section~\ref{section:space}, with the
    DECIGO and BBO sensitivity curves taken from models given in
    \cite{Yagi:2011}. At the high frequency the sensitivities are
    represented by three generations of laser interferometers: LIGO,
    Advanced LIGO and the Einstein Telescope (see
    Sections~\ref{section:construction}, \ref{subsection:aligo} and
    \ref{subsec:et}). Also included is a representative sensitivity
    for the AURIGA~\cite{AURIGA}, Allegro~\cite{Mauceli:1996} and
    Nautilus~\cite{NAUTILUS} bar detectors.}
  \label{fig:fullspectrum}
\end{figure}}

We recommend a number of excellent books for reference. For a popular account of
the development of the gravitational-wave field the reader should consult
Chapter~10 of \textit{Black Holes and Time Warps} by Kip S.\ Thorne~\cite{Thorne}, or
the more recent books, \textit{Einstein's Unfinished Symphony}, by Marcia
Bartusiak~\cite{Bartusiak:2000} and \textit{Gravity from the Ground Up}, by Bernard
Schutz~\cite{Schutz:2003}. A comprehensive review of developments toward laser
interferometer detectors is found in \textit{Fundamentals of Interferometric
Gravitational Wave Detectors} by Peter Saulson~\cite{Saulsonbook}, and
discussions relevant to the technology of both bar and interferometric detectors
are found in \textit{The Detection of Gravitational Waves} edited by David
Blair~\cite{Blair}.

In addition to the wealth of articles that can be found on the home site of
this journal, there are also various informative websites that can easily be
found, including the homepages of the various international collaborative
projects searching for gravitational waves, such as the LIGO Scientific
Collaboration~\cite{LSCweb}.

\newpage

%%%%%%%%%%%%%%%%%%%%%%%%%%%%%%%%%%%%%%%%%%%%%%%%%%%%%%%%%%%%%%%%%%%%%%%%%%%%%%%%
%%%%%%%%%%%%%%%%%%%%%%%%%%%%%%%%%%%%%%%%%%%%%%%%%%%%%%%%%%%%%%%%%%%%%%%%%%%%%%%%

\section{Gravitational Waves}
\label{section:gravwaves}

Some early relativists were sceptical about the existence of gravitational
waves; however, the 1993 Nobel Prize in Physics was awarded to Hulse and Taylor
for their experimental observations and subsequent interpretations of the
evolution of the orbit of the binary pulsar \epubtkSIMBAD{PSR~1913+16}~\cite{Hulse, Taylor},
the decay of the binary orbit being consistent with angular momentum and energy
being carried away from this system by gravitational waves~\cite{Will}.

Gravitational waves are produced when matter is accelerated in an asymmetrical
way; but due to the nature of the gravitational interaction, detectable levels
of radiation are produced only when very large masses are accelerated in very
strong gravitational fields. Such a situation cannot be found on Earth but is
found in a variety of astrophysical systems. Gravitational wave signals are
expected over a wide range of frequencies; from $\simeq$~10\super{-17}~Hz in the case
of ripples in the cosmological background to $\simeq$~10\super{3}~Hz from the formation
of neutron stars in supernova explosions. The most predictable sources are
binary star systems. However, there are many sources of much-greater
astrophysical interest associated with black-hole interactions and coalescences,
neutron-star coalescences, neutron stars in low-mass X-ray binaries such as
\epubtkSIMBAD[V818~Sco]{Sco-X1}, stellar collapses to neutron stars and black holes (supernova
explosions), pulsars, and the physics of the early Universe. For a full
discussion of sources refer to the material contained in~\cite{Sathyaprakash:2009,LISAsymposium, sources, Amaldiproc}.

Why is there currently such interest worldwide in the detection of gravitational
waves? Partly because observation of the velocity and polarisation states of the
signals will allow a direct experimental check of the wave predictions of
general relativity; but, more importantly, because the detection of the signals
should provide observers with new and unique information about astrophysical
processes. It is interesting to note that the gravitational wave signal from a
coalescing compact binary star system has a relatively simple form and the
distance to the source can be obtained from a combination of its signal strength
and its evolution in time. If the redshift at that distance is found, Hubble's
Constant -- the value for which has been a source of lively debate for many
years -- may then be determined with, potentially, a high degree of
accuracy~\cite{Schutz,Holtz:2005}.

Only now are detectors being built with the technology required to achieve the
sensitivity to observe such interesting sources.

%%%%%%%%%%%%%%%%%%%%%%%%%%%%%%%%%%%%%%%%%%%%%%%%%%%%%%%%%%%%%%%%%%%%%%%%%%%%%%%%
%%%%%%%%%%%%%%%%%%%%%%%%%%%%%%%%%%%%%%%%%%%%%%%%%%%%%%%%%%%%%%%%%%%%%%%%%%%%%%%%

\newpage

\section{Detection of Gravitational Waves}
\label{section:Detection}

Gravitational waves are most simply thought of as ripples in the curvature of
space-time, their effect being to change the separation of adjacent masses on
Earth or in space; this tidal effect is the basis of all present detectors.
Gravitational wave strengths are characterised by the gravitational-wave
amplitude $h$, given by
\begin{equation}
  h = \frac{2 \Delta L} L,
  \label{equation:h}
\end{equation}
where $\Delta L$ is the change in separation of two masses a distance $L$ apart;
for the strongest-allowed component of gravitational radiation, the value of $h$
is proportional to the third time derivative of the quadrupole moment of the
source of the radiation and inversely proportional to the distance to the
source. The radiation field itself is quadrupole in nature and this shows up in
the pattern of the interaction of the waves with matter.

The problem for the experimental physicist is that the predicted magnitudes of
the amplitudes or strains in space in the vicinity of the Earth caused by
gravitational waves even from the most violent astrophysical events are
extremely small, of the order of 10\super{-21} or lower~\cite{Sathyaprakash:2009,
LISAsymposium}. Indeed, current theoretical models on the event rate and strength
of such events suggest that in order to detect a few events per year -- from
coalescing neutron-star binary systems, for example, an amplitude sensitivity
close to 10\super{-22} over timescales as short as a millisecond is required. If
the Fourier transform of a likely signal is considered it is found that the
energy of the signal is distributed over a frequency range or bandwidth, which is
approximately equal to 1/timescale.  For timescales of a millisecond the
bandwidth is approximately 1000~Hz, and in this case the spectral density of the
amplitude sensitivity is obtained by dividing 10\super{-22} by the square root of
1000. Thus, detector noise levels must have an amplitude spectral density lower
than $\simeq$~10\super{-23}~\Hz over the frequency range of the signal.
Signal strengths at the Earth, integrated over appropriate time intervals, for a
number of sources are shown in Figure~\ref{figure:sourcestrengths}.

\epubtkImage{fig1.png}{%
\begin{figure}[htbp]
  \centerline{\includegraphics[width=0.8\textwidth]{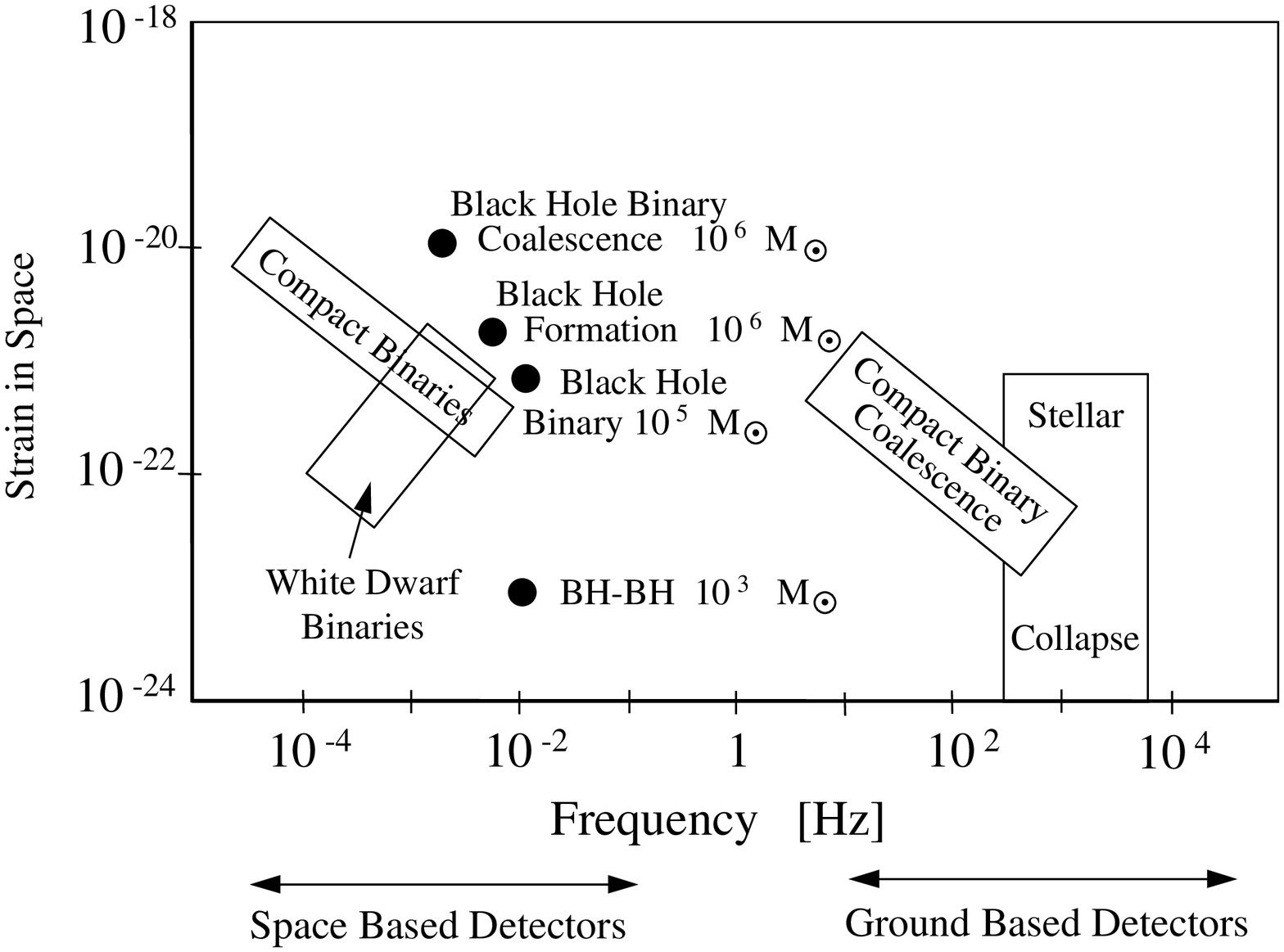}}
  \caption{Some possible sources for ground-based and space-borne
    detectors.}
  \label{figure:sourcestrengths}
\end{figure}}

The weakness of the signal means that limiting noise sources like the thermal
motion of molecules in the critical components of the detector (thermal noise),
seismic or other mechanical disturbances, and noise associated with the detector
readout, whether electronic or optical, must be reduced to an extremely low
level. For signals above $\simeq$~10~Hz ground based experiments are possible,
but for lower frequencies where local fluctuating gravitational gradients and
seismic noise on Earth become a problem, it is best to consider developing
detectors for operation in space~\cite{LISA}.

%%%%%%%%%%%%%%%%%%%%%%%%%%%%%%%%%%%%%%%%%%%%%%%%%%%%%%%%%%%%%%%%%%%%%%%%%%%%%%%%
%%%%%%%%%%%%%%%%%%%%%%%%%%%%%%%%%%%%%%%%%%%%%%%%%%%%%%%%%%%%%%%%%%%%%%%%%%%%%%%%

\subsection{Initial detectors and their development}
\label{subsection:initdet}

The earliest experiments in the field were ground based and were carried out by
Joseph Weber of the University of Maryland in the 1960s. With colleagues he
began by looking for evidence of excitation of the normal modes of the Earth by
very low frequency gravitational waves~\cite{Forward2}. Efforts to detect gravitational
waves via the excitation of Earth's normal modes was also pursued by Weiss and Block~\cite{Weiss:1965}.
Weber then moved on to look for tidal strains in aluminium bars, which were at room temperature and were
well isolated from ground vibrations and acoustic noise in the
laboratory~\cite{Weber1, Weber2}. The bars were resonant at $\simeq$~1600~Hz, a
frequency where the energy spectrum of the signals from collapsing stars was
predicted to peak. Despite the fact that Weber observed coincident excitations
of his detectors placed up to 1000~km apart, at a rate of approximately one
event per day, his results were not substantiated by similar experiments carried
out in several other laboratories in the USA, Germany, Britain and Russia. It
seems unlikely that Weber was observing gravitational-wave signals because,
although his detectors were very sensitive, being able to detect strains of the
order of 10\super{-16} over millisecond timescales~\cite{Weber1}, their sensitivity
was far away from what was predicted to be required theoretically. Development
of Weber bar type detectors continued with significant emphasis on cooling to
reduce the noise levels, although work in this area is now subsiding with
efforts continuing on Auriga~\cite{AURIGA}, Nautilus~\cite{NAUTILUS},
MiniGRAIL~\cite{MiniGRAIL, Gottardi:2007} and M\'{a}rio Schenberg
\cite{Schenberg, Aguiar:2006}.  In around 2003, the sensitivity of km-scale
interferometric gravitational-wave detectors began to surpass the peak
sensitivity of these cryogenic bar detectors ($\simeq$~10\super{-21})
and, for example, the LIGO detectors reached their design sensitivities at
almost all frequencies by 2005 (peak sensitivity
$\simeq$~2~\texttimes~10\super{-23} at
$\simeq$~200~Hz)~\cite{Whitcomb:2008}, see
Section~\ref{subsection:runs} for more information on science runs of
the recent generation of detectors.  In addition to gaining better
strain sensitivities, interferometric detectors have a marked
advantage over resonant bars by being sensitive to a broader range of
frequencies, whereas resonant bar are inherently sensitive only to
signals that have significant spectral energy in a narrow band around
their resonant frequency. The concept and design of gravitational-wave
detectors based on laser interferometers will be introduced in the
following Section~\ref{subsection:earth}.

%%%%%%%%%%%%%%%%%%%%%%%%%%%%%%%%%%%%%%%%%%%%%%%%%%%%%%%%%%%%%%%%%%%%%%%%%%%%
%%%%%%%%%%%%%%%%%%%%%%%%%%%%%%%%%%%%%%%%%%%%%%%%%%%%%%%%%%%%%%%%%%%%%%%%%%%%

\subsection{Long baseline detectors on Earth}
\label{subsection:earth}

An interferometric design of gravitational-wave detector offers the possibility
of very high sensitivities over a wide range of frequency. It uses test masses,
which are widely separated and freely suspended as pendulums to isolate against
seismic noise and reduce the effects of thermal noise; laser interferometry
provides a means of sensing the motion of these masses produced as they interact
with a gravitational wave (Figure~\ref{figure:schematicdetector}).

\epubtkImage{fig2.png}{%
\begin{figure}[htbp]
  \centerline{\includegraphics[width=0.5\textwidth]{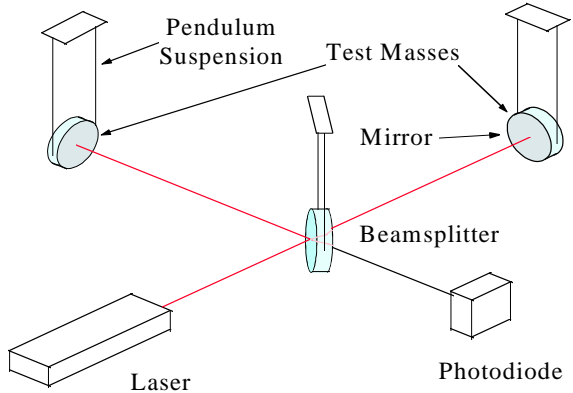}}
  \caption{Schematic of gravitational-wave detector using laser
    interferometry.}
  \label{figure:schematicdetector}
\end{figure}}

This technique is based on the Michelson interferometer and is particularly
suited to the detection of gravitational waves as they have a quadrupole nature.
Waves propagating perpendicular to the plane of the interferometer will result
in one arm of the interferometer being increased in length while the other arm
is decreased and vice versa. The induced change in the length of the
interferometer arms results in a small change in the intensity of the light
observed at the interferometer output.

As will be explained in detail in the next Section~\ref{section:noise},
the sensitivity of an interferometric gravitational-wave detector is
limited by noise from various sources. Taking this frequency-dependent
noise floor into account, a design goal can be estimated for a
particular detector design. For example, the design sensitivity for
initial LIGO is show in Figure~\ref{figure:LIGOsens} plotted alongside
the achieved sensitivities of the three individual interferometers
during the fifth science run (see Section~\ref{subsection:runs}). Such
strain sensitivities are expected to allow a reasonable probability
for detecting gravitational wave sources. However, in order to
guarantee the observation of a full range of sources and to initiate
gravitational-wave astronomy, a sensitivity or noise performance
approximately ten times better in the mid-frequency range and several
orders of magnitude better at 10~Hz, is desired. Therefore, initial
detectors will be upgraded to an advanced configuration, such as
Advanced LIGO, which will be ready for operation around 2015.

\epubtkImage{fig2.png}{%
\begin{figure}[htbp]
  \centerline{\includegraphics[width=0.9\textwidth]{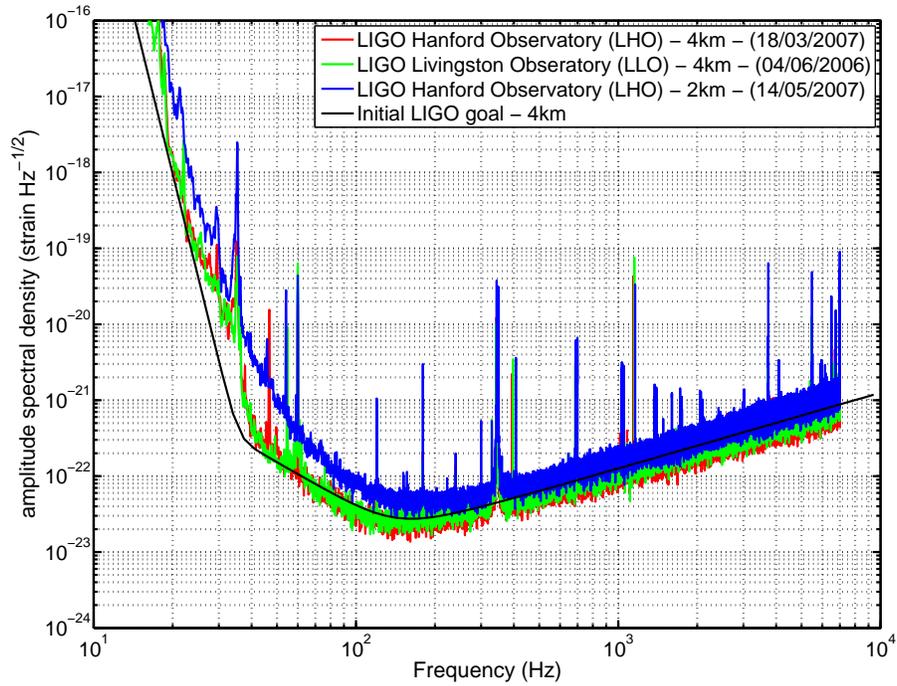}}
  \caption{Measured sensitivity of the initial LIGO interferometers
    during the S5 science run (see
    Section~\ref{sec:ligoruns}). Reproduced with permission from
    \cite{LIGOcurves}.}
  \label{figure:LIGOsens}
\end{figure}}

For the initial interferometric detectors, a noise floor in strain
close to 2~\texttimes~10\super{-23}~\Hz was achieved. Detecting a
strain in space at this level implies measuring a residual motion of
each of the test masses of about 8~\texttimes~10\super{-20}~m/\Hz over
part of the operating range of the detector, which may be from
$\simeq$~10~Hz to a few kHz. Advanced detectors will push this
target down further by another factor of 10\,--\,15. This sets a
formidable goal for the optical detection system at the output of the
interferometer.

%%%%%%%%%%%%%%%%%%%%%%%%%%%%%%%%%%%%%%%%%%%%%%%%%%%%%%%%%%%%%%%%%%%%%%%%%%%%%%
%%%%%%%%%%%%%%%%%%%%%%%%%%%%%%%%%%%%%%%%%%%%%%%%%%%%%%%%%%%%%%%%%%%%%%%%%%%%%%
%%%%%%%%%%%%%%%%%%%%%%%%%%%%%%%%%%%%%%%%%%%%%%%%%%%%%%%%%%%%%%%%%%%%%%%%%%%%%%

\newpage

\section{Main Noise Sources}
\label{section:noise}

In this section we discuss the main noise sources, which limit the sensitivity of
interferometric gravitational-wave detectors. Fundamentally it should be
possible to build systems using laser interferometry to monitor strains in space,
which are limited only by the Heisenberg Uncertainty Principle; however there
are other practical issues, which must be taken into account. Fluctuating
gravitational gradients pose one limitation to the interferometer sensitivity
achievable at low frequencies, and it is the level of noise from this source,
which dictates that experiments to look for sub-Hz gravitational-wave signals
have to be carried out in space~\cite{Spero, Saulson1, Beccaria, Thorne:1998}.
In general, for ground-based detectors the most important limitations to
sensitivity result from the effects of seismic and other ground-borne mechanical
noise, thermal noise associated with the test masses and their suspensions, and
quantum noise, which appears at high frequency as shot noise in the photocurrent
from the photodiode, which detects the interference pattern and can appear at low
frequency as radiation pressure noise due to momentum transfer to the test
masses from the photons when using high laser powers. The significance of each
of these sources will be briefly reviewed.

%%%%%%%%%%%%%%%%%%%%%%%%%%%%%%%%%%%%%%%%%%%%%%%%%%%%%%%%%%%%%%%%%%%%%%%%%%%%%
%%%%%%%%%%%%%%%%%%%%%%%%%%%%%%%%%%%%%%%%%%%%%%%%%%%%%%%%%%%%%%%%%%%%%%%%%%%%%

\subsection{Seismic noise}
\label{subsection:seismic}

Seismic noise at a reasonably quiet site on the Earth follows a
spectrum in all three dimensions close to 10\super{-7}{\it
  f}\super{-2}~m/Hz\super{1/2} (where here and elsewhere we measure
\textit{f} in Hz) and thus if the disturbance to each test mass must
be less than 3~\texttimes~10\super{-20}~m/Hz\super{1/2} at, for
example, 30~Hz, then the reduction of seismic noise required at that
frequency in the horizontal direction is greater than
10\super{9}. Since there is liable to be some coupling of vertical
noise through to the horizontal axis, along which the gravitational-wave--induced strains are to be sensed, a significant level of
isolation has to be provided in the vertical direction also. Isolation
can be provided in a relatively simple way by making use of the fact
that, for a simple pendulum system, the transfer function to the
pendulum mass of the horizontal motion of the suspension point falls
off as 1/(frequency)\super{2} above the pendulum resonance. In a
similar way isolation can be achieved in the vertical direction by
suspending a mass on a spring. In the case of the Virgo detector
system the design allows operation to below 10~Hz and here a
seven-stage horizontal pendulum arrangement is adopted with six of the
upper stages being suspended with cantilever springs to provide vertical
isolation~\cite{Braccini}, with similar systems developed in
Australia~\cite{Ju1} and at Caltech~\cite{DeSalvo}. For the GEO600
detector, where operation down to 50~Hz was planned, a triple pendulum
system is used with the first two stages being hung from cantilever
springs to provide the vertical isolation necessary to achieve the
desired performance. This arrangement is then hung from a plate
mounted on passive `rubber' isolation mounts and on an active
(electro-mechanical) anti-vibration system~\cite{Plissi1, Torrie}. The
upgraded seismic isolation for Advanced LIGO will also adopt a
variety of active and passive isolation stages. The total isolation
will be provided by one external stage (hydraulics), two stages of
in-vacuum active isolation, and being completed by the test mass
suspensions~\cite{Abbott:2002, Harry:2010}. For clarity, the two
stages of in-vacuum isolation are shown in
Figure~\ref{figure:LIGOseismic}, whereas the test-mass suspensions are
shown separately in Figure~\ref{figure:LIGOquad}.

%LivRev: do not use eps for html processing!
\epubtkImage{fig3.png}{%
\begin{figure}[htbp]
  \centerline{\includegraphics[width=340pt]{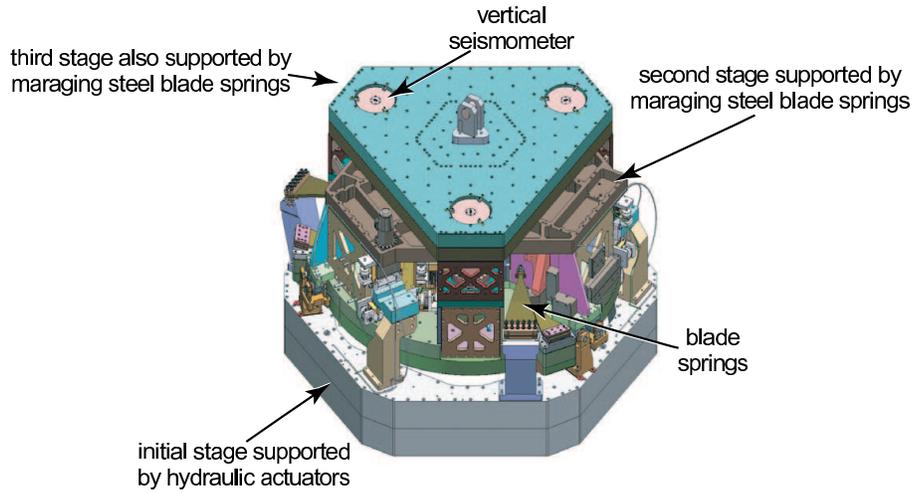}}
  \caption{Internal stages of the large chamber seismic isolation system
for Advanced LIGO (image is inverted for clarity).}
  \label{figure:LIGOseismic}
\end{figure}}

%LivRev: do not use eps for html processing!
\epubtkImage{fig4.png}{%
\begin{figure}[htbp]
  \centerline{\includegraphics[width=250pt]{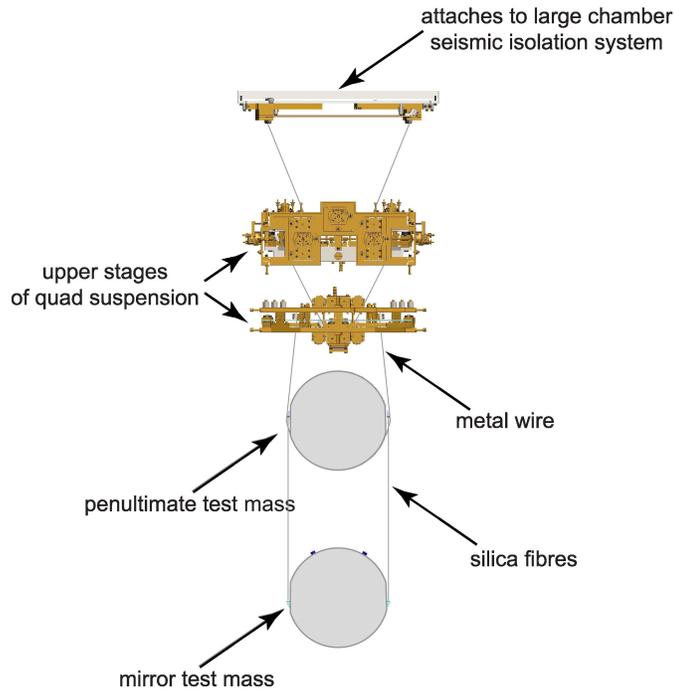}}
  \caption{CAD drawing of quad suspension system for Advanced LIGO, showing
the mirror test mass at the bottom and where the uppermost section is attached
to the third stage platform of the large chamber seismic isolation system shown
in Figure~\ref{figure:LIGOseismic}.}
  \label{figure:LIGOquad}
\end{figure}}

In order to cut down motion at the pendulum frequencies, active damping of the
pendulum modes has to be incorporated, and to reduce excess motion at low
frequencies around the micro-seismic peak, low-frequency isolators have to be
incorporated. These low-frequency isolators can take different forms -- tall
inverted pendulums in the horizontal direction and cantilever springs whose
stiffness is reduced by means of attractive forces between magnets for the
vertical direction in the case of the Virgo system~\cite{Losurdo},
Scott~Russell mechanical linkages in the horizontal and torsion bar arrangements
in the vertical for an Australian design~\cite{Winterflood}, and a
seismometer/actuator (active) system as shown here for Advanced
LIGO~\cite{Abbott:2002} and also used in GEO600~\cite{Plissi2}.  Such schemes
can provide sufficiently-large reduction in the direct mechanical coupling of
seismic noise through to the suspended mirror optic to allow operation down to
3~Hz~\cite{Braccini:1993,ETweb}. However, it is also possible for this
vibrational seismic noise to couple to the suspended optic through the
gravitational field.

%%%%%%%%%%%%%%%%%%%%%%%%%%%%%%%%%%%%%%%%%%%%%%%%%%%%%%%%%%%%%%%%%%%%%%%%%%%%%%
%%%%%%%%%%%%%%%%%%%%%%%%%%%%%%%%%%%%%%%%%%%%%%%%%%%%%%%%%%%%%%%%%%%%%%%%%%%%%%

\subsection{Gravity gradient (Newtonian) noise}
\label{subsection:gravitygradient}

Gravity gradients, caused by direct gravitational coupling of mass density
fluctuations to the suspended mirrors, were identified as a potential source of
noise in ground-based gravitational-wave detectors in 1972~\cite{Weiss}. The
noise associated with gravity gradients was first formulated by
Saulson~\cite{Saulson1} and Spero~\cite{Spero}, with later developments by
Hughes and Thorne~\cite{Thorne:1998} and Cella and Cuoco~\cite{Beccaria}.
These studies suggest that the dominant source of gravity gradients arise from
seismic surface waves, where density fluctuations of the Earth's surface are
produced near the location of the individual interferometer test masses, as
shown in Figure~\ref{figure:GGN}.

\epubtkImage{GGN.png}{%
\begin{figure}[htbp]
  \centerline{\includegraphics[width=\textwidth]{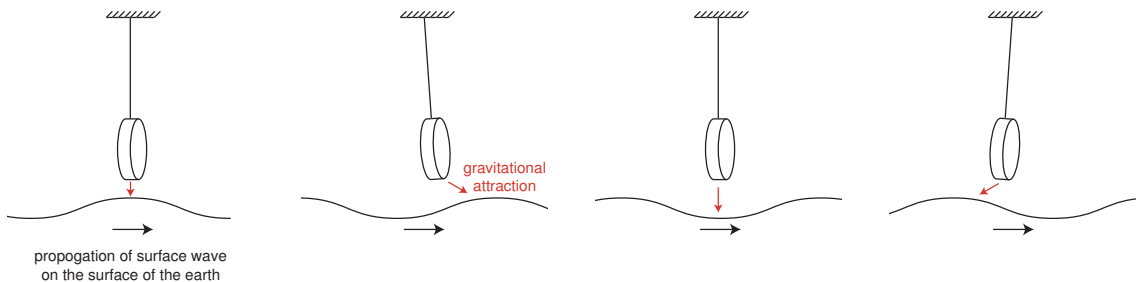}}
  \caption{Time-lapsed schematic illustrating the fluctuating
    gravitational force on a suspended mass by the propagation of a
    surface wave through the ground.}
  \label{figure:GGN}
\end{figure}}

The magnitude of the rms motion of the interferometer test masses,
$\tilde{x}(\omega)$, can be shown to be~\cite{Thorne:1998}
\begin{equation}
  \tilde{x}(\omega) = \frac{4 \pi G \rho}{\omega^{2}} \beta(\omega)
\tilde{W}(\omega),
  \label{equation:GGN}
\end{equation}
where $\rho$ is the Earth's density near the test mass, $G$ is
Newton's constant, $\omega$ is the angular frequency of the seismic
spectrum, $\beta(\omega)$ is a dimensionless reduced transfer function
that takes into account the correlated motion of the interferometer
test masses in addition to the reduction due to the separation between
the test mass and the Earth's surface, and $\tilde{W}(\omega)$ is the
displacement rms-averaged over 3-dimensional directions. In order to
eliminate noise arising from gravity gradients, a detector would have
to be operated far from these density fluctuations, that is, in space.
Proposed space missions are discussed in Section~\ref{section:space}.

However, there are two proposed approaches for reducing the level of gravity-gradient noise in future ground-based detectors. A monitor and subtraction
method can be used, where an array of seismometers can be distributed
strategically around each test mass to monitor the relevant ground motion (and
ground compression) that would be expected to couple through local gravity. A
subtraction signal may be developed from knowing how the observed density
fluctuations couple to the motion of each test mass, and can potentially allow a
significant reduction in gravity-gradient noise.

Another approach is to choose a very quiet location, or better still, to also go
underground, as is already going ahead for LCGT~\cite{Miyoki:2005}. Since the
dominant source of gravity-gradient noise is expected to arise from surface
waves on the Earth, the observed gravity-gradient noise will decrease with depth
into the Earth. Current estimates suggest that gravity-gradient noise can be
suppressed down to around 1~Hz by careful site selection and going $\sim$~150~m
underground~\cite{Beker:2011}. The most promising approach (or likely only
approach) to detecting gravitational waves whose frequency is below 1~Hz is to
build an interferometer in space.

%%%%%%%%%%%%%%%%%%%%%%%%%%%%%%%%%%%%%%%%%%%%%%%%%%%%%%%%%%%%%%%%%%%%%%%%%%%%%%
%%%%%%%%%%%%%%%%%%%%%%%%%%%%%%%%%%%%%%%%%%%%%%%%%%%%%%%%%%%%%%%%%%%%%%%%%%%%%%

\subsection{Thermal noise}
\label{subsection:thermal}

Thermal noise associated with the mirror masses and the last stage of their
suspensions is the most significant noise source at the low frequency end of the
operating range of initial long baseline gravitational wave
detectors~\cite{Saulson2}. Advanced detector configurations are also expected to
be limited by thermal noise at their most sensitive frequency
band~\cite{Levin, Nakagawa:2002, Harry:2002, Crooks:2002}. Above the operating
range there are the internal resonances of the test masses. The thermal noise in
the operating range comes from the \emph{tails} of these resonant modes. For any
simple harmonic oscillator such as a mass hung on a spring or hung as a pendulum,
the spectral density of thermal motion of the mass can be expressed
as~\cite{Saulson2}
\begin{equation}
  x^{2}(\omega) = \frac{4 k_{\mathrm{B}} T \omega_{0}^{2}
  \phi(\omega)}{\omega m [{(\omega_{0}^{2} - \omega^{2})^2 +
  \omega_{0}^{4} \phi^{2}(\omega)}]},
  \label{equation:thnoise}
\end{equation}
where $k_{\mathrm{B}}$ is Boltzmann's constant, $T$ is the temperature, $m$ is the
mass and  $\phi(\omega)$ is the loss angle or loss factor of the
oscillator of angular resonant frequency $\omega_0$. This loss factor is the
phase lag angle between the displacement of the mass and any force applied to
the mass at a frequency well below $\omega_0$. In the case of a mass on a spring,
the loss factor is a measure of the mechanical loss associated with the material
of the spring. For a pendulum, most of the energy is stored in the lossless
gravitational field. Thus, the loss factor is lower than that of the material,
which is used for the wires or fibres used to suspend the pendulum. Indeed,
following Saulson~\cite{Saulson2} it can be shown that for a pendulum of mass
$m$, suspended on four wires or fibres of length $l$, the loss factor of the
pendulum is related to the loss factor of the material by
\begin{equation}
  \phi_{\mathrm{pend}}(\omega) = \phi_{\mathrm{mat}}(\omega)\frac{4 \sqrt{TEI}}{mgl},
  \label{equation:pend}
\end{equation}
where $I$ is the moment of the cross-section of  each wire, and $T$ is the
tension in each wire, whose material has a Young's modulus $E$. In general, for
most materials, it appears that the intrinsic loss factor is essentially
independent of frequency over the range of interest for gravitational-wave
detectors (although care has to be taken with some materials in that a form of
damping known as thermo-elastic damping can become important for wires of small
cross-section~\cite{Nowick} and for some bulk crystalline
materials~\cite{Bragthermo}). In order to estimate the internal thermal noise of
a test mass, each resonant mode of the mass can be regarded as a harmonic
oscillator. When the detector operating range is well below the resonances of
the masses, following Saulson~\cite{Saulson2}, the effective spectral density of
thermal displacement of the front face of each mass can be expressed as the
summation of the motion of the various mechanical resonances of the mirror as
also discussed by Gillespie and Raab~\cite{Gillespie}. However, this intuitive
approach to calculating the thermally-driven motion is only valid when the
mechanical loss is distributed homogeneously and, therefore, not valid for real
test-mass mirrors. The mechanical loss is known to be inhomogeneous due to, for
example, the localisation of structural defects and stress within the bulk
material, and the mechanical loss associated with the polished surfaces is
higher than the levels typically associated with bulk effects.  Therefore, Levin
suggested using a direct application of the fluctuation-dissipation theorem to
the optically-sensed position of the mirror substrate surface~\cite{Levin}.
This technique imposes a notional pressure (of the same spatial profile as the
intensity of the sensing laser beam) to the front face of the substrate and
calculates the resulting power dissipated in the substrate on its elastic
deformation under the applied pressure.  Using such an approach we find that
$S_x(f)$ can then be described by the relation
\begin{equation}
 S_x(f) = \frac{2k_\mathrm{B}T}{\pi^2 f^2} \frac{W_{\mathrm{diss}}}{F_0^2},
 \label{eqn:S-x_Levin}
\end{equation}
where $F_0$ is the peak amplitude of the notional oscillatory force and
$W_{\mathrm{diss}}$ is the power dissipated in the mirror described
as,
\begin{equation}
 W_{\mathrm{diss}} = \omega \int{\epsilon(r)\phi(r)\partial V},
 \label{eqn:S-x_Levin2}
\end{equation}
where $\epsilon(r)$ and $\phi(r)$ are the strain and mechanical loss located at
specific positions within the volume $V$. This formalisation highlights the
importance of where mechanical dissipation is located with respect to the
sensing laser beam.  In particular, the thermal noise associated with the
multi-layer dielectric mirror coatings, required for high reflectivity, will in
fact limit the sensitivity of second-generation gravitational-wave detectors at
their most sensitive frequency band, despite these coatings typically being only
$\sim$~4.5~\mum in thickness~\cite{Harry:2002}. Identifying coating
materials with lower mechanical loss, and trying to understand the sources of
mechanical loss in existing coating materials, is a major R\&D effort targeted
at enhancements to advanced detectors and for third generation
instruments~\cite{Martin:2008}.

In order to keep thermal noise as low as possible the mechanical loss factors of
the masses and pendulum resonances should be as low as possible. Further, the
test masses must have a shape such that the frequencies of the internal
resonances are kept as high as possible, must be large enough to accommodate the
laser beam spot without excess diffraction losses, and must be massive enough to
keep the fluctuations due to radiation pressure at an acceptable level. Test
masses currently range in mass from 6~kg for GEO600 to 40~kg for Advanced LIGO.
To approach the best levels of sensitivity discussed earlier the loss factors of
the test masses must be $\simeq$~3~\texttimes~10\super{-8} or lower,
and the loss factor of the pendulum resonances should be smaller than
10\super{-10}.

Obtaining these values puts significant constraints on the choice of material
for the test masses and their suspending fibres. GEO600 utilises very-low--loss
silica suspensions, a technology, which should allow detector sensitivities to
approach the level desired for second generation instruments~\cite{Braginsky1,
Rowan1, Rowan2}, since the intrinsic loss factors in samples of synthetic fused
silica have been measured down to around
5~\texttimes~10\super{-9}~\cite{Ageev:2004}. Still, the use of other
materials such as sapphire is being seriously considered for future
detectors~\cite{Braginsky2, Ju2, Rowan1} such as in
LCGT~\cite{Miyoki:2005, Ohashi:2008}.

The technique of hydroxy-catalysis bonding provides a method of jointing oxide
materials in a suitably low-loss way to allow `monolithic' suspension systems to
be constructed~\cite{Rowan3}. A recent discussion on the level of mechanical
loss and the associated thermal noise in advanced detectors resulting from
hydroxy-catalysis bonds is given by Cunningham et al.~\cite{Cunningham:2010}.
Images of the GEO600 monolithic mirror suspension and of the prototype Advanced
LIGO mirror suspension are shown in Figure~\ref{figure:monolithic}.

%LivRev: do not use eps for html processing!
\epubtkImage{monolithic.png}{%
\begin{figure}[htbp]
  \centerline{\includegraphics[width=\textwidth]{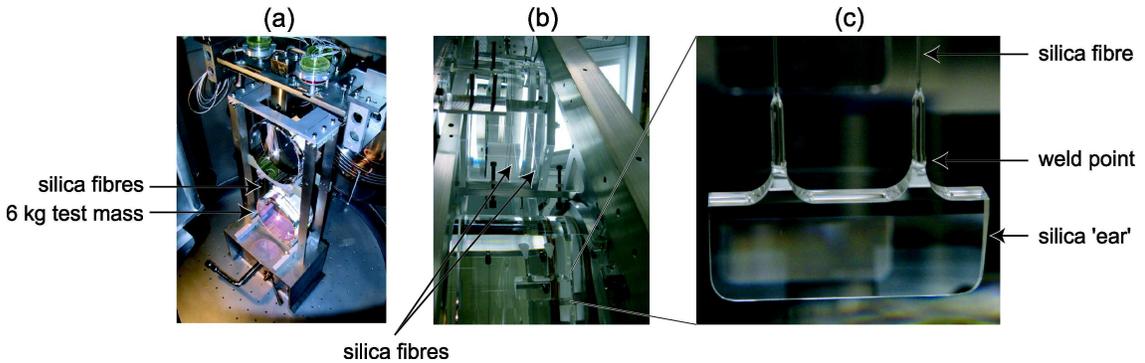}}
  \caption{Monolithic silica suspension of (a) GEO600 6~kg mirror test mass
suspended from 4 fibres of thickness 250~\mum and (b) prototype monolithic
suspension for Advanced LIGO at LASTI (mirror mass of 40~kg, silica fibre
thickness 400~\mum).}
  \label{figure:monolithic}
\end{figure}}

%%%%%%%%%%%%%%%%%%%%%%%%%%%%%%%%%%%%%%%%%%%%%%%%%%%%%%%%%%%%%%%%%%%%%%%%%%%%%%
%%%%%%%%%%%%%%%%%%%%%%%%%%%%%%%%%%%%%%%%%%%%%%%%%%%%%%%%%%%%%%%%%%%%%%%%%%%%%%

\subsection{Quantum noise}
\label{subsection:quantumnoise}

\subsubsection{Photoelectron shot noise}
\label{subsubsection:shotnoise}

For gravitational-wave signals to be detected, the output of the interferometer
must be held at one of a number of possible points on an interference fringe. An
obvious point to choose is halfway up a fringe since the change in photon number
produced by a given differential change in arm length is greatest at this
point (in practice this is not at all a sensible option and interferometers
are operated at, or near, a dark fringe -- see
Sections~\ref{subsection:powerrec} and \ref{sec:readout}). The
interferometer may be stabilised to this point by sensing any changes
in intensity at the interferometer output with a photodiode and
feeding the resulting signal back, with suitable phase, to a
transducer capable of changing the position of one of the
interferometer mirrors.  Information about changes in the length of
the interferometer arms can then be obtained by monitoring the signal
fed back to the transducer.

As mentioned earlier, it is very important that the system used for sensing the
optical fringe movement on the output of the interferometer can resolve strains
in space of 2~\texttimes~10\super{-23}~\Hz or lower, or differences in the
lengths of the two arms of less than $10^{-19} \mathrm{\ m/Hz}^{1/2}$,
a minute displacement compared to the wavelength of light
$\simeq$~10\super{-6}~m. A limitation to the sensitivity of the
optical readout scheme is set by shot noise in the detected photocurrent. From
consideration of the number of photoelectrons (assumed to obey Poisson
statistics) measured in a time $\tau$ it can be shown~\cite{HoughMG5}
that the detectable strain sensitivity depends on the level of laser
power $P$ of wavelength $\lambda$ used to illuminate the
interferometer of arm length $L$, and on the time $\tau$, such that:
\begin{equation}
  \mathrm{detectable\ strain\ in\ time\ } \tau = \frac 1{L}\left[\frac{\lambda h
  c}{2 \pi^{2} P \tau}\right]^{1/2},
  \label{equation:shot1}
\end{equation}
or
\begin{equation}
  \mathrm{detectable\ strain\ }(\mathrm{Hz})^{-1/2} = \frac
  1{L}\left[\frac{\lambda h c}{\pi^{2} P }\right]^{1/2},
  \label{equation:shot2}
\end{equation}
where $c$ is the velocity of light, $h$ is Planck's constant and we assume
that the photodetectors have a quantum efficiency $\simeq$~1. Thus, achievement
of the required strain sensitivity level requires a laser, operating at a
wavelength of 10\super{-6}~m, to provide 6~\texttimes~10\super{6}
power at the input to a simple Michelson interferometer. This is a
formidable requirement; however, there are a number of techniques which
allow a large reduction in this power requirement and these will be
discussed in Section~\ref{section:interferometry}.

%%%%%%%%%%%%%%%%%%%%%%%%%%%%%%%%%%%%%%%%%%%%%%%%%%%%%%%%%%%%%%%%%%%%%%%%%%%%%%

\subsubsection{Radiation pressure noise}
\label{subsubsection:radiationnoise}

As the effective laser power in the arms is increased, another phenomenon
becomes increasingly important arising from the effect on the test masses of
fluctuations in the radiation pressure. One interpretation on the origin of this
radiation pressure noise may be attributed to the statistical uncertainty in how
the beamsplitter divides up the photons of laser light~\cite{Edelstein}. Each
photon is scattered independently and therefore produces an anti-correlated
binomial distribution in the number of photons, $N$, in each arm, resulting in a
$\propto\sqrt{N}$ fluctuating force from the radiation pressure. This is more
formally described as arising from the vacuum (zero-point) fluctuations in the
amplitude component of the electromagnetic field. This additional light entering
through the dark-port side of the beamsplitter, when being of suitable phase,
will increase the intensity of laser light in one arm, while decreasing the
intensity in the other arm, again resulting in anti-correlated variations in
light intensity in each arm~\cite{Caves1, Caves2}. The laser light is
essentially in a noiseless ``coherent state''~\cite{Glauber:1963} as it splits
at the beamsplitter and fluctuations arise entirely from the addition
of these vacuum fluctuations entering the unused port of the beamsplitter. Using
this understanding of the coherent state of the laser, shot noise arises from
the uncertainty in the phase component (quadrature) of the interferometer's
laser field and is observed in the quantum fluctuations in the number of
detected photons at the interferometer output. Radiation pressure noise arises
from uncertainty in the amplitude component (quadrature) of the interferometer's
laser field. Both result in an uncertainty in measured mirror positions.

For the case of a simple Michelson, shown in
Figure~\ref{figure:schematicdetector}, the power spectral density of the
fluctuating motion of each test mass $m$ resulting from fluctuation in the
radiation pressure at angular frequency $\omega$ is given
by~\cite{Edelstein},
\begin{equation}
\delta x^2(\omega) = \biggl(\frac{4 P h}{m^2 \omega^4 c
\lambda}\biggr),
 \label{equ:radiation-pressure1}
\end{equation}
where $h$ is Planck's constant, $c$ is the speed of light and $\lambda$ is the
wavelength of the laser light. In the case of an interferometer with Fabry--P\'{e}rot
cavities, where the typical number of reflections is 50, displacement noise
$\delta x$ due to radiation pressure fluctuations scales linearly with the
number of reflections, such that,
\begin{equation}
\delta x^2(\omega) = 50^2 \times \biggl(\frac{4 P h}{m^2 \omega^4
c \lambda}\biggr).
 \label{equ:radiation-pressure2}
\end{equation}
Radiation pressure may be a significant limitation at low frequency and is
expected to be the dominant noise source in Advanced LIGO between around 10 and
50~Hz~\cite{Harry:2010}. Of course the effects of the radiation pressure
fluctuations can be reduced by increasing the mass of the mirrors, or by
decreasing the laser power at the expense of degrading sensitivity at higher
frequencies.

\subsubsection{The standard quantum limit}
\label{subsubsection:SQL}

Since the effect of photoelectron shot noise decreases when increasing the laser
power as the radiation pressure noise increases, a fundamental limit to
displacement sensitivity is set. For a particular frequency of operation, there
will be an optimum laser power within the interferometer, which minimises the
effect of these two sources of optical noise, which are assumed to be
uncorrelated. This sensitivity limit is known as the Standard Quantum Limit
(SQL) and corresponds to the Heisenberg Uncertainty Principle, in its position
and momentum formulation; see~\cite{Edelstein, Caves1, Caves2, Loudon:1981}.

Firstly, it is possible to reach the SQL at a tuned range of frequencies, when
dominated by either radiation-pressure noise or shot noise, by altering the
noise distribution in the two quadratures of the vacuum field. This effect can
be achieved ``by squeezing the vacuum field''. There are a number of proposed
designs for achieving this in future interferometric detectors, such as a
``squeezed-input interferometer''~\cite{Caves2, Unruh:1983}, a
``variational-output interferometer''~\cite{Vyatchanin:1993} or a
``squeezed-variational interferometer'' using a combination of both techniques.
This technique may be of importance in allowing an interferometer to reach the
SQL at a particular frequency, for example, when using lower levels of laser
power and otherwise being dominated by shot noise. Experiments are under way to
incorporate squeezed-state injection as part of the upgrades to current
gravitational-wave detectors, and where a squeezing injection bench has already
been installed in the GEO600 gravitational-wave detector, which expects to be
able to achieve an up-to-6~dB reduction in shot noise using the current
interferometer configuration~\cite{Vahlbruch:2006}. Similar experiments are also
under way to demonstrate variational readout, where ponderomotive squeezing
arises from the naturally-occurring correlation of radiation-pressure noise to
shot noise upon reflection of light from a
mirror~\cite{Corbitt:2006, Sakata:2006}

Secondly, if correlations exist between the radiation-pressure noise and the
shot-noise displacement limits, then it is possible to bypass the SQL, at least
in principle~\cite{Loudon:1981}.  There are at least two ways by which such
correlations may be introduced into an interferometer.  One scheme is where an
optical cavity is constructed, where there is a strong optical spring effect,
coupling the optical field to the mechanical system.  This is already the case
for the GEO600 detector, where the addition of a signal recycling cavity creates
such correlation, where signal recycling is described in
Section~\ref{subsection:sigrec}. Other schemes of optical springs have been studied,
such as optical bars and optical levers~\cite{Braginsky:1996, Braginsky:1997}.
Another method is to use suitable filtering at optical frequencies of the
output signal, by means of long Fabry--P\'{e}rot cavities, which effectively
introduces correlation~\cite{Kimble:2001, Corbitt:2004}.

%%%%%%%%%%%%%%%%%%%%%%%%%%%%%%%%%%%%%%%%%%%%%%%%%%%%%%%%%%%%%%%%%%%%%%%%%%%%%%%%
%%%%%%%%%%%%%%%%%%%%%%%%%%%%%%%%%%%%%%%%%%%%%%%%%%%%%%%%%%%%%%%%%%%%%%%%%%%%%%%%

%\newpage

\section{Laser Interferometric Techniques for Gravitational-Wave Detectors}
\label{section:interferometry}

As explained in Section~\ref{subsubsection:shotnoise}, high-power laser
light is needed to overcome limitations of a detector's sensitivity due to
photoelectron shot noise. The situation can be helped greatly if a multi-pass
arrangement is used in the arms of the interferometer as this multiplies up the
apparent movement by the number of bounces the light makes in the arms. The
multiple beams can either be separate, as in an optical delay line~\cite{Weiss,
Billing}, or may lie on top of each other as in a Fabry--P\'{e}rot resonant
cavity~\cite{Drever2}, as shown in Figure~\ref{figure:Michelsons}.

\epubtkImage{fig5e.png}{%
\begin{figure}[htbp]
  \centerline{\includegraphics[width=\textwidth]{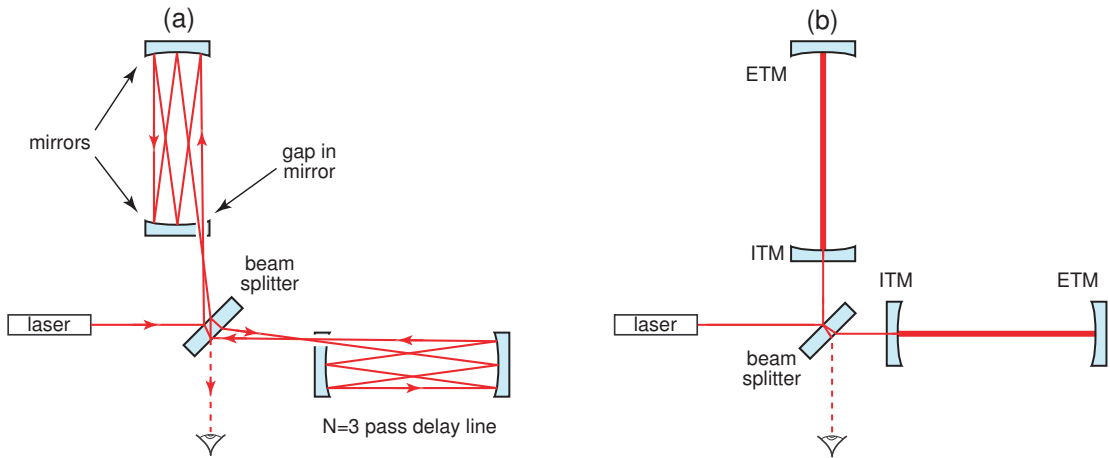}}
  \caption{Michelson interferometers with (a) delay lines and (b)
Fabry--P\'{e}rot cavities in the arms of the interferometer.}
  \label{figure:Michelsons}
\end{figure}}

Optimally, the light should be stored for a time comparable to the
characteristic timescale of the signal. Thus, if signals of characteristic
timescale 1~msec are to be searched for, the number of bounces should be
approximately 50 for an arm length of 3~km. With 50 bounces the required laser
power is reduced to 2.4~\texttimes~10\super{3}~W, still a formidable
requirement.

%%%%%%%%%%%%%%%%%%%%%%%%%%%%%%%%%%%%%%%%%%%%%%%%%%%%%%%%%%%%%%%%%%%%%%%%%%%%%%%%
%%%%%%%%%%%%%%%%%%%%%%%%%%%%%%%%%%%%%%%%%%%%%%%%%%%%%%%%%%%%%%%%%%%%%%%%%%%%%%%%

\subsection{Power recycling}
\label{subsection:powerrec}

It can be shown that an optimum signal-to-noise ratio in a Michelson interferometer
can be obtained when the arm lengths are such that the output light is very
close to a minimum (this is not intuitively obvious and is discussed more fully
in~\cite{Edelstein}). Thus, rather than lock the interferometer to the side of a
fringe as discussed above in Section~\ref{subsubsection:shotnoise}, it is usual
to make use of a modulation technique to operate the interferometer close to a
null in the interference pattern. An electro-optic phase modulator placed in
front of the interferometer can be used to phase modulate the input laser light.
If the arms of the interferometer are arranged to have a slight mismatch in
length this results in a detected signal, which, when demodulated, is zero with
the cavity exactly on a null fringe and changes sign on different sides of the
null providing a bipolar error signal; this can be fed back to the transducer
controlling the interferometer mirror to hold the interferometer locked near to
a null fringe (this is the RF readout scheme discussed in Section~\ref{sec:readout}).

In this situation, if the mirrors are of very low optical loss, nearly all of the
light supplied to the interferometer is reflected back towards the laser. In
other words the laser is not properly impedance matched to the interferometer.
The impedance matching can be improved by placing another mirror of correctly
chosen transmission -- a power recycling mirror -- between the laser and the
interferometer so that a resonant cavity is formed between this mirror and the
rest of the interferometer; in the case of perfect impedance matching, no light
is reflected back towards the laser~\cite{Drever3, Schilling}. There is then a
power build-up inside the interferometer as shown in 
Figure~\ref{figure:Michelsons2a}. This can be high enough to create the required
kilowatts of laser light at the beamsplitter, starting from an input laser light
of only about 10~W.

\epubtkImage{fig6a.png}{%
\begin{figure}[htbp]
  \centerline{\includegraphics[width=0.5\textwidth]{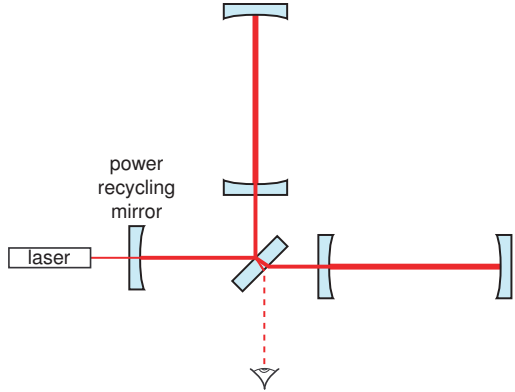}}
  \caption{The implementation of power recycling on a
    Michelson interferometer with Fabry--P\'{e}rot cavities.}
  \label{figure:Michelsons2a}
\end{figure}}

To be more precise, if the main optical power losses are those associated with
the test mass mirrors -- taken to be A per reflection -- the intensity inside
the whole system considered as one large cavity is increased by a factor given
by $(\pi L)/(c A \tau)$, where the number of bounces, or light storage time, is
optimised for signals of timescale $\tau$ and the other symbols have their usual
meaning. Then:
\begin{equation}
  \mathrm{detectable\ strain\ in\ time\ } \tau = \left( \frac{\lambda h
  A}{4 \pi L P \tau^2} \right)^{1/2}.
  \label{equation:shotpower}
\end{equation}

%%%%%%%%%%%%%%%%%%%%%%%%%%%%%%%%%%%%%%%%%%%%%%%%%%%%%%%%%%%%%%%%%%%%%%%%%%%%%%%%
%%%%%%%%%%%%%%%%%%%%%%%%%%%%%%%%%%%%%%%%%%%%%%%%%%%%%%%%%%%%%%%%%%%%%%%%%%%%%%%%

\subsection{Signal recycling}
\label{subsection:sigrec}

To enhance further the sensitivity of an interferometric detector and to allow
some narrowing of the detection bandwidth, which may be valuable in searches for
continuous wave sources of gravitational radiation, another technique known as
signal recycling can be implemented~\cite{Meers, Strain, Heinzel}. This relies
on the fact that sidebands created on the light by gravitational-wave signals
interacting with the arms do not interfere destructively and so do appear at the
output of the interferometer. If a mirror of suitably-chosen reflectivity is put
at the output of the system as shown in Figure~\ref{figure:Michelsons2b}, then the
sidebands can be recycled back into the interferometer, where they resonate, and
hence the signal size over a given bandwidth (set by the mirror reflectivity) is
enhanced.

\epubtkImage{fig6b.png}{%
\begin{figure}[htbp]
  \centerline{\includegraphics[width=0.5\textwidth]{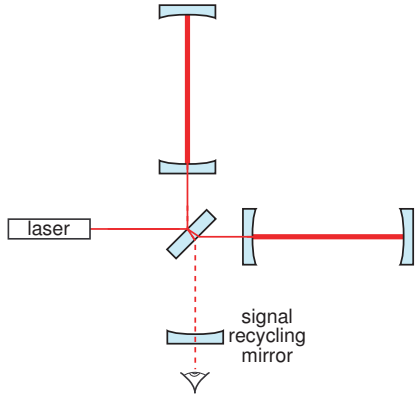}}
  \caption{The implementation of signal recycling on a
    Michelson interferometer with Fabry--P\'{e}rot cavities.}
  \label{figure:Michelsons2b}
\end{figure}}

The centre of this frequency band is set by the precise length of the cavity
formed by the signal recycling mirror and the cavities in the interferometer
arms. Thus, control of the precise position of the signal recycling mirror allows
tuning of the frequency at which the performance is peaked.

Often signal recycling will be used to provide a narrow bandwidth to search for
continuous wave sources as mentioned above, however it may also be used with a
relatively broad bandwidth, centred away from zero frequency, and this
application is useful for matching the frequency response of the detector to
expected spectral densities of certain broadband or ``chirping'' signals.

%%%%%%%%%%%%%%%%%%%%%%%%%%%%%%%%%%%%%%%%%%%%%%%%%%%%%%%%%%%%%%%%%%%%%%%%%%%%%%%%

\subsection{Application of these techniques}
\label{subsection:application}

Using appropriate optical configurations that employ power and signal recycling
as described in Sections~\ref{subsection:powerrec} and~\ref{subsection:sigrec}, the required laser power may thus be reduced
to a level (in the range of 10 to 100~W) where laser sources are now available;
however stringent requirements on technical noise must be satisfied.

%%%%%%%%%%%%%%%%%%%%%%%%%%%%%%%%%%%%%%%%%%%%%%%%%%%%%%%%%%%%%%%%%%%%%%%

\subsubsection{Technical noise requirements}
\label{subsubsection:lasernoise}

\begin{itemize}
\item \textbf{Power fluctuations} \\
As described above in Section~\ref{subsection:powerrec} gravitational-wave
interferometers are typically designed to operate with the length of the
interferometer arms set such that the output of the interferometer is close to a
minimum in the output light. If the interferometer were operated exactly at the
null point in the fringe pattern, then, in principle, it would be insensitive to
power fluctuations in the input laser light. However in practice there will be
small offsets from the null position causing some sensitivity to this noise
source.  In this case, it can be shown~\cite{Hough} that the required power
stability of the laser in the frequency range of interest for gravitational-wave
detection may be estimated to be
\begin{equation}
  \frac{\delta P}{P} \simeq h (\delta L/L)^{-1},
  \label{equation:intnoise1}
\end{equation}
where $\delta P/P$ are the relative power fluctuations of the laser, and $\delta
L$ is the offset from the null fringe position for an interferometer of arm
length $L$.  From calculations of the effects of low-frequency seismic noise for
the initial designs of long baseline detectors~\cite{Hough} it can be estimated
that the rms motion will be on the order of 10\super{-13}~m when the system is
operating. Taking strains of around 3~\texttimes~10\super{-24}~\Hz at
300~Hz, as is the case for Advanced LIGO~\cite{Harry:2010}, requires
power fluctuations of the laser not to exceed
\begin{equation}
  \frac{\delta P}{P} \leq 10^{-7} \mathrm{\ Hz}^{-1/2} \quad \mathrm{at}
  \simeq 300 \mathrm{\ Hz}.
  \label{equation:intnoise2}
\end{equation}
To achieve this level of power fluctuation typically requires the use of active
stabilisation techniques of the type developed for argon ion
lasers~\cite{Mangan}.
However, it should be noted that the most stringent constraint on laser-power fluctuations
in future ground-based detectors, where circulating laser powers will approach 1~MW or even beyond,
will ultimately arise from classical radiation pressure on the mirrors, as described in
Section~\ref{subsubsection:radiationnoise}.

\item \textbf{Frequency fluctuations} \\
For a simple Michelson interferometer it can be shown that a change $\delta x$
in the differential path length, $x$, of the interferometer arms causes a phase
change $\delta \phi$ in the light at the interferometer output given by $\delta
\phi = (2\pi/c) (\nu \delta x + x \delta\nu)$ where $\delta \nu$ is a change in
the laser frequency $\nu$ and $c$ is the speed of light. It follows that if
the lengths of the interferometer arms are exactly equal (i.e.,\ $x = 0$), the
interferometer output is insensitive to fluctuations in the frequency of the
input laser light, provided that, in the case of Fabry--P\'{e}rot cavities in the
arms, the fluctuations are not so great that the cavities cannot remain on
resonance.  However, in practice, differences in the optical properties of the
interferometer mirrors result in  slightly different effective arm lengths, a
difference of perhaps a few tens of metres. Then the relationship between
the limit to detectable gravitational-wave amplitude and the fluctuations
$d\nu$ of the laser frequency $\nu$ is given by~\cite{Hough}
\begin{equation}
  \frac{\delta \nu}{\nu} \simeq h(x/L)^{-1}.
  \label{equation:frequnoise}
\end{equation}
Hence, to achieve the target sensitivity used in the above calculation using a
detector with arms of length 4~km, maximal fractional frequency fluctuations of,
\begin{equation}
  \frac{\delta \nu}{\nu} \leq 10^{-21} \mathrm{\ Hz}^{-1/2},
  \label{equation:freqspecification}
\end{equation}
are required. This level of frequency noise may be achieved by the use of
appropriate laser-frequency stabilisation systems involving high finesse
reference cavities~\cite{Hough}.

Although the calculation here is for a simple Michelson
interferometer, similar arguments apply to the more sophisticated
systems with arm cavities, power recycling and signal recycling
discussed earlier and lead to the same conclusions. Frequency
stabilisation is also important in other applications such as in high
resolution optical spectroscopy~\cite{Rafac:2000}, optical frequency
standards~\cite{Ludlow:2006, Webster:2004} and fundamental quantum
measurements~\cite{Schmidt-Kaler:2003}. The best reference cavities,
such as those developed by the Ye~\cite{Notcutt:2006} and
H\"{a}nsch~\cite{Alnis:2008} groups, have reported a frequency stability
performance of around 10\super{-16}, which is broadly equivalent to that
achieved for ground-based gravitational-wave detectors when scaling by
cavity length.

\item \textbf{Beam geometry fluctuations} \\
Fluctuations in the lateral or angular position of the input laser beam, along
with changes in its size and variations in its phase-front curvature may all
couple into the output signal of an interferometer and reduce its sensitivity.
These fluctuations may be due to intrinsic laser mechanical noise (from water
cooling for example) or from seismic motion of the laser with respect to the
isolated test masses. As an example of their importance, fluctuations in the
lateral position of the beam may couple into interferometer measurements through
a misalignment of the beamsplitter with respect to the interferometer mirrors. A
lateral movement $\delta z$ of the beam incident on the beamsplitter, coupled
with an angular misalignment of the beamsplitter of $\alpha/2$ results in a
phase mismatch $\delta \phi$ of the interfering beams, such that~\cite{Rudiger}
\begin{equation}
  \delta \phi = (4 \pi/\lambda) \alpha \delta z.
  \label{equation:beamgeomfluc}
\end{equation}
A typical beamsplitter misalignment of $\simeq$~10\super{-7} radians means that to
achieve sensitivities of the level described above using a detector with 3 or
4~km arms, and 50 bounces of the light in each arm, a level of beam geometry
fluctuations at the beamsplitter of close to 10\super{-12}~m/\Hz at
300~Hz is required.

Typically, and ignoring possible ameliorating effects of the power recycling
cavity on beam geometry fluctuations, this will mean that the beam positional
fluctuations of the laser need to be suppressed by several orders of magnitude.
The two main methods of reducing beam geometry fluctuations are 1) passing the
input beam through a single mode optical fibre~\cite{Meersphd} and 2) using a
resonant cavity as a mode cleaner~\cite{Rudiger, Skeldon, Willke, Araya}.

Passing the beam through a single mode optical fibre helps to eliminate beam
geometry fluctuations, as deviations of the beam from a Gaussian TEM\sub{00} mode are
equivalent to higher-order spatial modes, which are thus attenuated by the
optical fibre.  However, there are limitations to the use of optical fibres,
mainly due to the limited power-handling capacity of the fibres; care must also
be taken to avoid introducing extra beam geometry fluctuations from movements of
the fibre itself.

A cavity may be used to reduce beam geometry fluctuations if it is adjusted to
be resonant only for the TEM\sub{00} mode of the input light. Any higher order modes
should thus be suppressed~\cite{Rudiger}. The use of a resonant cavity should
allow the handling of higher laser powers and has the additional benefits of
acting as a filter for fast fluctuations in laser frequency and
power~\cite{Skeldon, Willke}. This latter property is extremely useful for the
conditioning of the light from some laser sources as will be discussed below.
\end{itemize}

%%%%%%%%%%%%%%%%%%%%%%%%%%%%%%%%%%%%%%%%%%%%%%%%%%%%%%%%%%%%%%%%%%%%%%%%%%%%%%%%

\subsubsection{Laser design}
\label{subsubsection:laserdesign}

From Equation~(\ref{equation:shot1}) it can be seen that the photon-noise
limited sensitivity of an interferometer is proportional to $\sqrt{P}$ where $P$
is the laser power incident on the interferometer, and $\sqrt{\lambda}$ where
$\lambda$ is the wavelength of the laser light. Thus, single frequency lasers of
high output power and short wavelength are desirable. With these constraints in
mind, laser development started on argon-ion lasers and Nd:YAG lasers.
Argon-ion lasers emitting light at 514~nm were used to illuminate several
interferometric gravitational-wave detector prototypes, see, for
example,~\cite{Shoemaker, Robertson}.  However, their efficiency, reliability,
controllability and noise performance has ruled them out as suitable laser
sources for current and future gravitational wave detectors.

Nd:YAG lasers, emitting at 1064~nm or frequency doubled to 532~nm, present an
alternative. The longer (infrared) wavelength may initially appear less
desirable than the 514~nm of the argon-laser and the frequency doubled
532~nm, as more laser power is needed to obtain the same sensitivity.  In
addition, the resulting increase in beam diameter leads to a need for larger
optical components. For example in an optical cavity the diameter of the beam at
any point is proportional to the square root of the wavelength~\cite{Kogelnik}
and to keep diffractive losses at each test mass below
1~\texttimes~10\super{-6}, it can be shown that the diameter of each
test mass must be greater than 2.6~times the beam diameter at the test
mass. Thus, the test masses for gravitational wave detectors have to be
1.4 times larger in diameter for infrared than for green light. However, Nd:YAG
sources at 1064~nm have demonstrated some compelling
advantages, in particular the demonstration of scaling the power up to
levels suitable for second generation interferometers ($\sim$~200~W)
combined with their superior
efficiency~\cite{Shine,Vogt,Kerr}. Frequency-doubled Nd:YVO lasers at
532~nm have currently only been demonstrated to powers approaching
20~W and have not been actively stabilised to the levels needed for
gravitational-wave detectors~\cite{Mavalvala:2010}. An additional
problem associated with shorter wavelength operation is the potential
for increased absorption, possibly leading to photochemistry (damage)
in the coating materials, in addition to increased scatter. For this
reason, all the initial long-baseline interferometer projects,
along with their respective upgrades, have chosen some form of Nd:YAG
light source at 1064~nm.

As an example, the laser power is being upgraded from 10~W in initial LIGO to
180~W for Advanced LIGO to improve the SNR of the shot-noise--limited regime.
This power will be delivered by a three stage injection-locked oscillator
scheme~\cite{Cregut, Nabors, Golla, Frede:2005}.  The first stage uses a
monolithic non-planar ring oscillator (NPRO) to initially produce 2~W of output
power.  This output is subsequently amplified by a four-head Nd:YVO laser
amplifier to a power of 35~W~\cite{Frede:2007}, which is in turn delivered into
an injection locked Nd:YAG oscillator to produce 200~W of output
power~\cite{Wilke:2008}.

Other laser developments are being pursued, such as high-power--fibre
lasers, which are currently being investigated by the AEI in
Germany~\cite{Schnabel:2010} and prototyped for Advanced VIRGO by
Gr\'{e}verie et al.\ in France~\cite{Greverie:2010}. Fibre amplifiers
show great potential for extrapolation to higher laser powers in
addition to lower production costs.

Third-generation interferometric gravitational-wave detectors, such as the
Einstein Telescope, require input laser powers of around 500~W at 1064~nm in
order to achieve their high-frequency shot-noise--limited
sensitivities~\cite{Hild:2010}.  Low-frequency sensitivity is expected to be
achieved through the use of separate low-power interferometers with silicon
optics operating at cryogenic temperatures~\cite{Rowan:2003, Punturo:2010}.
Longer wavelengths are proposed here due to excessive absorption in silicon at
1064~nm and the expected low absorption (less than 0.1~ppm/cm) at
around 1550~nm~\cite{Green:1995}. Worldwide laser developments may
provide new baseline light sources that can provide different
wavelength and power options for future detectors. However, the
stringent requirements on the temporal and spatial stability for
gravitational-wave detectors are beyond that sought in other laser
applications. Therefore, a dedicated laser-development program will be
required to continue beyond the second-generation interferometers in
order to design and build a laser system that meets third-generation
requirements, as discussed in more detail in~\cite{Mavalvala:2010}.

Another key area of laser development, targeted at improving the sensitivity of
future gravitational-wave detectors, is in the use of special optical modes to reduce thermal noise. It can be shown that the amplitude of thermal
noise associated with the mirror coatings is inversely proportional to the beam
radius~\cite{Nakagawa:2002}. The configuration within current interferometers is
designed to inject and circulate TEM\sub{00} optical modes, which have a
Gaussian beam profile.  To keep diffraction losses suitably low for this case
($<$~1~ppm), a beam radius of a maximum size $\sim$~35\% of the radius of the test mass
mirror can be used. The thermal noise could be further reduced if optical modes
are circulated that have a larger effective area, yet not increasing the level
of diffraction losses. This would be possible through the use of higher-order
Laguerre-Gauss beams, and other ``exotic'' beams, such as mesa or conical beams.
A more in-depth discussion of how these optical schemes can be implemented and
the potential increase in detector sensitivity attainable can be found
in~\cite{Vinet:2009}.

\subsubsection{Thermal compensation and parametric instabilities}
\label{subsubsection:thermalcomp}

Despite the very-low levels of optical absorption in fused silica at 1064~nm,
thermal loading due to high-levels of circulating laser powers within
advanced gravitational-wave detectors will cause significant thermal loading. In
the case of Advanced LIGO, thermal lensing will be most significant in the input
test masses of the Fabry--P\'{e}rot cavities, where the beam must transmit through the
substrate in addition to the high-power within the cavity being incident on the
coating surfaces. Thermal distortion in the optics will be sensed by Hartmann
sensors and coupled to two schemes of thermal compensation. Firstly, ring
resistance heaters will be installed around the barrel of the input mass in
order to compensate for the beam heating the central region of the optics, as
demonstrated for radius of curvature tuning in GEO600~\cite{Luck:2004}.
Secondly, a flexible CO\sub{2} laser based system will be used to deposit heat onto
the reaction mass (otherwise called the compensation plate) for the input test
mass, as demonstrated in initial LIGO~\cite{Lawrence:2002,Waldman:2006}. The
laser beam shape and intensity can be easily modified from outside with the vacuum
system and can therefore adapt to non-uniformities in the absorption and other
changes in the interferometer's thermal state.

It should also be noted that energy can couple from the optical modes resonating
in the interferometer Fabry--P\'{e}rot cavities and the acoustic modes of the test
masses. When there is sufficient coupling between these optical and mechanical
modes, and the mechanical modes have a suitably high-quality factor, then
mechanical resonances can be `rung-up' by the large circulating laser power to
the point where the interferometer is no longer stable, a phenomenon called
parametric instabilities~\cite{Braginsky:2001}. Mechanical dampers that are
tuned to damp at high-frequency yet not significantly increasing thermal noise
at low-frequency are being considered in possible upgrades to advanced
detectors, in addition to other schemes, such as active feedback to damp
problematic modes provided through the electrostatic actuators.

%%%%%%%%%%%%%%%%%%%%%%%%%%%%%%%%%%%%%%%%%%%%%%%%%%%%%%%%%%%%%%%%%%%%%%%%%%%%%%%%

\subsection{Readout schemes}
\label{sec:readout}

There are various schemes that can be applied to readout the
gravitational-wave signal from an interferometer. A good discussion of
some of these can be found in~\cite{Hild:2009}. If the interferometer
laser has a frequency of $f_{\mathrm{l}}$ then a passing
gravitational wave, with frequency $f_{\mathrm{gw}}$, will introduce
sidebands onto the laser with a frequency of $f_{\mathrm{s}} =
f_{\mathrm{l}} \pm f_{\mathrm{gw}}$. A readout scheme must be able to
decouple the gravitational-wave component, with frequencies of order
$\sim$~100~Hz, from the far higher laser frequency at hundreds of
tera-Hz. To do this it needs to be able to compare the sideband
frequency with a known stable optical local oscillator. Ideally this
oscillator would be the laser light itself (a homodyne scheme), but
the initial generation of gravitational-wave detectors are operated at
a dark fringe (i.e., the interferometer is held, so as the light from
the arms completely destructively interferes at the beam splitter), so
no light at the laser frequency exits (a gravitational wave will alter
the arms lengths and constructively interfere, causing light to exit,
but only at the sideband frequency).

The standard scheme used by the initial interferometers is a radio frequency (RF) heterodyne
readout. In this case the laser light is modulated at an RF
(called Schnupp modulation~\cite{Schnupp:1988}) prior to entering the
interferometer arms, giving rise to sidebands offset from the laser frequency at
the RF. The interferometer is set up to allow these RF sidebands to exit at
the output port. This can be used as a local optical oscillator with which to
demodulate the gravitational wave sidebands. However, the demodulation will
introduce a beat between the RF and the gravitational wave frequency, which must
be removed by a second (hence \textit{hetero}dyne) demodulation at the RF.

The preferred method for future detectors is a DC scheme
(see~\cite{Fritschel:2003, Ward:2008, Hild:2009} for motivations and
advantages of using such a scheme). In this no extra modulation has to
be applied to the light. Instead the interferometer is held just off
the dark fringe, so some light at the laser frequency reaches the
output to serve as the local oscillator.

%%%%%%%%%%%%%%%%%%%%%%%%%%%%%%%%%%%%%%%%%%%%%%%%%%%%%%%%%%%%%%%%%%%%%%%%%%%%%%%%
%%%%%%%%%%%%%%%%%%%%%%%%%%%%%%%%%%%%%%%%%%%%%%%%%%%%%%%%%%%%%%%%%%%%%%%%%%%%%%%%
%%%%%%%%%%%%%%%%%%%%%%%%%%%%%%%%%%%%%%%%%%%%%%%%%%%%%%%%%%%%%%%%%%%%%%%%%%%%%%%%

\newpage

\section{Operation of First-Generation Long-Baseline Detectors}
\label{section:construction}

Prior to the start of the 21st century there existed several prototype
laser interferometric detectors, constructed by various research groups around
the world -- at the Max-Planck-Instit\"ut f\"ur Quantenoptik in
Garching~\cite{Shoemaker}, at the University of Glasgow~\cite{Robertson}, at the
California Institute of Technology~\cite{Abramovici}, at the Massachusetts
Institute of Technology~\cite{Fritschel2}, at the Institute of Space and
Astronautical Science in Tokyo~\cite{Mizuno} and at the astronomical observatory
in Tokyo~\cite{Araya}. These detectors had arm lengths varying from 10~m to
100~m and had either multibeam delay lines or resonant Fabry--P\'{e}rot cavities in
their arms. The 10~m detector that used to exist at Glasgow is shown in
Figure~\ref{figure:Glasgowprototype}.

\epubtkImage{figpro.jpg}{%
\begin{figure}[htbp]
  \centerline{\includegraphics[scale=0.4]{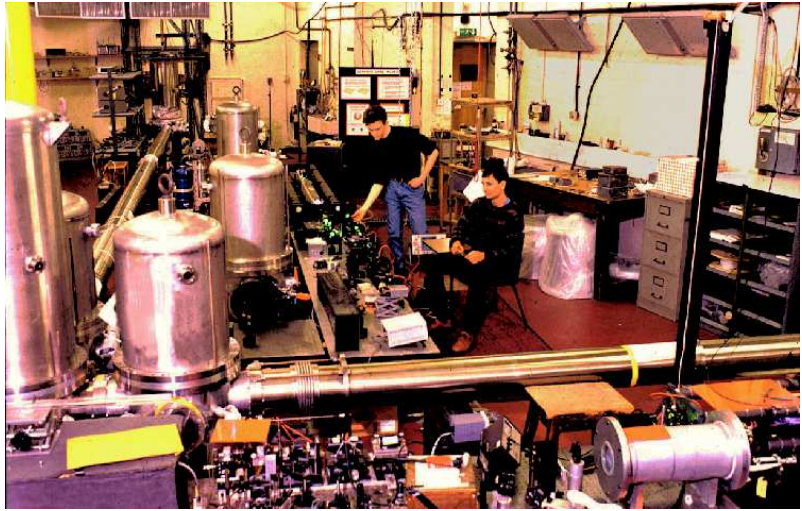}}
  \caption{The 10~m prototype gravitational wave detector
    at Glasgow.}
  \label{figure:Glasgowprototype}
\end{figure}}

The sensitivities of some of these detectors reached a level -- better than
10\super{-18} for millisecond bursts -- such that the technology could be
considered sufficiently mature to propose the construction of detectors of much
longer baseline that would be capable of reaching the performance required to
have a real possibility of detecting gravitational waves.  An international
network of such long baseline gravitational wave detectors has now been
constructed and commissioned, and science-quality data from these has been
produced and analysed since 2002 (see Section~\ref{subsection:runs} and
Section~\ref{subsection:results} for a review of recent science data runs and
results).

The American LIGO project~\cite{LIGOweb} comprises two detector systems with
arms of 4~km length, one in Hanford, Washington, and one in Livingston,
Louisiana (also known as the LIGO Hanford Observatory 4k [LHO~4k] and LIGO
Livingston Observatory 4k [LLO~4k], or H1 and L1, respectively). One half length,
2~km, interferometer was also contained inside the same evacuated enclosure at
Hanford (also known as the LHO~2k, or H2). The design goal of the 4~km
interferometers was to have a peak strain sensitivity between 100\,--\,200~Hz of
$\sim$~3~\texttimes~10\super{-23}~\Hz~\cite{LIGOSRD} (see
Figure~\ref{figure:LIGOstrains}), which was achieved during the fifth science run
(Section~\ref{subsection:runs}). A birds-eye view of the Hanford site showing the
central building and the directions of the two arms is shown in
Figure~\ref{figure:LIGOsite}. In October 2010 the LIGO detectors shut down and
decommissioning began in preparation for the installation of a more sensitive
instrument known as Advanced LIGO (see Section~\ref{subsection:aligo}).

\epubtkImage{fig7.jpg}{%
\begin{figure}[htbp]
  \centerline{\includegraphics[width=12cm]{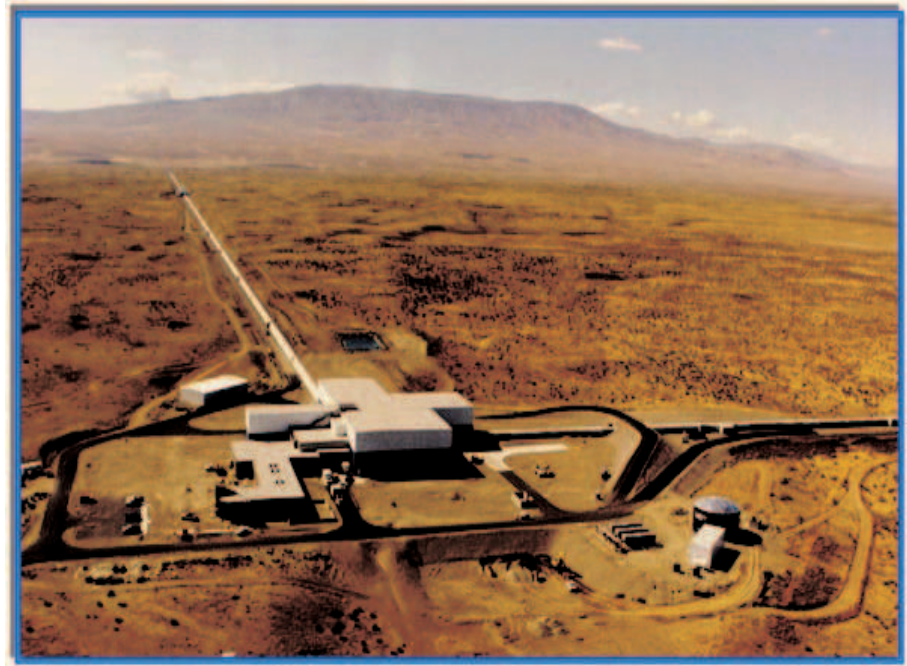}}
  \caption{A bird's eye view of the LIGO detector, sited in
    Hanford, Washington.}
  \label{figure:LIGOsite}
\end{figure}}

The French/Italian Virgo project~\cite{VIRGOweb} comprises a single
3~km arm-length detector at Cascina near Pisa. As mentioned earlier,
it is designed to have better performance than the other detectors,
down to 10~Hz.

The TAMA300 detector~\cite{TAMAweb}, which has arms of length 300~m, at the
Tokyo Astronomical Observatory was the first of the ``beyond-prototype''
detectors to become operational. This detector is built mainly underground and
partly has the aim of adding to the gravitational-wave detector network for
sensitivity to events within the local group of galaxies, but is primarily a
test bed for developing techniques for future larger-scale detectors. Initial
operation of the interferometer was achieved in 1999 and power recycling was
implemented for data taking in 2003~\cite{Arai:2003}.

All the systems mentioned above are designed to use resonant cavities in the
arms of the detectors and use standard wire-sling techniques for suspending the
test masses. The German/British detector, GEO600~\cite{GEOweb}, built near
Hannover, Germany, is somewhat different. It makes use of a four-pass delay-line
system with advanced optical signal-enhancement techniques, utilises very-low
loss-fused silica suspensions for the test masses, and, despite its smaller size,
was designed to have a sensitivity at frequencies above a few hundred Hz
comparable to the first phases of Virgo and LIGO during their initial operation.
It uses both power recycling (Section~\ref{subsection:powerrec}) and tunable signal
recycling (Section~\ref{subsection:sigrec}), often referred to together as dual
recycling.

To gain the most out of the detectors as a true network, data sharing and joint
analyses are required. In the summer of 2001 the LIGO and GEO600 teams signed a
Memorandum of Understanding (MoU), under the auspices of the LIGO Scientific
Collaboration (LSC)~\cite{LSCweb}, allowing complete data sharing between the
two groups. Part of this agreement has been to ensure that both LIGO and
GEO600 have taken data in coincidence (see below). Coincident data taking, and
joint analysis, has also occurred between the TAMA300 project and the LSC
detectors. The Virgo collaboration also signed an MoU with the LSC, which has
allowed data sharing since May 2006.

The operation and commissioning of these detectors is a continually-evolving
process, and the current state of this review only covers developments until
late-2010. For the most up-to-date information on detectors readers are advised
to consult the proceedings of the Amaldi meetings, GWDAW/GWPAW (Gravitational
Wave Data Analysis Workshops), and GWADW (Gravitational Wave Advanced Detectors
Workshops) -- see~\cite{confs} for a list of past conferences.

For the first and second generations of detector, much effort has gone into
estimating the expected number of sources that might be observable given their
design sensitivities. In particular, for what are thought to be the strongest
sources: the coalescence of neutron-star binaries or black holes (see
Section~\ref{sec:cbc} for current rates as constrained by observations). These
estimates, based on observation and simulation, are summarised in
Table~5 of~\cite{Abadie:2010e} and the \textit{realistic} rates
suggest initial detectors would expect to see 0.02, 0.004 and 0.007
events per year for neutron-star--binary, black-hole--neutron-star, and
black-hole--binary systems, respectively (it should be noted that there
is a range of possible rates consistent with current observations and
models)\epubtkFootnote{In terms of event rates the current best estimates
  for neutron-star--binary merger rates, based on the known population
  of neutron-star--binary systems, gives a 95\% confidence interval
  between 1\,--\,1000~\texttimes~10\super{-6} per year per Milky Way
  Equivalent Galaxy (MWEG), where MWEG is equivalent to a volume that
  contains a blue light luminosity with $L = 9\times10^9\,L_{\odot}$
  (MWEG was used in the S1 and S2 LIGO search, but was then changed to
  the $L_{10}$ unit, where $L_{10}$ is given as 10\super{10} times the
  blue-light luminosity of the sun, although there is only a 10\%
  difference between the two),~\cite{Abadie:2010e, Kalogera:2004a,
    Kalogera:2004b}, with a peak in the distribution at
  100~\texttimes~10\super{-6} per year per MWEG -- or $\approx$~0.02 per year
  for initial LIGO at design sensitivity. The expected rate of black-hole binary systems, or black-hole--neutron-star systems is far harder
  to infer as none have been observed, but estimates can be made on
  the population for a wide variety of models and give a 95\%
  confidence range of 0.05\,--\,100~\texttimes~10\super{-6} per year per
  MWEG and 0.01\,--\,30~\texttimes~10\super{-6} per year per MWEG
  respectively~\cite{Abadie:2010e, OShaughnessy:2005,
    OShaughnessy:2008, Abbott:2008a}. As an example of how to convert
  from rates to event numbers, cumulative blue-light luminosities with
  respect to distance from the Earth in Mpcs, and the horizon
  distances of the LIGO detectors from S2 through to S4, can be seen
  in Figure~3 of~\cite{Abbott:2008a}.}. Second generation detectors
(see Section~\ref{subsection:aligo}), which can observe approximately
1000 time more volume than the initial detectors might, expect to see
40, 10, and 20 per year for the same sources. With such rates a great
deal of astrophysics could be possible (see~\cite{Sathyaprakash:2009}
for examples).

\subsection{Science runs}
\label{subsection:runs}

Over the last decade the commissioning and improvement of the various
gravitational-wave detectors has been suspended at various stages to take data
for astrophysical analysis. These have been times when it was considered that
the detectors were sensitive and stable enough (or had made sufficient
improvements over earlier states) to make astrophysical searches worthwhile.
Within the LSC these have been called the \textit{Science} (S) runs, for Virgo they
have been the \textit{Virgo Science Runs} (VSR), and for TAMA300 they have been
the \textit{Data Taking} (DT) periods. A time-line of science runs for the various
interferometric detectors, can be seen in Figure~\ref{figure:runtimes}.

\epubtkImage{runtimes.png}{%
\begin{figure}[htbp]
  \centerline{\includegraphics[width=\textwidth]{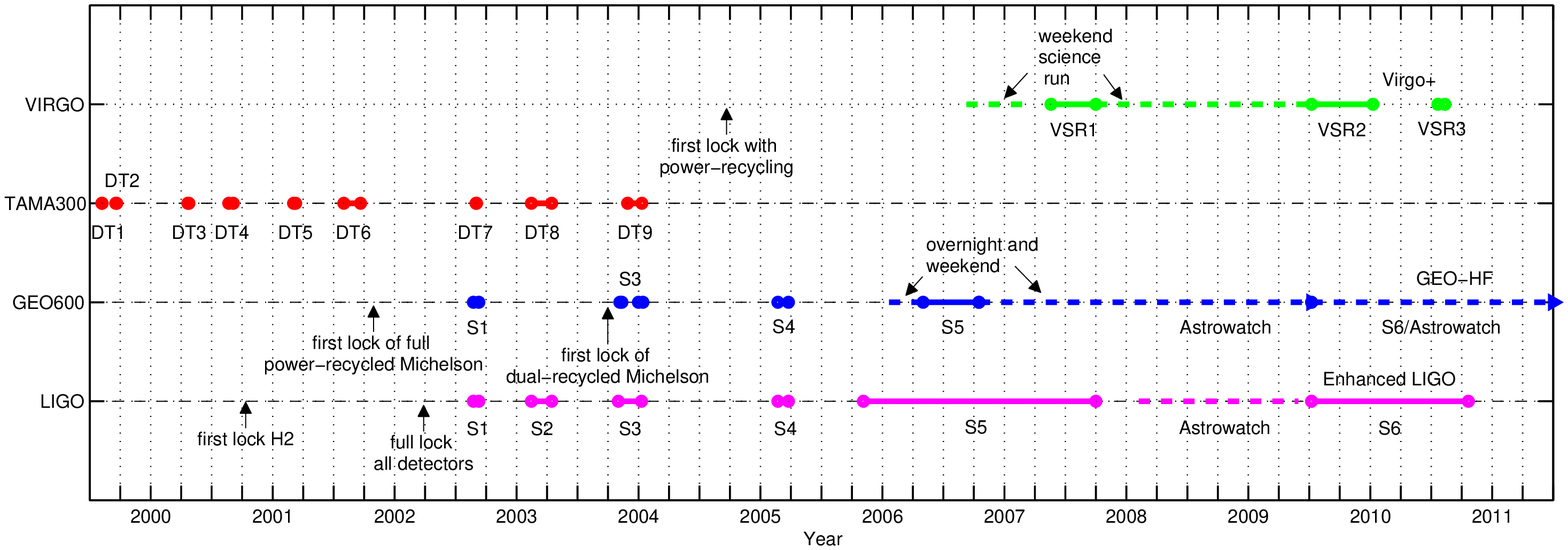}}
  \caption{A time-line of the science runs of the first generation
interferometric gravitational-wave detectors, from their first lock to
mid-2011.}
  \label{figure:runtimes}
\end{figure}}

A figure of merit for the sensitivity of a detector is to calculate
its \textit{horizon distance}. This is the maximum range out to which
it could see the coalescence of two $1.4\,M_{\odot}$ neutron stars
that are optimally oriented and located (i.e., with the orbital plane
perpendicular to the line-of-sight, and with this plane parallel to
the detector plane, so that the antenna response is at its maximum) at
a signal-to-noise ratio of 8~\cite{Abbott:2005b}. The horizon distance
can be converted to a range that is an average over all sky locations
and source orientations (i.e.\, not the best case scenario) by dividing
it by 2.26~\cite{Sutton:2003}) -- we shall use this angle averaged
range throughout the rest of this review.

\subsubsection{TAMA300}

The first interferometric detector to start regular data taking with sufficient
sensitivity and stability to enable it to potentially detect gravitational waves
from the the galactic centre was TAMA300~\cite{Ando:2001}. Over the period
between August 1999 to January 2004 TAMA had nine data-taking periods
(denominated DT1--9) over which time its typical strain noise sensitivity, in
its most sensitive frequency band improved from
$\sim$~3~\texttimes~10\super{-19}~\Hz to
$\sim$~1.5~\texttimes~10\super{-21}~\Hz~\cite{Akutsu:2006}. TAMA300
operated in coincidence with the LIGO and GEO600 detectors for two of
the science data-taking periods. More recently focus has shifted to
the Cryogenic Laser Interferometer Observatory (CLIO) prototype
detector~\cite{Yamamoto:2008, CLIOweb}, designed to test technologies
for a future \textit{second-generation} Japanese detector called the
Large-scale Cryogenic Gravitational-Wave Telescope (LCGT) (see
Section~\ref{subsection:aligo}).

\subsubsection{LIGO}
\label{sec:ligoruns}

The first LIGO detector to achieve lock (meaning having the interferometer
stably held on a dark fringe of the interference pattern, with light resonating
throughout the cavity) was H2 in late 2000. By early 2002 all three detectors
had achieved lock and have since undergone many periods of commissioning and
science data taking. Over the period from mid-2001 to mid-2002 the
commissioning process improved the detectors' peak sensitivities by several
orders of magnitude, with L1 going from
$\sim$~10\super{-17}\,--\,10\super{-20}~\Hz at 150~Hz. In summer 2002
it was decided that the detectors were at a sensitivity, and had a
good enough lock stability, to allow a science data-taking run. This
was potentially sensitive to local galactic burst events. From 23
August to 9 September 2002 the three LIGO detectors, along with GEO600
(and, for some time, TAMA300), undertook their first coincident
science run, denoted S1 (see~\cite{Abbott:2004a} for the state of the
LIGO and GEO600 detectors at the time of S1). At this time the most
sensitive detector was L1 with a peak sensitivity at around 300~Hz of
2\,--\,3~\texttimes~10\super{-21}~\Hz. The best strain
amplitude sensitivity curve for S1 (and the subsequent LIGO science runs) can be seen in
Figure~\ref{figure:LIGOstrains}. The amount of time over the run that
the detectors were said to be in science mode, i.e., stable and with
the interferometer locked, called their duty cycle, or duty factor,
was 42\% for L1, 58\% for H1 and 73\% for H2. For the most sensitive
detector, L1, the inspiral range was typically 0.08~Mpc.

% LIGO S1 through S5 best strain sensitivities
\epubtkImage{LIGOSrunASDs.png}{%
\begin{figure}[htbp]
  \centerline{\includegraphics[scale=0.45]{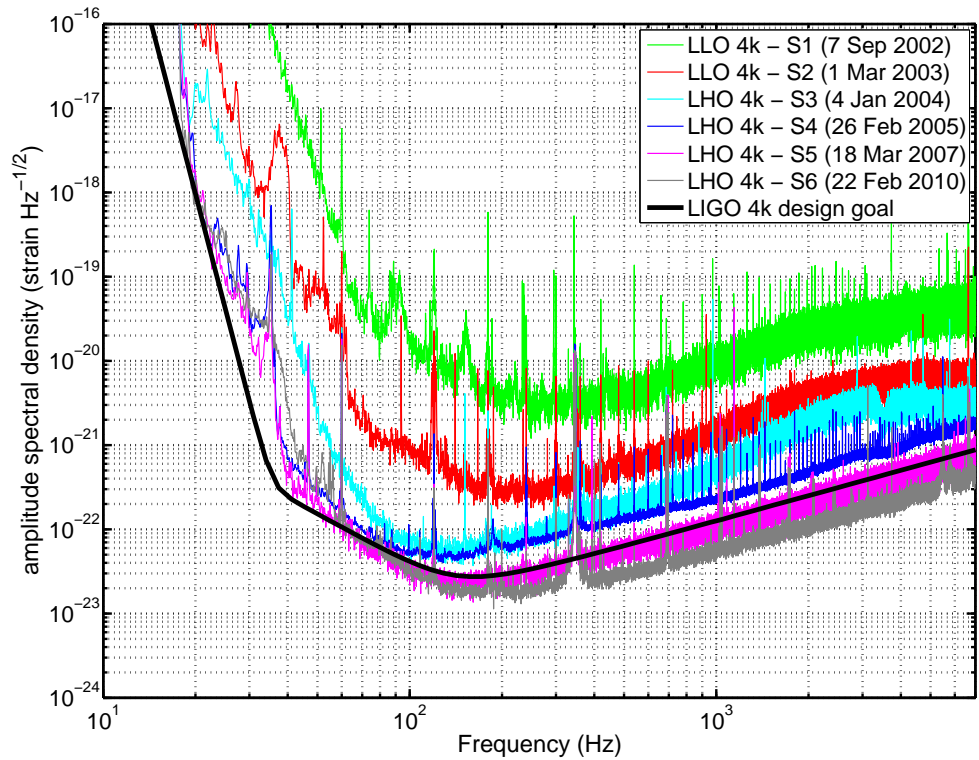}}
  \caption{The best strain sensitivities from the LIGO science runs S1
through S6~\cite{LIGOcurves}. The S6 curve is preliminary and based on h(t) data
that has not been completely reviewed and may be subject to change. Also shown
is the LIGO 4~km design sensitivity.}
  \label{figure:LIGOstrains}
\end{figure}}

For the second science run (S2), from 14 February to 14 April 2003, the noise
floor was considerably improved over S1 by several upgrades including: improving
and stabilising the optical levers used to measure the mirror orientation to
reduce the low frequency ($\lesssim$~50~Hz) noise; replacing the coil drivers
that are used as actuators to control the position and orientation of suspended
mirrors, to improve the mid-frequency ($\sim$~50\,--\,200~Hz) noise floor; and
increasing the laser power in the interferometer to reduce shot noise and
improve the high frequency ($\gtrsim$~200~Hz) sensitivity (see
Section~IIA of~\cite{Abbott:2005a} for a more thorough description of
the detector improvements made for S2). These changes improved the
sensitivities by about an order of magnitude across the frequency band
with a best strain, for L1, of $\sim$~3~\texttimes~10\super{-22}~\Hz
between 200\,--\,300~Hz. The duty factor during S2 was 74\% for H1,
58\% for H2 and 38\% for L1, with a triple coincidence time when all
three detectors were in lock of 22\% of the run. The average inspiral
ranges during the run were approximately 0.9, 0.4 and 0.3~Mpc for L1,
H1 and H2 respectively. This run was also operated in coincidence with
the TAMA300 DT8 run. 

For the the third science run (S3), from 31 October 2003 to 9 January 2004,
the detectors were again improved, with the majority of sensitivity increase in
the mid-frequency range. This run was also operated partially in coincidence
with GEO600. The best sensitivity, which was for H1, was
$\sim$~5~\texttimes~10\super{-23}~\Hz between 100\,--\,200~Hz. The duty factors
were 69\% for H1, 63\% for H2 and only 22\% for L1, with a 16\% triple
coincidence time. L1's poor duty factor was due to large levels of anthropogenic
seismic noise near the site during the day.

The fourth science run (S4), from 22 February to 23 March 2005, saw less-drastic
improvements in detector sensitivity across a wide frequency band, but did make
large improvements for frequencies $\lesssim$~70~Hz. Between S3 and S4 a better
seismic isolation system, which actively measured and countered for ground
motion, was installed in L1, greatly reducing the amount of time it was thrown
out of lock. For H1 the laser power was able to be increased to its full design
power of 10~W~\cite{Abbott:2007b}. The duty factors were 80\% for H1, 81\% for
H2 and 74\% for L1, with a 56\% triple coincidence time. The most sensitive
detector, H1, had an inspiral range of 7.1~Mpc.

By mid-to-late 2005 the detectors had equaled their design sensitivities over
most of the frequency band and were also maintaining good stability and high
duty factors. It was decided to perform a long science run with the aim of
collecting one year's worth of triple coincident data, with an angle-averaged
inspiral range of equal to, or greater than, 10~Mpc for L1 and H1, and 5~Mpc
or better for H2. This run, S5, spanned from 4 November 2005 (L1 started
slightly later on 14 November) until 1 October 2007, and the performance of the
detectors during it is summarised in~\cite{LIGOS5}. One year of triple
coincidence was achieved on 21 September 2007, with a total triple coincidence
duty factor of 52.5\% for the whole run. The average insprial range over S5
was $\sim$~15~Mpc for H1 and L1, and $\sim$~8~Mpc for H2.

After the end of S5 the LIGO H2 detector and GEO600 were kept operational while
possible in an evening and weekend mode called Astrowatch. This observing mode
continued until early 2009, after which H2 was retired. During this time
commissioning of some upgrades to the 4~km LIGO detectors took place for the
sixth and final initial LIGO science run (S6) -- some of which are
summarised in~\cite{Whitcomb:2008}. The aim of these upgrades, called
Enhanced LIGO~\cite{EnhancedLIGO}, was to try and  increase
sensitivity by a factor of two. Enhanced LIGO involved the direct
implementation of technologies and techniques designed for the later
upgrade to Advanced LIGO (see Section~\ref{subsection:aligo}) such as,
most notably, higher-powered lasers, a DC readout scheme (see
Section~\ref{sec:readout}), the addition of output mode cleaners and
the movement of some hardware into the vacuum system. The lasers,
supplied by the Albert Einstein Institute and manufactured by Laser
Zentrum Hannover, give a maximum power of $\approx$~30~W, which is
around 3 times the initial LIGO power. The upgrade to higher power
required that several of the optical components needed to be
replaced. These upgrades were only carried out on the 4~km H1 and L1
detectors due to the H2 detector being left in Astrowatch mode during
the commissioning period. The upgrades were able to produce 1.5\,--\,2
times sensitivity increases at frequencies above $\approx$~200~Hz, but
generally at lower frequencies various sources of noise meant
sensitivity increases were not possible. S6 took place from July 2009
until 20 October 2010, at which point decommissioning started for the
full upgrade to Advanced LIGO. Typically the detectors ran with laser
power at $\approx$~10~W during the day (at higher power the detector
was less stable and the higher level of anthropogenic noise during the
day meant that achieving and maintaining lock required lower power)
and $\approx$~20~W at night, leading to inspiral ranges from
$\approx$~10\,--\,20~Mpc.

\subsubsection{GEO600}

GEO600 achieved first lock as a power-recycled Michelson (with no signal
recycling) in late 2001. Commissioning over the following year,
detailed in~\cite{Hewitson:2003}, included increases in the laser
power, installation of monolithic suspensions for the end test masses
(although not for the beam splitter and inboard mirrors),
rearrangement of the optical bench to reduce scattered light and
implementation of an automatic alignment system. For the S1 run,
carried out in coincidence with LIGO (and, in part, TAMA300), the
detector was kept in this configuration (see~\cite{Abbott:2004a} for
the status of the detector during S1). It had a very high duty factor
of $\sim$~98\%, although its strain sensitivity was $\sim$~2 orders of
magnitude lower than the LIGO instruments. The auto-alignment system
in GEO600 has since meant that it has been able to operate for long
periods without manual intervention to regain lock, as has been the
case for initial LIGO.

% GEO S1 through S5 typical strain sensitivities
\epubtkImage{GEOSrunASDs.png}{%
\begin{figure}[htbp]
  \centerline{\includegraphics[scale=0.45]{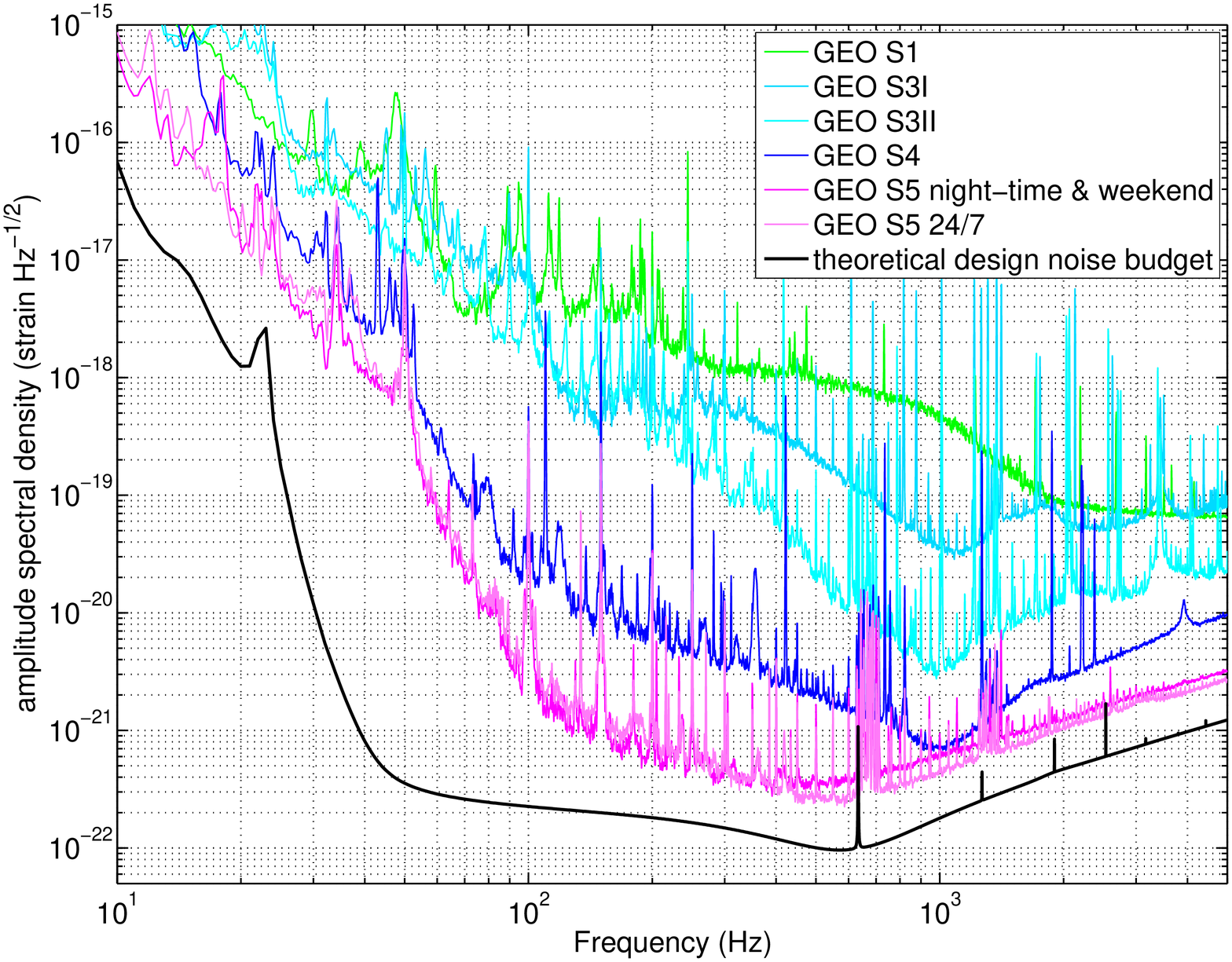}}
  \caption{The typical strain sensitivities from the GEO600 science runs S1
through S5~\cite{GEOcurves}. Also shown is the theoretical noise budget for the
detector when tuned to 550~Hz -- the operating position for the S5 run.}
  \label{figure:GEOstrains}
\end{figure}}

Following S1 the signal recycling mirror was installed and in late 2003 the
first lock of the fully dual-recycled system was achieved
(see~\cite{Smith:2004, Willke:2004, Grote:2005} for information on the
commissioning of GEO600 as a dual-recycled detector). Other upgrades included the
installation of the final mirrors, suspended as triple pendulums, and with
monolithic final stages. Once installed it was found that there was a radius of
curvature mismatch with one of the mirrors, which had to be compensated for by
carefully heating the mirror. Due to this commissioning effort GEO600 did not
participate in the S2 run. Very soon after the implementation of dual-recycling
GEO600 took part in the S3 run. This occurred over two time intervals from
5\,--\,11 November 2003, dubbed S3I, and from 30 December 2003 to 13 January 2004,
dubbed S3II. During S3I GEO600 operated with the signal-recycling cavity tuned
to $\sim$~1.3~kHz, and had a $\sim$~95\% duty factor, but was then taken
off-line for more commissioning work. In the period between S3I and II various
sources of noise and lock loss were diagnosed and mitigated, including noise
from a servo in the signal recycling cavity and electronic noise on a
photo-diode~\cite{Smith:2004}. This lead to improved sensitivity by up to an
order of magnitude at some frequencies (see Figure~\ref{figure:GEOstrains}). For
S3II the signal recycling cavity was tuned to 1~kHz and, due to the upgrades,
had an increased duty factor of $\sim$~99\%. GEO600 operated during the whole of
S4 (22 February to 24 March 2004), in coincidence with LIGO, with a $\sim$~97\%
duty factor. It used the same optical configuration as S3, but had sensitivity
improvements from a few times to up to an order of magnitude over the S3
values~\cite{Hild:2006a}.

The main changes to the detector after S4 were to shift the resonance condition
of the signal recycling cavity to a lower frequency, 350~Hz, allowing better
sensitivity in the few hundred Hz regime, and increasing the circulating laser
power, with an input power of 10~W. The pre-S5 peak sensitivity was
$\sim$~4~\texttimes~10\super{-22}~\Hz at around 400~Hz, with an inspiral
range of 0.6\,Mpc~\cite{Hild:2006b}. GEO600 did not join S5 at the start of the
LIGO run, but from 21 January 2006 was in a night-and-weekend data-taking mode
whilst noise hunting studies and commissioning were conducted. For S5 the signal
recycling cavity was re-tuned up to 550~Hz. It went into full-time data taking
from 1 May to 16 October 2006, with an instrumental duty factor of 94\%. The
average peak sensitivity during S5 was better than
3~\texttimes~10\super{-22}~\Hz (see~\cite{Willke:2007} for a
summary of GEO600 during S5). After this it was deemed more valuable
for GEO600 to continue more noise hunting and commissioning work, to
give as good a sensitivity as possible for when the LIGO detectors
went offline for upgrading. However, it did continue operating in night-and-weekend mode.

GEO600 continued operating in Astrowatch mode between November 2007 and July
2009 after which upgrades began. The plans for the GEO600 detector are to
continue to use it as a test-bed for more novel interferometric techniques
whilst focusing on increasing in sensitivity at higher frequencies (greater than
a few hundred Hz). This project is called GEO-HF~\cite{Willke:2006}. The
upgrading towards GEO-HF has been taking place since Summer
2009~\cite{Grote:2010}. The main upgrades started during 2009 were to
change the read-out scheme from an RF read-out to a DC read-out system~\cite{Hild:2009}
(also see Section~\ref{sec:readout}), install an output mode cleaner, place the
read-out system in vacuum, injecting squeezed light~\cite{Vahlbruch:2008,
Chelkowski:2007} into the output port, and finally increasing the input laser
power to 35~W. Running the interferometer with squeezed light will be the first
demonstration of a full-scale gravitational-wave detector operating beyond the
standard quantum limit. GEO-HF participated in S6 in an overnight and weekend
mode, alongside a commissioning schedule, and is continuing in this mode
following the end of S6.

\subsubsection{Virgo}

In summer 2002 Virgo completed the commissioning of the central area
interferometer, consisting of a power-recycled Michelson interferometer, but
without the 3~km Fabry--P\'{e}rot arm cavities. Over the next couple of years
various steps were made towards commissioning the full-size interferometer. In
early 2004 first lock with the 3~km arms was achieved, but without
power-recycling, and by the end of 2004 lock with power recycling was achieved.
During summer 2005 the commissioning runs provided order-of-magnitude
sensitivity improvements, with a peak sensitivity of
6~\texttimes~10\super{-22}~\Hz at 300~Hz, and an inspiral range
of over 1~Mpc. In late 2005 several major upgrades brought Virgo to
its final configuration. See~\cite{Acernese:2004, Acernese:2005,
  Acernese:2006, Acernese:2007} for more detailed information on the
commissioning of the detector.

Virgo joined coincident observations with the LIGO and GEO600 S5 run with 10
weekend science runs (WSRs) starting in late 2006 until March 2007. Over this
time improvements were made mainly in the mid-to-low frequency regime
($\lesssim$~300~Hz). Full-time data taking, under the title of Virgo
Science run 1 (VSR1), began on 18 May 2007 and ended with the end of
S5 on 1 October 2007. During VSR1, the science-mode duty factor was
81\% and by the end of the run maximum neutron-star--binary inspiral
range was frequently up to about 4.5~Mpc. The best sensitivity curves
for WSR1, WSR10 and VSR1 can be seen in Figure~\ref{figure:Virgostrains}.

At the same time as commissioning for Enhanced LIGO was taking place there was
also a similar effort to upgrade the Virgo detector, called Virgo+. The main
upgrade was to the lasers to increase their power from 10 to 25~W at the input
mode cleaner, with upgrades also to the thermal compensation system on the
mirrors, the control electronics, mode cleaners and injection optics
\cite{Acernese:2008b, AdvVirgoWhitepaper}. Virgo+ started taking data
with Enhanced LIGO for Virgo Science Run 2 (VSR2) and sensitivities of
Virgo+ close to the initial Virgo design sensitivity were
reached. VSR2 finished on 8 January 2010 to allow for further
commissioning and noise hunting. This was followed by VSR3, which
began on 11 August 2010 and ran until 20 October 2010. Further Virgo+
runs are expected during 2011. Following these the upgrades to
Advanced Virgo will begin.

% Virgo typical strain sensitivities
\epubtkImage{VirgoSrunASDs.png}{%
\begin{figure}[htbp]
  \centerline{\includegraphics[scale=0.45]{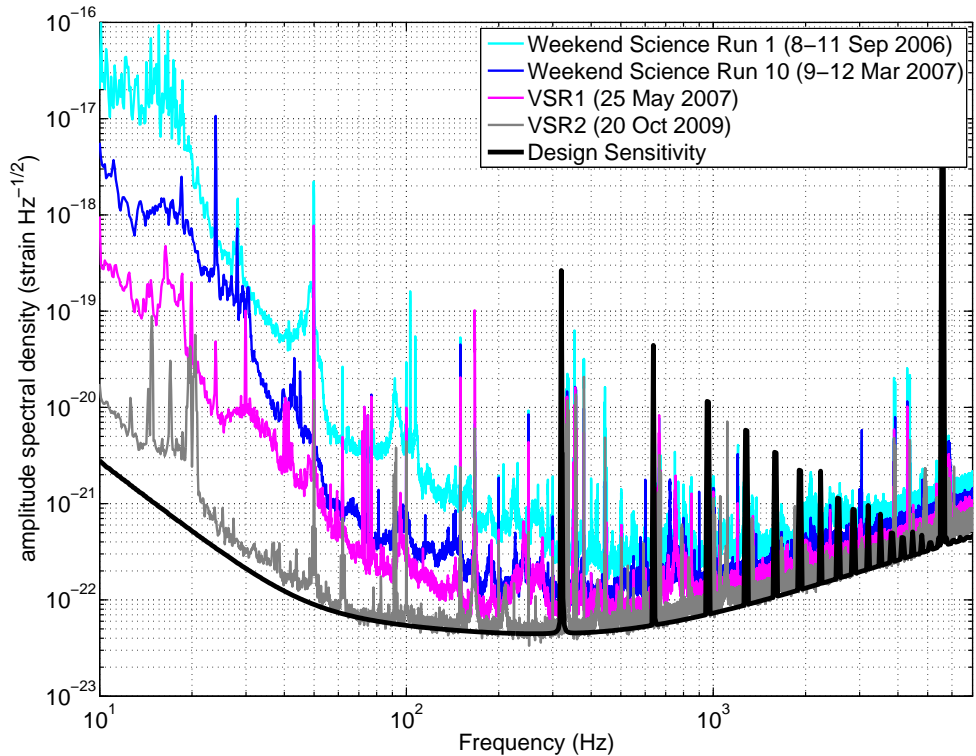}}
  \caption{The best strain sensitivities from the Virgo weekend and full
time science runs WSR1, WSR10, VSR1 and VSR2~\cite{VIRGOcurves, VSR2paper}.}
  \label{figure:Virgostrains}
\end{figure}}

\subsection{Astrophysics results}
\label{subsection:results}

Prior to the advent of the large scale interferometric detectors there had been
some limited effort to produce astrophysical results with the prototype
interferometers. The Caltech 40~m detector was used to search for, and set an
upper limit on, the gravitational wave emission from pulsar
\epubtkSIMBAD{PSR~J1939+2134}~\cite{Hereld:1984}, and on the rate of neutron star
binary inspirals in our galaxy, using coincident observations with the
University of Glasgow prototype~\cite{Smith:1988} and, more recently,
on its own~\cite{Allen:1999}. Coincident observations using the
prototype detectors at the University of Glasgow and Max Planck
Institute for Quantum Optics, in Garching, Germany, were used to set
an upper limit on the strain of gravitational wave bursts~\cite{Nicholson:1996}.
The Garching detector was used to search for periodic signals from pulsars, and
in particular set a limit on a potential source in
\epubtkSIMBAD{SN~1987A}~\cite{Niebauer:1993}. However since the start
of science data taking for the large scale detectors there has been a
rapid rise in the number, and scope, of science result papers being
published. With the vastly improved sensitivities pushing upper limits
on source populations and strengths towards astrophysically interesting areas.

The recent analysis efforts have generally been split into four broad areas
depending on the expected signal type: unmodelled transients or bursts, e.g.,
supernovae; modelled transients, e.g., inspirals and ring-downs (or more
specifically compact binary coalescences, CBC); continuous sources; stochastic
sources. Within each area a variety of different sources could exist and a
variety of analysis techniques have been developed to search for them. Some
electromagnetic sources, such as radio pulsars and $\gamma$-ray bursts, are also
used to enhance searches. A good review of the data analysis methods used in
current searches, and the astrophysical consequences of some of the results
described below, can be found in~\cite{Sathyaprakash:2009}.

Here we will briefly summarise the main astrophysics results from the science
runs. We will mainly focus on those produced by the LIGO Scientific
Collaboration~\cite{LSCweb} detectors LIGO and GEO600 from S1 to the S5 run.
At the time of writing not all of the data from the S6 run had been fully
analysed, with more results expected over the next year. Reviews of some early
S5, and prior science run, results can also be found in~\cite{Papa:2008,
Fairhurst:2009}. In none of the searches so far has convincing evidence for a
gravitational-wave signal been seen.

\subsubsection{Unmodelled bursts}
\label{subsubsection:unmodelled}

Searches for unmodelled bursts, e.g., from supernova core-collapse, are based on
looking for short duration periods of excess power in the detectors. Transients
are common features in the data, so to veto these events from true gravitational-wave signals they must be coincident in time, and to some extent amplitude and
waveform, between multiple detectors. Various methods to assess instrumental
excess power, and inter-detector correlations, are used, some examples of which
can be found here~\cite{Klimenko:2004, Anderson:2001, Searle:2008, McNabb:2004,
Cadonati:2004, Chatterji:2004, Chatterji:2006}. These algorithms will produce
\textit{triggers}, which are periods of excess power that cross a predetermined
signal-to-noise ratio threshold (determined by tuning the algorithms on a
section of \textit{playground} data, so that the output produces a desired false
alarm rate). The number of triggers are then compared to a background rate. Real
signals cannot be turned off, and detectors cannot be shielded from them, so the
background rate has to be approximated by time shifting one detector's data
stream with respect the the others. Time shifts should only leave triggers due
to random coincidences in detector noise and there should be no contribution
from real signals. Once a background is calculated, the statistical significance
of the foreground rate can be compared to it. To assess the sensitivity of these
searches, hardware (the interferometer mirrors are physically moved via the
control system) and software signals are injected into the data stream at
various strengths and the efficiency of the algorithms at detecting them is
measured. A good description of some of these techniques can be found in
\cite{Abbott:2004b} and~\cite{Abbott:2006a}.

Data from the LIGO S1 run was searched for gravitational-wave bursts of between
4 to 100~ms, and within the frequency band 150 to 3000~Hz~\cite{Abbott:2004b}.
Triple coincident data from all three detectors was used for the analysis. No
plausible candidate event was found, but a 90\% confidence upper limit on the
event rate of 1.6 events per day was set. The search was typically sensitive, at
a $\gtrsim$~50\% detection efficiency, to bursts with amplitudes of
$h_{\mathrm{rss}}\sim10^{-19}\mbox{\,--\,}10^{-17} \mathrm{\ Hz}^{-1/2}$ (defined in
terms of $h_{\mathrm{rss}} \equiv \sqrt{\int|h|^2 \mathrm{d}t}$, which is the root-sum-squared strain
amplitude spectral density). Due, in part, to its lower sensitivity GEO600 data
was not used in this analysis. The S2 data's improved sensitivity, and advances
in the analysis techniques, allowed a sensitivity to signals (in the frequency
range 100\,--\,1100~Hz) in the amplitude range
$h_{\mathrm{rss}}\sim10^{-20}\mbox{\,--\,}10^{-19}
\mathrm{\ Hz}^{-1/2}$~\cite{Abbott:2005a}. Interpreting the best
sensitivities astrophysically gave order of magnitude estimates on the
visible range of $\sim$~100~pc for a class of theoretical supernova waveforms,
and 1~Mpc for the merger of $50\,M_{\odot}$ black holes. Again no signal
was seen, but a 90\% upper limit of 0.26 events per day was set for strong
bursts. In the frequency range 700\,--\,2000~Hz TAMA300 data was also used in the
search giving amplitude sensitivities of
$h_{\mathrm{rss}}\sim1\mbox{\,--\,}3\times10^{-19}
\mathrm{\ Hz}^{-1/2}$ and decreasing the rate upper limit to 0.12
events per day~\cite{Abbott:2005c}.

The S3 run produced two searches for burst sources. One used the 8 days of
triple coincidence data from the three LIGO detectors to search for sub-second
bursts in the frequency range 100\,--\,1100~Hz~\cite{Abbott:2006a}. The search was
sensitive to signals with amplitudes over
$h_{\mathrm{rss}}\sim1\times10^{-20}\mathrm{\ Hz}^{-1/2}$, but did not
include an astrophysical interpretation of the limit or event rate
upper limit. This run included coincident operation with the Italian
AURIGA bar detector and this data has been
analysed~\cite{Baggio:2008}. This search looked for short bursts of
less than 20~ms within the 850\,--\,950~Hz band (around the bar's
sensitive resonant frequency). This had comparable sensitivity to the
LIGO-only S2 search and produced a 90\% confidence rate upper limit of
0.52 events/day.

For S4 15.5~days of LIGO data were searched for sub-second bursts in the
frequency range 64\,--\,1600~Hz~\cite{Abbott:2007b}. This was sensitive to signals
with $h_{\mathrm{rss}}\lesssim10^{-20}$ and set a 90\% confidence rate upper limit of
0.15 per day. The search results are also cast as astrophysical limits on source
ranges and energetics. These show that there would be a 50\% detection
efficiency to signals of sine-Gaussian nature (at the most sensitive frequency
of 153~Hz and quality factor $Q=8.9$) at a distance of 10~kpc for an energy of
$10^{-7}\,M_{\odot}c^2$, and would be sensitive to signals out to the Virgo
cluster ($\sim$~16~Mpc) for an energy release of $0.25\,M_{\odot}c^2$. See
\cite{Abbott:2007b} for a comparison of previous burst searches. There was also
a burst search combining S4 GEO600 and LIGO data for the first time. This
searched data between 768\,--\,2048~Hz where the sensitivities were most comparable
and used 257 hours of quadruple coincidence between the detectors and saw no
gravitational wave events~\cite{Abbott:2008b}.

For the analysis on the first year of S5 data the frequency range for the
all-sky burst search was split -- a low frequency search covered the most
sensitive region between 60\,--\,2000~Hz~\cite{Abbott:2009h}, and a high frequency
search covering 1\,--\,6~kHz (this being the first time an untriggered burst search
looked at frequencies above 3~kHz)~\cite{Abbott:2009i}. The high frequency
search set a 90\% upper limit on the rate of 5.4 events per year for strong
events. The low frequency search analysed more data than the high frequency and
set an event rate limit of 3.6 events per year. The second year of S5 LIGO data
was analysed with GEO600 and Virgo VSR1 data~\cite{Abadie:2010d} to search for
bursts over the whole 50\,--\,6000~Hz band. Combining this with the earlier S5
searches gave $h_{\mathrm{rss}}$ upper limits for a variety of
simulated waveforms of 6~\texttimes~10\super{-22}~\Hz to
2~\texttimes~10\super{-20}~\Hz, and a 90\% confidence event rate for
signals between 64\,--\,2048~Hz of less than two per year.

\subsubsection{Modelled bursts -- compact binary coalescence}
\label{sec:cbc}

Modeled bursts generally mean the inspiral and coalescence stage of binaries
consisting of compact objects, e.g., neutron stars and black holes. The signals
are generally well approximated by post-Newtonian expansions of the Einstein
equation, which give the amplitude and phase evolution of the orbit. More
recently signal models have started to include numerical relativity simulations
of the merger stage~\cite{Aylott:2009}. As mentioned in
Section~\ref{section:construction} the best estimate of the number of signals
observable with initial LIGO at design sensitivity (i.e.\ during S5) would be
0.02 per year (based on an event rate of 1~\texttimes~10\super{-6} per year per MWEG).

The majority of inspiral searches make use of matched filtering in which a
template bank of signal models is built~\cite{Owen:1996, Owen:1999}, with a
maximum mismatch between templates that is generally of order $\sim$~10\%. These
templates are then cross-correlated with the data and statistically significant
\textit{triggers} (i.e.\ times when the template and data are highly correlated)
from this are looked for. Triggers must be coincident between detectors and the
significance of any trigger is judged against a background calculated in the
same way as described in Section~\ref{subsubsection:unmodelled}. See
\cite{Abbott:2005b} for a good description of the search method.

The first search for an inspiral signal with data from the LIGO S1 run looked
for compact object coalescences with component masses between $1\mbox{\,--\,}3\,M_{\odot}$
and was sensitive to such sources within the Milky Way and Magellanic
Clouds~\cite{Abbott:2004c}. It gave a 90\% confidence level upper
limit on the rate of 170 per year per MWEG.

For the S2 LIGO analysis the search was split into 3 areas covering neutron-star
binaries, black-hole binaries and primordial black-hole binaries in the galactic
halo. The neutron-star--binary search~\cite{Abbott:2005b} used 15 days of data
with coincidence between either H1 and L1 or H2 and L1. It had a range of
$\sim$~1.5~Mpc, which spanned the Local Group of galaxies, and gave a 90\% event
rate upper limit on systems with component masses of $1\mbox{\,--\,}3\,M_{\odot}$ of 47 per
year per MWEG. The black-hole--binary search looked for systems with component
masses in the $3\mbox{\,--\,}20\,M_{\odot}$ range using the same data set as the
neutron-star--binary search~\cite{Abbott:2006a}. This search had a 90\% detection
efficiency for sources out to 1\,Mpc and set a 90\% rate upper limit of 38 per
year per MWEG. The third search looked for low mass ($0.2\mbox{\,--\,}1\,M_{\odot}$)
primordial black-hole binaries in a 50~kpc radius halo surrounding the Milky
Way~\cite{Abbott:2005e}. This placed a 90\% confidence-rate upper limit of 63
events per year per Milky Way halo. The S2 search was performed in coincidence
with the TAMA300 DT8 period and an inspiral search for neutron-star binaries
was
performed on data when TAMA300 and at least one of the LIGO sites was
operational. This gave a total of 584~hours of data for the analysis, which set
a 90\% rate upper limit of 49 per year per MWEG, although this search was only
sensitive to sources within the majority of the Milky Way~\cite{Abbott:2006b}.

The search for neutron-star--black-hole binaries in S3 LIGO data used techniques
designed specifically for systems with spinning components. It searched for
systems with component masses in the range $1\mbox{\,--\,}20\,M_{\odot}$ and analysed 167
hours of triple coincident data and 548 hours of H1-H2 data to set the upper
limits~\cite{Abbott:2008d}. For a typical system with neutron-star and black-hole mass distributions centred on $1.35\,M_{\odot}$ and $5\,M_{\odot}$ (from
the population statistics discussed in~\cite{Abbott:2008a}) this search produced
a 90\% confidence-rate upper limit of 15.9 per year per $L_{10}$.

The search for a wide range of binary systems with components consisting of
primordial black holes, neutron stars, and black holes with masses in the ranges
given above was conducted on the combined S3 and S4 data~\cite{Abbott:2008a}.
788 hours of S3 data and 576 hours of S4 data were used and no plausible
gravitational-wave candidate was found. The highest mass range for the black-hole--binary search was set at $40\,M_{\odot}$ for S3 and $80\,M_{\odot}$ for S4. At peak in the
mass distribution of these sources 90\% confidence-rate upper limits were set at
4.9 per year per $L_{10}$ for primordial black holes, 1.2 per year per $L_{10}$
for neutron-star binaries, and 0.5 per year per $L_{10}$ for black-hole--binaries.
S4 data has also been used to search for ring-downs from perturbed black holes,
for example following black-hole-binary coalescence~\cite{Abbott:2009g}. The
search was sensitive to ring-downs from $10\mbox{\,--\,}500\,M_{\odot}$ black holes out to
a maximum range of 300~Mpc, and produced a best 90\% confidence upper limit on
the rate of ring-downs to be 1.6~\texttimes~10\super{-3} per year per
$L_{10}$ for the mass range $85\mbox{\,--\,}390\,M_{\odot}$.

One other kind of modeled burst search is that looking for gravitational waves
produced by cusps in cosmic (super)strings. Just over two weeks of LIGO S4 data
were used to search for such signals~\cite{Abbott:2009j}. This was used to
constrain the rate and parameter space (string tension, reconnection
probability, and loop sizes), but was not able to beat limits set by Big Bang
nucleosynthesis.

Data from the first~\cite{Abbott:2009e} and second year of S5 (prior to Virgo
joining with VSR1)~\cite{Abbott:2009f} have been searched for low-mass binary
coalescences with total masses in the range $2\mbox{\,--\,}35\,M_{\odot}$. The second
year search results have produced the more stringent upper limits with 90\%
confidence rates for neutron-star-binaries, black-hole-binaries and neutron-star--black-hole systems respectively of 1.4~\texttimes~10\super{-2},
7.3~\texttimes~10\super{-4} and 3.6~\texttimes~10\super{-3} per year
per $L_{10}$. Five months of overlapping S5 and VSR1 data were also
searched for the same range of signals~\cite{Abadie:2010f} giving 90\%
confidence upper rates of 8.7~\texttimes~10\super{-3} per year per
$L_{10}$, 2.2~\texttimes~10\super{-3} per year per $L_{10}$, and
4.4~\texttimes~10\super{-4} per year per $L_{10}$. The whole 2 years
of LIGO S5 data were also used to search for higher mass binary
coalescences with component mass between $1\mbox{\,--\,}99\,M_{\odot}$
and total masses of $25\mbox{\,--\,}100\,M_{\odot}$. No signal was
seen, but a 90\% confidence upper limit rate on mergers of black-hole--binary systems with component masses between 19 and
$28\,M_{\odot}$, and with negligible spin, was set at
2.0~Mpc\super{-3}~Myr\super{-1} \cite{Abadie:2011a}.

\subsubsection{Externally-triggered burst searches}

Many gravitational wave burst sources will be associated with electromagnetic
(or neutrino) counterparts, for example short $\gamma$-ray bursts (GRBs) are
potentially caused by black-hole and neutron-star coalescences. Joint
observation of a source as both a gravitational wave and electromagnetic
event also greatly increases the confidence in a detection. Therefore many
searches have been performed to look for bursts coincident (temporally and
spatially) with external electromagnetic triggers, such as GRBs observed by
Swift for example. These searches have used both excess power and modeled
matched-filter methods to look for signals.

During S2 a particularly bright $\gamma$-ray burst event (\epubtkSIMBAD{GRB~030329}) occurred
and was specifically targeted using data from H1 and H2. The search looked for
signals with duration less than $\sim$~150~ms and in the frequency range
80\,--\,2048~Hz~\cite{Abbott:2005d}. This produced a best strain upper limit for an
unpolarised signal around the most sensitive region at  $\sim$~250~Hz
of $h_{\mathrm{rss}}=6\times10^{-21} \mathrm{\ Hz}^{-1/2}$.

For S4 there were two burst searches targeting specific sources. The first
target was the hyperflare from the Soft $\gamma$-ray Repeater \epubtkSIMBAD{SGR~1806--20}
(SGRs are thought to be ``magnetars'', neutron stars with extremely large
magnetic fields of order 10\super{15}~Gauss) on 27 December 2004
\cite{Hurley:2005} (this actually occurred before S4 in a period when only the
H1 detector was operating). The search looked for signals at frequencies
corresponding to short duration quasi-periodic oscillations (QPOs) observed in
the X-ray light curve following the flare~\cite{Abbott:2007c}. The most
sensitive 90\% upper limit was for the 92.5~Hz QPO at $h_{\mathrm{rss}} =
4.5\times10^{-22} \mathrm{\ Hz}^{-1/2}$, which corresponds to an energy emission limit
of $4.3\times10^{-8}\,M_{\odot}c^2$ (of the same order as the total
electromagnetic emission assuming isotropy). The other search used LIGO data
from S2, S3 and S4 to look for signals associated with 39 short duration
$\gamma$-ray bursts (GRBs) that occurred in coincidence with these
runs~\cite{Abbott:2008c}. The GRB triggers were provided by IPN,
Konus-Wind, HETE-2, INTEGRAL and Swift as distributed by the GRB
Coordinate Network~\cite{GCN}. The search looked in a 180-second
window around the burst peak time (120 seconds before and 60 seconds
after) and for each burst there were at least two detectors
contributing data. No signal coincident with a GRB was observed and
the sensitivities were not enough to give any meaningful astrophysical
constraints, although simulations suggest that for S4, as in the general burst
search, it would have been sensitive to sine-Gaussian signals out to tens of Mpc
for an energy release of order a solar mass.

The first search of Virgo data in coincidence with a GRB was performed on data
from a commissioning run in September 2005. The long duration \epubtkSIMBAD{GRB~050915a} was
observed by Swift on 15 September 2005 and Virgo data was used to search for an
unmodelled burst in a window of 180 seconds around (120~s before and 60~s after)
the GRB peak time~\cite{Acernese:2008a}. The search produced a strain upper
limit of order 10\super{-20} in the frequency range 200\,--\,1500~Hz, but was mainly
used as a test-bed for setting up the methodology for future searches, including
coincidence analysis with LIGO.

Data from the S5 run has been used to search for signals associated with even
more $\gamma$-ray bursts. One search looked specifically for emissions from
\epubtkSIMBAD{GRB~070201}~\cite{Golenetskii:2007a, Golenetskii:2007b}, which showed 
evidence of originating in the nearby Andromeda galaxy (\epubtkSIMBAD{M31}). The data
around the time of this burst was used to look for an unmodelled burst and an
inspiral signal as might be expected from a short GRB. The analysis saw no
gravitational-wave event associated with the GRB, but ruled out the event being
a neutron-star--binary inspiral located in \epubtkSIMBAD{M31} with a 99\% confidence
\cite{Abbott:2008g}. Again, assuming a neutron-star--binary inspiral, but located
outside \epubtkSIMBAD{M31}, the analysis set a 90\% confidence limit that the source must be at
a distance greater than 3.5~Mpc. Assuming a signal again located in M31, the
unmodelled burst search set an upper limit on the energy emitted via
gravitational waves of $4.4\times10^{-4}\,M_{\odot}c^2$, which was
well within the allowable range for this being an SGR hyper-flare in \epubtkSIMBAD{M31}.
Searches for 137 GRBs (both short and long GRBs) that were observed, mainly with
the Swift satellite, during S5 and VSR1 have been performed again using
unmodelled burst methods~\cite{Abbott:2009d} and for (22 short bursts) inspiral
signals~\cite{Abadie:2010b}. No evidence for a gravitational-wave signal
coincident with these events was seen. The unmodeled burst observations were
used to set lower limits on the distance to each GRB, with typical limits,
assuming isotropic emission, at
$D\sim15\mathrm{\ Mpc}(E^{\mathrm{iso}}_{\mathrm{GW}}/0.01\,M_{\odot}c^2)^{1/2}$. The
inspiral search, which was sensitive to CBCs with total system masses
between $2\,M_{\odot}$ and $40\,M_{\odot}$, was able to exclude with
90\% confidence any bursts being neutron-star--black-hole mergers
within 6.7~Mpc, although the peak distance distribution of GRBs is
well beyond this.

Another search has been to look for gravitational waves associated with flares
from known SGRs and anomalous X-ray pulsars (AXPs), both of which are thought to
be \textit{magnetars}. During the first year of S5 there were 191 (including the
December 2004 \epubtkSIMBAD{SGR~1806--20} event) observed flares from SGRs \epubtkSIMBAD[SGR~1806--20]{1806--20} and
\epubtkSIMBAD[SGR~1900+14]{1900+14} for which at least one LIGO detector was online~\cite{Abbott:2008h}, and
1279 flare events if extending that to six known galactic magnetars and
including all S5 and post-S5 Astrowatch data including Virgo and
GEO600~\cite{Abadie:2010c}. The data around each event was searched
for ring-down signals in the frequency range 1\,--\,3~kHz and with
decay times 100\,--\,400~ms as might be expected from \textit{f}-mode
oscillations in a neutron star. It was also searched for unmodeled
bursts in the 100\,--\,1000~Hz range. No gravitational bursts were
seen from any of the events. For the earlier
search~\cite{Abbott:2008h} the lowest 90\% upper limit on the
gravitational-wave energy from the ring-down search was $E_{\mathrm{GW}}^{90\%} =
2.4\times10^{48}\mathrm{\ erg}$ for an \epubtkSIMBAD{SGR~1806--20} burst on 24 August 2006. The
lowest 90\% upper limit on the unmodeled search was
$E_{\mathrm{GW}}^{90\%} = 2.9\times10^{45}\mathrm{\ erg}$ for an \epubtkSIMBAD{SGR~1806--20}
burst on 21 July 2006. The smallest limits on the ratio of energy
emitted via gravitational waves to that emitted in the electromagnetic
spectrum were of order 10\,--\,100, which are into a theoretically-allowed range. The latter search~\cite{Abadie:2010c} gave the lowest
gravitational-wave emission-energy upper limits for white noise bursts
in the detector-sensitive band, and for \textit{f}-mode ring-downs (at
1090~Hz), of 3.0~\texttimes~10\super{44}~erg and
1.4~\texttimes~10\super{47}~erg respectively, assuming a distance of
1~kpc. The \textit{f}-mode energy limits approach the range seen emitted
electromagnetically during giant flares. One of these flares, on 29
March 2006, was actually a ``storm'' of many flares from
\epubtkSIMBAD{SGR~1900+14}. For this event a more sensitive search has been performed
by stacking data around the time of each
flare~\cite{Abbott:2009c}. Waveform dependent upper limits of the
gravitational-wave energy emitted were set between
2~\texttimes~10\super{45}~erg and 6~\texttimes~10\super{50}~erg, which
are an order of magnitude lower than the previous upper limit for this
storm (included in the search of~\cite{Abbott:2008h}) and overlap with
the range of electromagnetic energies emitted in SGR giant flares.

Another possible source of gravitational waves associated with
electromagnetically-observed phenomenon are pulsar glitches. During these it is
possible that various gravitational-wave--emitting vibrational modes of the
pulsar may be excited. A search has been performed for fundamental modes
(\textit{f}-modes) in S5 data following a glitch observed in the
timing of the \epubtkSIMBAD{Vela} pulsar in August
2006~\cite{Abadie:2010a}. Over the search frequency range of
1\,--\,3~kHz this provided upper limits on the peak strain of
0.6\,--\,1.4~\texttimes~10\super{-20} depending on the spherical
harmonic that was excited. 

Already efforts are under way to invert this process of searching gravitational-wave data for external triggers, and instead supplying gravitational-wave burst
triggers for electromagnetic follow-up. This is being investigated across the
range of the electromagnetic spectrum from radio~\cite{Predoi:2010}, through
optical (e.g.,~\cite{Kanner:2008, Coward:2010}) and X-ray/$\gamma$-ray, and even
looking for coincidence with neutrino detectors~\cite{Aso:2008, Pradier:2010,
Chassande:2010}. Having \textit{multi-messenger} observations can have a
large impact on the amount of astrophysical information that can be learnt about
an event~\cite{Phinney:2009}.

\subsubsection{Continuous sources}

Searches for continuous waves focus on rapidly-spinning neutron stars as
sources. There are fully targeted searches, which look for gravitational waves
from known radio pulsars in which the position and spin evolution of the objects
are precisely known. There are semi-targeted searches, which look at potential
sources in which some, but not all, the source signal parameters are known, for
example neutron stars in X-ray binary systems, or sources in supernova remnants
where no pulses are seen, which have known position, but unknown frequency.
Finally, there are all-sky broadband searches in which none of the signal
parameters are known. The targeted searches tend to be most sensitive as they
are able to perform coherent integration over long stretches of data with
relatively low computational overheads, and have a much smaller parameter space
leading to fewer statistical outliers. Due to various neutron-star population
statistics, creation rates and energetics arguments, there is an estimate that
the amplitude of the strongest gravitational-wave pulsar observed at Earth will
be $h_0 \lesssim 4\times10^{-24}$~\cite{Abbott:2007a} (a more thorough
discussion of this argument can be found in~\cite{Knispel:2008}), although this
does not rule out stronger sources.

The various search techniques used to produce these results all look for
statistically-significant excess power in narrow frequency bins that have been
Doppler demodulated to take into account the signal's shifting frequency caused
by the Earth's orbital motion with respect to the source (or also including the
modulations to the signal caused by the source's own motion relative to the
Earth, such as for a pulsar in a binary system). The statistical significance of
a measured level of excess power is compared to what would be expected from data
that consisted of Gaussian noise alone. A selection of the searches are
summarised in~\cite{Prix:2006}, but for more detailed descriptions of the
various methods see~\cite{Brady:2000, Krishnan:2004, Jaranowski:1998,
Abbott:2008e, Abbott:2007a, Dupuis:2005}.

In S1 a fully-coherent targeted search for gravitational waves from the then-fastest millisecond pulsar J1939+2134 was performed~\cite{Abbott:2004d}. This
analysis and the subsequent LSC known-pulsar searches assume that the star is
triaxial and emitting gravitational waves at exactly twice its rotation
frequency. All the data from LIGO and GEO600 was analysed and no evidence of a
signal was seen. A 95\% degree-of-belief upper limit on the gravitational-wave
strain amplitude was set using data from the most sensitive detector, L1, giving
a value of 1.4~\texttimes~10\super{-22}. This result was also interpreted as an
ellipticity of the star given a canonical moment of inertia of
10\super{38}~kg~m\super{2} at
$\epsilon$~=~2.9~\texttimes~10\super{-4}.  However, this was still of
order 100\,000 times higher than the limit that can be set by equating
the star's rate of loss of rotational kinetic energy with that emitted
via gravitational radiation -- called the ``spin-down limit''.

In S2 the number of known pulsar sources searched for with LIGO data increased
from 1 to 28, although all of these were isolated pulsars (i.e.\, not in binary
systems, although potentially still associated with supernova remnants or
globular clusters). This search used pulsar timing data supplied by Lyne
and Kramer from Jodrell Bank Observatory to precisely reconstruct the
phase of the gravitational-wave signal over the period of the run. The lowest
95\% upper limit on gravitational-waves amplitude was
1.7~\texttimes~10\super{-24} for \epubtkSIMBAD{PSR~J1910--5959D}, and the smallest
upper limit on ellipticity (again assuming the canonical moment of
inertia) was 4.5~\texttimes~10\super{-6} for the relatively-close
pulsar \epubtkSIMBAD{PSR~J2124--3358}~\cite{Abbott:2005f}, at a distance of
0.25~kpc. The pulsar closest to its inferred spin-down limit was the
Crab pulsar (\epubtkSIMBAD{PSR~J0534+2200}) with an upper limit 30 times greater than
that from spin-down. S2 also saw the use of two different all-sky--wide
frequency band searches that focused on isolated sources, but also
including a search for gravitational waves from the low mass X-ray
binary Scorpius~X1 (\epubtkSIMBAD[V818~Sco]{Sco-X1}). The first search used a semi-coherent
technique to search $\sim$~60~days of S2 data in the frequency band
between 200\,--\,400~Hz and with signal spin-downs between
--1.1~\texttimes~10\super{-9} and
0~Hz~s\super{-1}~\cite{Abbott:2005g}. This gave a lowest gravitational-wave strain 95\% upper limit of 4.4~\texttimes~10\super{-23} for the
L1 detector at around 200~Hz. The other all-sky search was fully
coherent and as such was computationally limited to only use a few
hours of the most sensitive S2 data. It searched frequencies between
160\,--\,728.8~Hz and spin-downs less than
--4~\texttimes~10\super{-10}~Hz~s\super{-1} for isolated sources and
gave a 95\% upper limit across this band from
6.6~\texttimes~10\super{-23} to 1~\texttimes~10\super{-21}~\cite{Abbott:2007a}. The search for gravitational waves from \epubtkSIMBAD[V818~Sco]{Sco-X1} used the
same period of data. It did not have to search over sky position as this is well
known, but did have to search over two binary orbital parameters -- the projected
semi-major axis and the orbital phase reference time. The frequency ranges of
this search relied on estimates of the spin-frequency from quasi-periodic
oscillations in the X-rays from the source and covered two 20~Hz bands from
464\,--\,484~Hz and 604\,--\,624~Hz (it should be noted that it is now
thought that these estimates of the spin-frequency are unreliable). In
these two ranges upper limits of 1.7~\texttimes~10\super{-22} and
1.3~\texttimes~10\super{-21} were found respectively.

One search that was carried out purely on LIGO S3 data was the coherent all-sky
wide-band isolated pulsar search using the distributed computing project
Einstein@Home~\cite{eath}. The project is built upon the Berkeley Open
Infrastructure for Network Computing~\cite{BOINC} and allows the computational
workload to be distributed among many computers generally contributed by the
general public who sign up to the project. This used the most sensitive 600
hours of data from H1 and cut it into 60 ten hour stretches on each of which
a coherent search could be performed. The data was farmed out to computers owned
by participants in the project and ran as a background process or screen saver.
The search band spanned the range from 50\,--\,1500.5~Hz. The search saw no
plausible gravitational-wave candidates and the result is described
at~\cite{eathS3}, but it was not used to produce an upper limit.

In the known pulsar search the number of sources searched for using the combined
LIGO data from S3 and S4 was increased to 78. This included many pulsars within
binary systems. For many of the pulsars that overlapped with the previous S2
analysis results were improved by about an order of magnitude. The lowest 95\%
upper limit on gravitational-waves amplitude was 2.6~\texttimes~10\super{-25} for
\epubtkSIMBAD{PSR~~J1603--7202}, and the smallest ellipticity was again for \epubtkSIMBAD{PSR~J2124--3358}
at just less than 10\super{-6}~\cite{Abbott:2007d}. The upper limit for the Crab
pulsar was found to be only 2.2 times above that from the spin-down limit. Three
different, but related, semi-coherent all-sky continuous wave searches were
performed on S4 LIGO data, looking for isolated neutron stars in the frequency
range from 50\,--\,1000~Hz and the spin-down range from --1~\texttimes~10\super{-8}
to 0~Hz~s\super{-1}~\cite{Abbott:2008e}. The best 95\% upper limit based on an
isotropically-distributed, randomly-oriented, population of neutron stars was
4.3~\texttimes~10\super{-24} near 140~Hz. This is approaching the amplitude of the
strongest potential signal discussed above. For one of the searches, which
combined data from the different detectors, an isolated pulsar emitting at near
100~Hz, and with an extreme ellipticity of 10\super{-4} could have been seen at a
distance of 1~kpc, although for a more realistic ellipticity of 10\super{-8} only
a distance of less than 1~pc would be visible over the entire LIGO band. The
Einstein@Home project~\cite{eath} was also used to search the most sensitive
data from S4, which consisted of 300 hours of H1 data and 210 hours of L1 data.
The search performed a coherent analysis on 30-hour stretches of this data and
covered the frequency range of 50\,--\,1500~Hz~\cite{Abbott:2008f}. The range
of spin-downs $\dot{f}$ was chosen by using a minimum spin-down age $\tau$ and
having $-f/\tau < \dot{f} < 0.1f/\tau$ (small spin-ups are allowed as some
pulsars in globular clusters exhibit this due to their Doppler motions within
the clusters), with $\tau$~= 1000~years for signals below 300~Hz and
$\tau$~= 10\,000~years above 300~Hz. Approximately 6000 years of computational time
spread over about 100\,000 computers were required to perform the analysis. No
plausible gravitational-wave candidates were found, although the results suggest
that 90\% of sources with strain amplitudes greater than 10\super{-23} would have
been detected by the search. A search designed to produce a sky map of the
stochastic background was also used to search for gravitational waves from
\epubtkSIMBAD[V818~Sco]{Sco-X1} using a method of cross-correlating H1 and L1 data~\cite{Abbott:2007f}.
This produced a 90\% root-mean-squared upper limit on gravitational wave strain
of $h = 3.4\times10^{-24}(f/200 \mathrm{\ Hz})$ for frequencies greater than 200 Hz.

The first 8 months of S5 have been used to perform an all-sky search for
periodic gravitational waves. This search used a semi-coherent method to look
in the frequency range 50\,--\,1100~Hz and spin-down range
--5~\texttimes~10\super{-9}\,--\,0~Hz~s\super{2} and used data from
the H1 and L1 detectors~\cite{Abbott:2008i}. It obtained 95\% strain
upper limits of less than 10\super{-24} over a frequency band of
200~Hz. The search would have been sensitive to a neutron star with
equatorial ellipticity greater than 10\super{-6} within around
500~pc. Einstein@Home~\cite{eath} has been used to search for periodic
waves of 50\,--\,1500~Hz in 860 hours of data from a total span
of 66 days of S5 data \cite{Abbott:2009a}. This search looked for
young pulsars, but saw no significant candidates. It would have been
sensitive to 90\% of sources in the 125\,--\,225~Hz band with
amplitudes greater than 3~\texttimes~10\super{-24}. The first
approximately 9~months of S5 data was used for a coherent search for
gravitational waves from the Crab pulsar~\cite{Abbott:2008j}. In this
search two methods were used: the first followed the method of the targeted
search and assumed that the gravitational waves are phase locked to
the electromagnetic pulses; the second allowed for some mechanism,
which would cause a small mismatch between the two phases. Two 95\%
upper limits were set, one using astrophysical
constraints on the pulsar orientation angle and polarisation angle~\cite{Ng:2008}
and the other applying no such constraints. With the first method these 95\%
upper limits were 3.4~\texttimes~10\super{-25} and 2.7~\texttimes~10\super{-25} respectively,
which correspond to ellipticities of 1.8~\texttimes~10\super{-4} and 1.4~\texttimes~10\super{-4}
(assuming the canonical moment of inertia). These beat the Crab pulsar's
spin-down limit by 4 to 5 times and can be translated into the amount of the
available spin-down power that is emitted via gravitational waves, with the
lower of these limits showing that less than 4\% of power is going into
gravitational waves. For the second search the uniform and restricted prior analyses gave
upper limits of 1.7~\texttimes~10\super{-24} and 1.2~\texttimes~10\super{-24}
respectively. The whole of S5 was used to search for emissions from 116 known
pulsars~\cite{Abbott:2010a}. During this search the Crab limit was further
brought down to be less than a factor of 7 below the spin-down limit, and the
spin-down limit is reached for one other pulsar \epubtkSIMBAD{PSR~J0537--6910}. Of the other
pulsars, the best (lowest) upper limit on gravitational-wave amplitude was
2.3~\texttimes~10\super{-26} for \epubtkSIMBAD{PSR~J1603--7202} and our best (lowest) limit on the
inferred pulsar ellipticity is 7.0~\texttimes~10\super{-8} for \epubtkSIMBAD{PSR~J2124--3358}.

A semi-targeted search was performed with 12 days of S5 data, although this time
searching for a source with a known position in the Cassiopeia~A (\epubtkSIMBAD{Cas~A})
supernova remnant, but for which there is no known frequency. The
search~\cite{Abadie:2010g} looked in the frequency band between
100\,--\,300~Hz and covered a wide range of first and second frequency
derivatives and no signal was seen, but it gave 95\% amplitude and
ellipticity upper limits over the band of
(0.7\,--\,1.2)~\texttimes~10\super{-24} and
(0.4\,--\,4)~\texttimes~10\super{-4} respectively. These results beat
indirect limits on the emission based on energy-conservation arguments
(similar, but not the same as the spin-down limits) and were also the
first results to be cast as limits on the \textit{r}-mode amplitude~\cite{Owen:2010}.

The Vela pulsar has a spin frequency of $\sim$~11~Hz and was not accessible with
current LIGO data.  However, Virgo VSR2 data had sensitivity in the low
frequency band that made a search for it worthwhile. Using $\sim$~150~days of
Virgo data, three semi-independent methods were used to search for the
Vela pulsar \cite{Abadie:2011b}. No signal was seen, but a 95\% upper limit on
the amplitude of $\sim$~2~\texttimes~10\super{-24} was set, which beat the spin-down limit
by $\sim$~1.6 times. Other than the Crab pulsar, this is currently the only other
object for which the spin-down limit has been beaten.

\subsubsection{Stochastic sources}

Searches are conducted for a cosmological, or astrophysical, background of
gravitational waves that would show up as a coherent stochastic noise source
between detectors. This is done by performing a cross-correlation of data from
two detectors as described in~\cite{Allen:1999b}.

In S1 the most sensitive detector pair for this correlation was H2--L1 (the H1--H2
pair are significantly hampered by local environmental correlations) and they
gave a 90\% confidence upper limit of $\Omega_{\mathrm{gw}} < 44\pm9$\epubtkFootnote{The
result published in~\cite{Abbott:2004e} give an upper limit value of
$\Omega_{\mathrm{gw}} < 23$, but this is for a Hubble constant
of 100~km~s\super{-1}~Mpc\super{-1}, so for consistency with later results it has
been converted to use a Hubble constant of 72~km~s\super{-1}~Mpc\super{-1} as in
\cite{Abbott:2005h}.} within the 40\,--\,314~Hz band, where the upper limit is in
units of closure density of the universe and for a Hubble constant in units of
72~km~s\super{-1}~Mpc\super{-1}~\cite{Abbott:2004e}. This limit was several times
better than previous direct-detector limits, but still well above the
concordance $\Lambda$CDM cosmology value of the \textit{total} energy density of
the universe of $\Omega_0\approx1$ (see, e.g.,~\cite{Jarosik:2010}).

No published stochastic background search was performed on S2 data, but S3 data
was searched and gave an upper limit that improved on the S1 result by a factor
of $\sim$~10\super{5}. The most sensitive detector pair for this search was H1--L1 for
which 218 hours of data were used~\cite{Abbott:2005h}. Upper limits were set for
three different power-law spectra of the gravitational-wave background. For a
flat spectra, as predicted by some inflationary and cosmic string models, a 90\%
confidence upper limit of $\Omega_{\mathrm{gw}}(f) = 8.4\times10^{-4}$ in the
69\,--\,156~Hz range was set (again for a Hubble constant of
72~km~s\super{-1}~Mpc\super{-1}). This is still about 60 times greater than a
conservative bound on primordial gravitational waves set by big-bang
nucleosynthesis (BBN). For a quadratic power law, as predicted for a
superposition of rotating neutron-star signals, an upper limit of
$\Omega_{\mathrm{gw}}(f) = 9.4\times10^{-4}(f/100 \mathrm{\ Hz})^2$ was set in the range 73\,--\,244~Hz,
and for a cubic power law, from some pre-Big-Bang cosmology models, an upper
limit of $\Omega_{\mathrm{gw}}(f) = 8.1\times10^{-4}(f/100 \mathrm{\ Hz})^3$ in the range
76\,--\,329~Hz was produced.

For S4 $\sim$~354~hours of H1--L1 data and $\sim$~333~hours of H2--L1 data were
used to set a 90\% upper limit of $\Omega_{\mathrm{gw}}(f) < 6.5\times10^{-5}$ on
the stochastic background between 51\,--\,150~Hz, for a flat spectrum and Hubble
constant of 72~km~s\super{-1}~Mpc\super{-1}~\cite{Abbott:2007e}. This result is
still several times higher than BBN limits. About 20 days of H1 and L1 S4 data
was also used to produce an upper limit map on the gravitational wave background
across the sky as would be appropriate if there was an anisotropic background
dominated by distinct sources~\cite{Abbott:2007f}. This search covered a
frequency range between 50\,--\,1800~Hz and had spectral \textit{power} limits (which
come from the square of the amplitude $h$) ranging from
$1.2\times10^{-48} \mathrm{\ Hz}^{-1} (100 \mathrm{\ Hz}/f)^3$ and
$1.2\times10^{-47} \mathrm{\ Hz}^{-1} (100 \mathrm{\ Hz}/f)^3$ for an
$f^{-3}$ source power spectrum, and limits of
8.5~\texttimes~10\super{-49}~Hz\super{-1} and
6.1~\texttimes~10\super{-48}~Hz\super{-1} for a flat spectrum.

Data from S4 was also used to perform the first cross-correlation between an
interferometric and bar detector to search for stochastic backgrounds. L1 data
and data from the nearby ALLEGRO bar detector were used to search in the
frequency range 850\,--\,950~Hz, several times higher than the LIGO only
searches~\cite{Abbott:2007g}. A 90\% upper limit on the closure
density of $\Omega_{\mathrm{gw}}(f) \leq 1.02$ (for the above Hubble
constant) was set, which beat previous limits in that frequency range
by two orders of magnitude. This limit beats what would be achievable
with LLO-LHO cross correlation of S4 data in this frequency range by a
factor of several tens, due to the physical proximity of LLO and ALLEGRO.

The entire two years of S5 data from the LIGO detectors has been used to set a
limit on the stochastic background around 100~Hz to be $\Omega_{\mathrm{gw}}(f) <
6.9\times10^{-6}$ at 95\% confidence (for a flat gravitational-wave spectrum)
\cite{Abbott:2009b}. This now beats the indirect limits provided by BBN and cosmic microwave background observations.

\subsection{Detector upgrades}

All the current detectors have upgrades planned over the next several years.
These upgrades will give rise to the second generation of gravitational-wave
detectors, which should start to open up gravitational-wave astronomy as a
real observational tool. There are also currently plans being made for third
generation detectors, which could provide the premier gravitational-wave
observatories for the first half of the century. A brief summary of the planned
upgrades to current and future detectors is given below. An overview can be also
be found here~\cite{Whitcomb:2008}. Some of the technologies for these upgrades
are discussed earlier in this review (e.g.\, Section~\ref{section:interferometry}).

\subsubsection{Advanced LIGO, Advanced Virgo and LCGT}
\label{subsection:aligo}

Advanced LIGO (aLIGO)~\cite{Harry:2010, AdvLIGO, AdvLIGOweb} and Advanced Virgo
(AdvVirgo)~\cite{AdvVirgoDesign, AdvVirgoweb} are the second generation
detectors. They are planned to have a sensitivity increase over the levels of
the initial detectors by a factor of 10\,--\,15 times. These increased sensitivity
levels would expand the volume of space observed by the detectors by $\sim$~1000
times meaning that there is a realistic detection rate of neutron-star--binary
coalescences of around 40~yr\super{-1}~\cite{Abadie:2010e, Kopparapu:2008}.
The technological issues required to reach these sensitivities, such as choice
of test mass and mirror coating materials, suspension design, interferometric
layout, control and readout, would need a separate review article to themselves,
but we shall very briefly summarise them here.

Advanced LIGO will consist of three 4~km detectors in the current LIGO vacuum
system; two at the Hanford site\epubtkFootnote{There is currently a plan that
has been approved by the LIGO Laboratory and the NSF to potentially construct
one of the Hanford detectors at a site in Australia~\cite{Marx:2010}, although
this is reliant on construction and running costs being provided by the
Australian government. Such an observatory in the southern hemisphere would
greatly improve sky localisation of any transient sources and enhance
electromagnetic follow-up observations (e.g.,~\cite{Barriga:2010}).} and one at
Livingston. It will apply some of the technologies from the GEO600
interferometer, such as the use of a signal recycling mirror at the output port
and monolithic silica suspensions for the test masses, rather than the current
steel wire slings. Larger test masses will be used with an increase from 11 to
40 kg, although the masses will still be made from fused silica. The mirror
coating is likely to consist of multiple alternating layers of silica and
tantala, with the tantala layers doped with titania to reduce the coating
thermal noise~\cite{Agresti:2006}. The seismic isolation systems will be
replaced with improved versions offering a seismic cut-off frequency
of $\sim$~10~Hz as opposed to the current cut-off of $\sim$~40~Hz. As
stated for Enhanced LIGO (in Section~\ref{sec:ligoruns}), the laser
power will be greater than for initial LIGO and a DC readout scheme
will be used. Initial/Enhanced LIGO was shut down to begin the
installation of these upgrades on 20 October 2010. The design strain
amplitude sensitivity curve for aLIGO (and AdvVirgo and LCGT) is shown
in Figure~\ref{fig:advcurves}.

% second generation detector design strain sensitivities
\epubtkImage{advcurves.png}{%
\begin{figure}[htbp]
  \centerline{\includegraphics[scale=0.4]{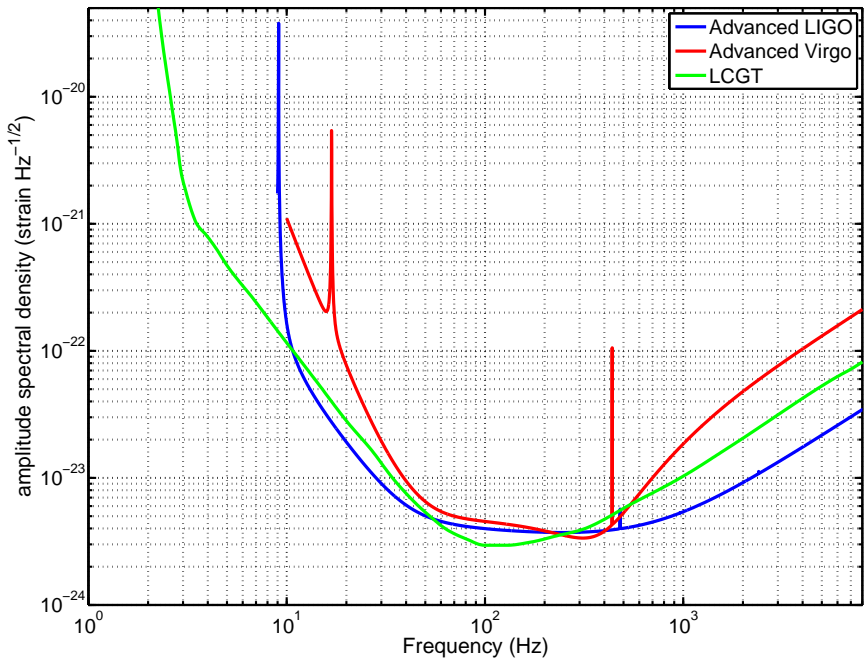}}
  \caption{Design sensitivity curves for the Advanced LIGO, Advanced Virgo
and LCGT second-generation detectors. The Advanced LIGO curve comes from
\cite{Harry:2010}, the Advanced Virgo curve comes
from~\cite{AdvVirgoweb}, and the LCGT curve comes
from~\cite{Arai:2009}. These curves are based on specific
configurations of the detectors and are therefore subject to change.}
  \label{fig:advcurves}
\end{figure}}

AdvVirgo will apply similar upgrades to those for aLIGO and over a similar
timescale (for details see~\cite{AdvVirgoWhitepaper} and~\cite{AdvVirgoDesign}). Plans are
to add a signal recycling mirror, monolithic suspensions, increased laser power
to $\sim$~200~W, improved coatings, and to potentially use non-Gaussian beams (see,
e.g.,~\cite{Freise:2010}), although this option is unlikely. The seismic isolation
system will not be changed. Virgo will shut down to begin these upgrades in July
2011.

The Large-scale Cryogenic Gravitational-Wave Telescope (LCGT)
\cite{Miyoki:2005, Ohashi:2008, Kuroda:2010} is a planned Japanese detector to
be sited underground in the Kamioka mine. The LGCT will consist of a detector
with 3~km arms, using sapphire mirrors and sapphire suspensions. Initially it
will operate at room temperature, but will later be cooled to cryogenic
temperatures. This detector is planned to have similar sensitivities
to aLIGO and AdvVirgo, with a reach for binary coalescences of about 200~Mpc
with SNR of 10. There currently exists a technology demonstrator called the
Cryogenic Laser Interferometer Observatory (CLIO)~\cite{Yamamoto:2008, CLIOweb},
which has a 100~m baseline and is also sited in the Kamioka mine. This is to
demonstrate the very stable conditions (i.e.,\ low levels of seismic noise)
existing in the mine and also the cryogenically-cooled sapphire mirrors
suspended from aluminium wires. In experiments with CLIO at room temperature
(i.e.\, 300~K), using a metallic glass called Bolfur for its wire suspensions, it
has already been used to produce an astrophysics result by looking for
gravitational waves from the Vela pulsar~\cite{Akutsu:2008}, giving a 99.4\%
confidence upper limit of $h$~=~5.3~\texttimes~10\super{-20}. Tests with the cryogenic
system activated and using aluminium suspensions allowed two mirrors to be
cooled to $\sim$~14~K.

Having a network of comparably-sensitive detectors spread widely across the
globe is vital to gain the fullest astrophysical insight into transient sources.
Position reconstruction for sources relies on triangulating the location based
on time-of-flight delays observed between detectors. Therefore, having long
baselines, and different planes between as many detectors as possible, gives the
best positional reconstruction -- in~\cite{Fairhurst:2010} it is shown that for
the 2 US aLIGO sites sky localisation will be on the order of 1000 square degrees,
whereas this can be brought down to a few square degrees with the inclusion of
more sites and detectors. Observation with multiple detectors also provides the
best way to give confidence that a signal is a real gravitational wave rather
than the accidental coincidence of background noise. Finally, multiple,
differently-oriented, detectors will increase the ability to reconstruct a
transient sources waveform and polarisation.

\subsubsection{Third-generation detectors}
\label{subsec:et}

Currently design studies are under way for a third-generation gravitational-wave
observatory called the Einstein Telescope (ET)~\cite{ETweb}. This is a European
Commission funded study with working groups looking into various aspects of the
design including the site location and characteristics (e.g.\, underground),
suspensions technologies; detector topology and geometry (e.g.\, an equilateral
triangle configuration); and astrophysical aims. The preliminary plan is to
aim for an observatory, which improves upon the second-generation detectors by
an order of magnitude over a broad band. There are many technological challenges
to be faced in attempting to make this a reality and research is currently under
way into a variety of these issues.

Investigations into the interferometric configuration have already been studied
(see~\cite{Freise:2008, Hild:2008, Hild:2010}), with suggestions including a
triple interferometer system made up from an equilateral triangle, an
underground location, and potentially a xylophone configuration (two independent
detectors covering different frequency ranges, i.e., ultimately giving six
detectors in total, although constructed over a period of years). Three
potential sensitivity curves are plotted in Figure~\ref{fig:etsens} for different
configurations of detectors.

\epubtkImage{etcurve.png}{%
\begin{figure}[htbp]
  \centerline{\includegraphics[scale=0.4]{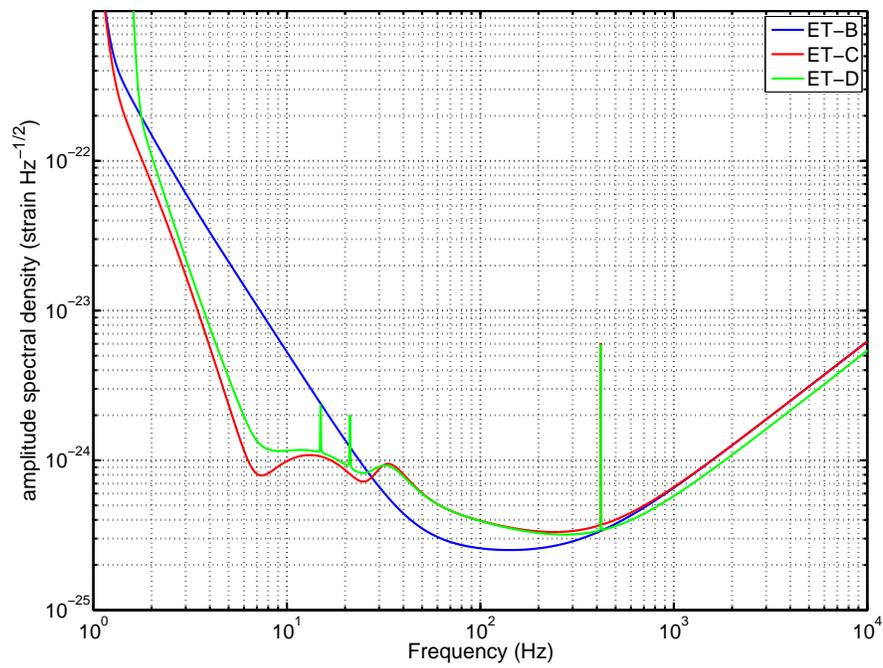}}
  \caption{Potential sensitivities of the Einstein Telescope for 3 different
design concepts: ET-B~\cite{Hild:2008}, ET-C~\cite{Hild:2010b} and ET-D
\cite{Hild:2010}. The curves are available from~\cite{etcurves}}
  \label{fig:etsens}
\end{figure}}

%%%%%%%%%%%%%%%%%%%%%%%%%%%%%%%%%%%%%%%%%%%%%%%%%%%%%%%%%%%%%%%%%%%%%%%%%%%%%%%%
%%%%%%%%%%%%%%%%%%%%%%%%%%%%%%%%%%%%%%%%%%%%%%%%%%%%%%%%%%%%%%%%%%%%%%%%%%%%%%%%
%%%%%%%%%%%%%%%%%%%%%%%%%%%%%%%%%%%%%%%%%%%%%%%%%%%%%%%%%%%%%%%%%%%%%%%%%%%%%%%%

\newpage

\section{Longer Baseline Detectors in Space}
\label{section:space}

Some of the most interesting gravitational-wave signals, resulting from the
mergers of supermassive black holes in the range 10\super{3} to
$10^{6}\,M_{\odot}$ and cosmological stochastic backgrounds, will lie
in the frequency region below that of ground-based detectors. The most
promising way of looking for such signals is to fly a laser
interferometer in space, i.e.\, to launch a number of drag-free
spacecraft into orbit and to compare the distances between test masses
in these craft using laser interferometry.

\subsection{Laser Interferometer Space Antenna (LISA)}

Until early 2011, the Laser Interferometer Space Antenna (LISA) -- see,
for example,~\cite{LISA, LISAsymposium, NASAweb, ESAweb} --  was under
consideration as a joint ESA/NASA mission as one L-class candidate
within the ESA Cosmic Visions program \cite{ESACosmicVisions}. Funding
constraints within the US now mean that ESA must examine the
possibility of flying an L-class mission with European-only funding
\cite{LISAESAstatement}. Accordingly all three L-class candidates are
undergoing a rapid redesign phase with the goal of meeting the new
European-only cost cap. Financial, programmatic and scientific issues
will be reassessed following the redesigns and it is currently
expected that the selection of the first L-class mission will take place in 2014.

However, for the rest of this article we will discuss the plans for LISA prior
to these developments. More concrete information is expected to emerge very
soon. 

LISA would consist of an array of three drag-free spacecraft at the vertices of
an equilateral triangle of length of side 5~\texttimes~10\super{6}~km, with the cluster
placed in an Earth-like orbit at a distance of 1~AU from the Sun, 20\textdegree\
behind the Earth and inclined at 60\textdegree\ to the ecliptic. A current review
of LISA technologies, with expanded discussion of, and references for, topics
touched upon below, can be found in \cite{Jennrich:2009}. Here we will focus
upon a couple of topics regarding the interferometry needed to give the required
sensitivity.

\epubtkImage{lisacurve.png}{%
\begin{figure}[htbp]
  \centerline{\includegraphics[scale=0.4]{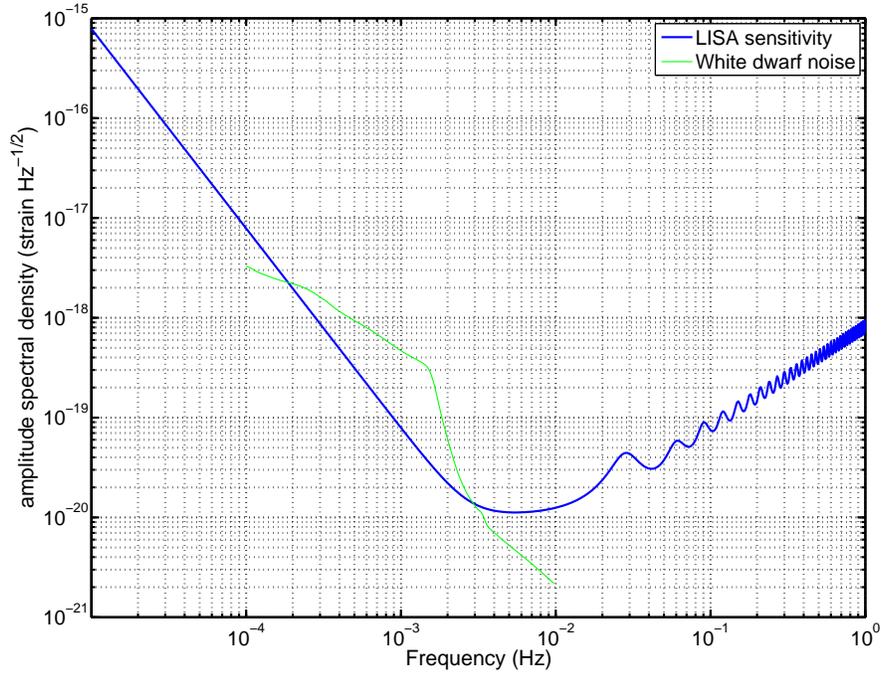}}
  \caption{A design sensitivity amplitude spectral density curve for LISA
created using the standard parameters in the online generator
at~\cite{lisasens}. The curve assumes equal length arms, sensitivity
averaged over the whole sky and all polarisations, and an SNR of
1. Also included is a curve showing the expected background noise from
galactic white-dwarf--binary systems, which will dominate over the
instrumental noise in the range from $\approx$~0.1\,--\,1~mHz.}
  \label{fig:lisasens}
\end{figure}}

Proof masses inside the spacecraft (two in each spacecraft) form the end points
of three separate, but not independent, interferometers. Each single two-arm
Michelson-type interferometer is formed from a vertex (actually consisting of
the proof masses in a `central' spacecraft), and the masses in two remote
spacecraft as indicated in Figure~\ref{figure:LISA}. The three-interferometer
configuration provides redundancy against component failure, gives better
detection probability, and allows the determination of the polarisation of the
incoming radiation. The spacecraft, which house the optical benches, are
essentially there as a way to shield each pair of proof masses from external
disturbances (e.g.,~solar radiation pressure). Drag-free control servos enable
the spacecraft to follow the proof masses to a high level of precision, the drag
compensation being effected using proportional electric thrusters. Illumination
of the interferometers is by highly-stabilised laser light from Nd:YAG lasers at
a wavelength of 1.064 microns, laser powers of $\simeq$~2~W being available from
monolithic, non-planar ring oscillators, which are diode pumped.  For LISA to
achieve its design performance strain sensitivity of around
10\super{-20}~\Hz, adjacent arm lengths have to be sensed
to an accuracy of about 10~pm(Hz)\super{-1/2}. Because of
the long distances involved and the spatial extent of the laser beams
(the diffraction-limited laser spot size, after travelling
5~\texttimes~10\super{6}~km, is approximately 50~km in diameter), the low
photon fluxes make it impossible to use standard mirrors for
reflection; thus, active mirrors with phase locked laser transponders
on the spacecraft will be implemented. Telescope mirrors will be used
to reduce diffraction losses on transmission of the beam and to
increase the collecting area for reception of the beam. With the given
laser power, and using arguments similar to those already discussed
for ground-based detectors with regard to photoelectron shot noise
considerations, means that for the required sensitivity the
transmitting and receiving telescope mirrors on the spacecraft will
have diameters of 40~cm.

\epubtkImage{fig8.png}{%
\begin{figure}[htbp]
  \centerline{\includegraphics[scale=0.4]{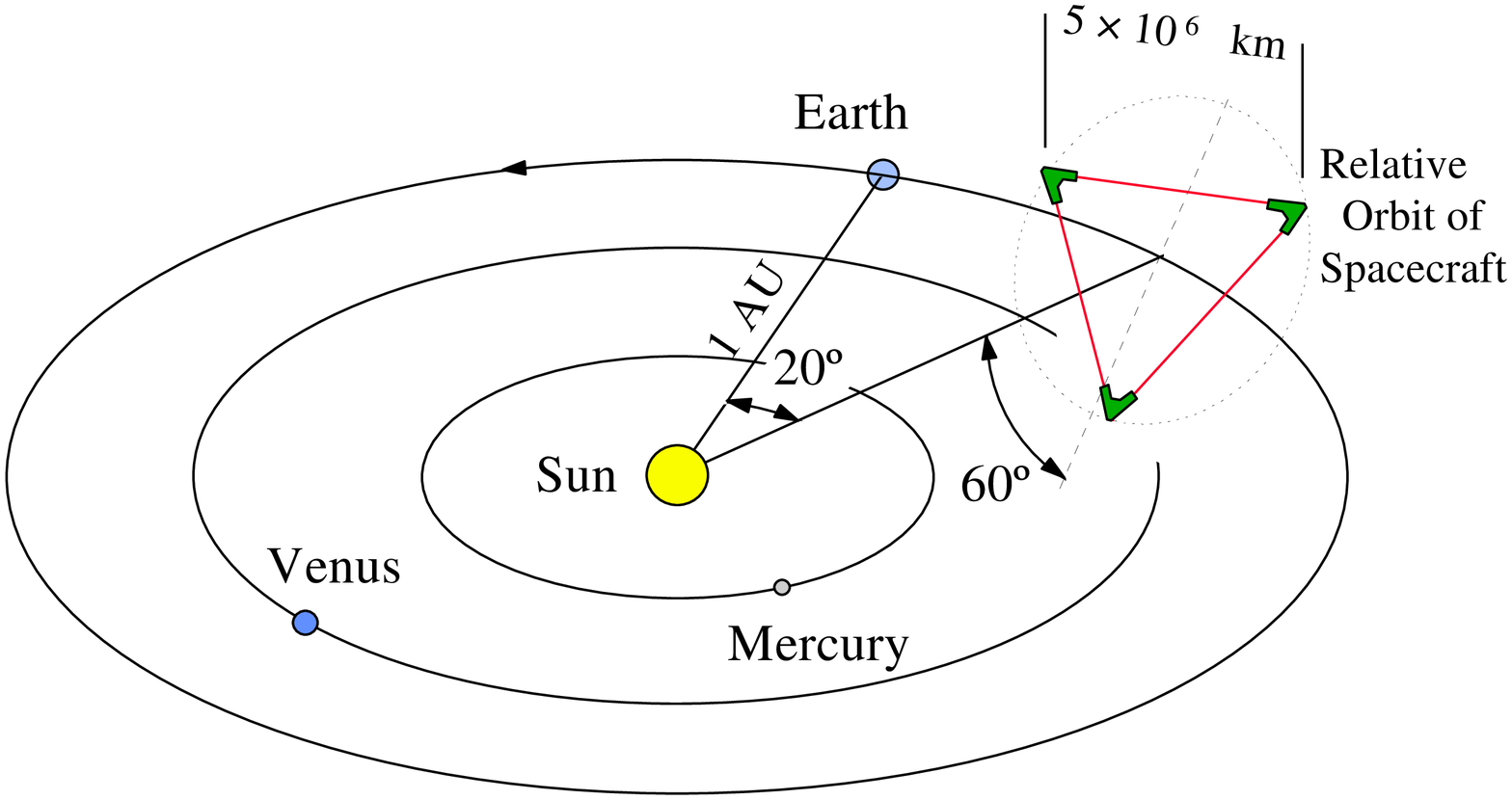}}
  \caption{The proposed LISA detector.}
  \label{figure:LISA}
\end{figure}}

Further, just as in the case of the ground-based detectors, the presence of
laser frequency noise is a limiting factor. It leads to an error in the
measurement of each arm length. If the arms are equal, these errors cancel out,
but if they are unequal, the comparison of lengths used to search for
gravitational waves may be dominated by frequency noise. For the 5~\texttimes~10\super{9}~m
long arms of LISA, a difference in arm length of 10\super{8}~m is likely. Then, for a
relative arm length measurement of 2~\texttimes~10\super{-12}~m~Hz\super{-1/2}
(the error budget level allowed in the LISA design for this noise source),
Equation~(\ref{equation:frequnoise}) suggests that a laser stability
of $\simeq$~6~\texttimes~10\super{-6}~\HzHz is required, a level much
better than can be achieved from the laser on its own. Thus, frequency stabilisation has
to be provided. The first method of stabilisation is to lock the frequency of
one laser in the system on to a local frequency reference, e.g., a Fabry--P\'{e}rot
cavity mounted on one of the craft (see, for example,~\cite{McNamara}), and then
to effectively transfer this stability to other lasers in the system by phase
locking techniques. With the temperature fluctuations inside each craft limited
in the region of 3~mHz to approximately 10\super{-6}~K~\Hz by
three stages of thermal insulation, a cavity formed of material of low expansion
coefficient such as ULE allows a stability level of approximately
30~\HzHz (again at 3~mHz). This level of laser
frequency noise is clearly much worse than the required
1.2~\texttimes~10\super{-6}~\HzHz (at 3~mHz) and a further correction
scheme is needed. A second possible stage of frequency stabilisation
is arm-locking~\cite{Sheard:2003}, which relies on the fact that, by
design, the fractional stability of the LISA arms is of order
$\delta{}l/L \sim 10^{-21} \mathrm{\ Hz}^{-1/2}$ to derive an error
signal from the phase difference between the local laser and the
received light. As the received light is phase locked with the local
laser from the craft that sent it, it caries a replica of the frequency
noise of the local laser noise delayed by one round trip time $\tau =
33\mathrm{\ s}$. Using this fact, this noise can be suppressed at
frequencies smaller than the round trip frequency $f= 1/\tau =
30\mathrm{\ mHz}$. This scheme requires no additional hardware and can
be completely implemented in software, but it will still leave
frequency noise that is several orders of magnitude above required
levels. A third stage frequency stabilisation scheme, which is a
post-processing step, is time-delay interferometry (TDI). This makes
use of the fact that, because the beams coming down each arm are not
combined, the phase of each beam can be measured and
recorded. Therefore, correlations in the frequency noise can be
calculated and subtracted by algebraically combining phase
measurements from different craft delayed by the multiples of the time
delay between the spacecraft. The accuracy of this is set by the phase
measurement accuracy, which allows frequency noise subtraction to
below the required level. A simple TDI scheme, for a much simplified
constellation, was first based in the frequency
domain~\cite{Giampieri}, but due to complexities in taking into
account changing arm lengths and a more complex interferometric scheme,
subsequent implementations have been in the time domain. A
mathematical overview of the TDI scheme, along with moving spacecraft
and unequal arm lengths, can be found in~\cite{Tinto:2005}.

One of the major components of LISA is the disturbance reduction system (DRS),
which is responsible for making sure the test masses follow, as far as
possible, purely gravitational orbits. This consists of the gravitational
reference sensor (GRS) and the control and propulsion systems used to keep the
spacecraft centred on the test mass. The test masses for LISA are 1.96~kg
cubes, with sides of 46~mm and made of an alloy of 75\% gold and 25\% platinum,
chosen because of its very small magnetic susceptibility. The masses are housed
in a cube of electrodes designed to capacitively sense their position and
to have measurement noise levels of 1.8~nm~\Hz. The masses need to be
tightly held in place during launch and then released, so a caging mechanism
has been designed consisting of 8 hydraulic fingers (one for each corner of the
mass) pushing with 1200~N of force. There will be adhesion between the fingers
and the masses, which will require about 10~N of force per finger to break. To
provide this force two plungers will push on the the top and bottom surfaces of
the masses releasing them from the fingers, followed by pushing smaller release
tips in each plunger, and quickly retracting them, to overcome their adhesion to
the masses. Charged particles produced by cosmic radiation interacting with the
surrounding spacecraft can cause the test masses to become charged at a rate of
about 50 electrons per second. Current plans are to use UV light from mercury
lamps (or potentially UV LEDs) to discharge the masses. Another key technology
for the DRS are the micro-Newton thrusters, which provide the fine control
needed for drag-free flight. These will mainly be used to counteract solar
radiation pressure on the spacecraft, which requires about 10~\muN per
relevant thruster. Thrust noise as a function of frequency is required to be
smaller than 
\[
0.1\,\mu{\mathrm{N\ Hz}}^{-1/2}\times\sqrt{1 +
\left(\frac{10\mathrm{\ mHz}}{f}\right)^4}.
\]
Two types of system, both of
which meet the requirements, will be tested on LISA Pathfinder: the US colloid
micro-Newton thruster (CMNT); and the European field emission electric
propulsion system (FEEP). The CMNT uses small drops of a colloid, which it
ionises through field emission, accelerates and ejects from the thruster. Two
designs of FEEP currently exist, one using Indium and the other Caesium and with
different geometries, which, instead of ionising a droplet of colloid, just use
single ions. This means FEEPs have a better charge to mass ratio. The current
baseline is to use caesium FEEPs. Many of the systems above are being tested
in the LISA Pathfinder mission (see below) and use the nominal LISA designs.

There are many other issues associated with laser interferometry, and other
aspects of the mission mentioned above, for LISA, which are not dealt with here
and the interested reader should refer to~\cite{Hough, et.al., Jennrich:2009,
Johann:2008} for a discussion of some of these.

For LISA the baseline mission design was finalised in 2005. An industrial
contract was awarded to Astrium GmbH for the LISA Mission Formulation
study~\cite{Johann:2008}. Within the current ESA Science Programme
LISA is in the Cosmic Vision 2015\,--\,2025
Programme~\cite{ESACosmicVisions}, and launch after 2020 seems
likely. In 2007 the National Research Council report on the NASA
Beyond Einstein Program (soon to become the Physics of the Cosmos
Program) gave LISA the highest scientific ranking, and it has been
rated very highly in the Astro2010 decadal
survey~\cite{astro2010}. However, as stated above, as of earlier this
year (2011) ESA is considering the feasibility of LISA as a single agency
mission~\cite{LISAESAstatement}. Recent technical reports for LISA can
be found at~\cite{LISATechReports}.

Several of the key technologies for LISA are being testing on the LISA
Pathfinder mission (formerly SMART-2). Details of the current status of this
mission can be found in~\cite{Armano:2009}. LISA Pathfinder will fly the LISA
Technology Package (LTP), which essentially consists of a downscaled version of
one LISA arm compressed from 5 million km to 38~cm. The LTP \textit{arm} contains
two test masses (an emitter and a receiver) with a Doppler link between them.
The three main things it will measure are: the acceleration phase noise caused
by the relative motion of the emitter and receiver from non-gravitational
forces; the readout noise; and noise caused by the departure of the Doppler link
from the ideal scheme, due to the fact that we are not truly measuring the
relative accelerations of two point particles, but instead a more complex
system of multiple Doppler links and extended masses. It is designed to test
the accuracy of these to within an order  of magnitude of that required by
the full LISA. Other aspects of the mission that will be tested are the
discharging of the test masses, the caging and release of the masses following
launch and the micro-Newton thrusters. As much as possible the nominal LISA
systems and hardware are being used. LISA Pathfinder is currently scheduled for
launch in mid 2013, after which it will orbit the L1 point, with a 180-day
mission plan.

\subsection{Other missions}

LISA is the most advanced space-based project, but there exist concepts for at
least two more detectors. DECIGO (DECi-hertz Interferometer Gravitational Wave
Observatory)~\cite{Sato:2009, Kawamura:2011} is a Japanese project designed to
fill the gap in frequency between ground-based detectors and LISA, i.e.\, the
0.1\,--\,10~Hz band. It would have a similar configuration to LISA with three
drag-free spacecraft, but have far shorter arm lengths at 1000~km. Although
still early in its design there are plans for two precursor technology
demonstration missions (DECIGO Pathfinder~\cite{Ando:2009} and Pre-DECIGO), with
a the main mission having a launch date in the mid-2020s.

A similar mission, in terms of the frequency band it seeks to cover, is the US
Big Bang Observer (BBO) (see~\cite{Crowder:2005, Cutler:2009, Harry:2006} for
overviews of the proposal). One of its main aims will be to detect the
stochastic background from the early universe, but it can also be used for high
precision cosmology~\cite{Cutler:2009}. The current configuration would consist
of three LISA-like constellations of three spacecraft each, with 50\,000~km arm
lengths, and separated in their orbit by 120\textdegree. The launch of this
mission would be after DECIGO, but is designed to be 2\,--\,3 times as sensitive.

At lower frequencies than LISA, $\sim$~0.1~\muHz\,--\,1~mHz, there are Chinese
proposals for Super-ASTROD (Super Astrodynamical Space Test of
Relativity using Optical Devices)~\cite{Ni:2009}.

%%%%%%%%%%%%%%%%%%%%%%%%%%%%%%%%%%%%%%%%%%%%%%%%%%%%%%%%%%%%%%%%%%%%%%%%%%%%%%%%
%%%%%%%%%%%%%%%%%%%%%%%%%%%%%%%%%%%%%%%%%%%%%%%%%%%%%%%%%%%%%%%%%%%%%%%%%%%%%%%%
%%%%%%%%%%%%%%%%%%%%%%%%%%%%%%%%%%%%%%%%%%%%%%%%%%%%%%%%%%%%%%%%%%%%%%%%%%%%%%%%

%\newpage

\section{Conclusion}
\label{section:conclusion}

Significant effort worldwide has been invested, and is continuing to be
invested, in the development of both ground and spaced based gravitational-wave
detectors. Over the coming years, ground-based detectors will reach
sensitivities that will enable the direct observation of gravitational waves as
predicted by Einstein's General Theory of Relativity and open the exciting new
field of gravitational-wave astronomy. Studying the gravitational-wave
sky could radically change our understanding of the Universe, expanding our
knowledge of fundamental physics, cosmology and relativistic astrophysics.
Gravitational-wave observations will allow us to gaze into the heart of the most
violent events in the Universe, and will help answer some of the biggest
questions, particularly in cosmology when combined with other astronomical
observations.

%%%%%%%%%%%%%%%%%%%%%%%%%%%%%%%%%%%%%%%%%%%%%%%%%%%%%%%%%%%%%%%%%%%%%%%%%%%%%%%%
%%%%%%%%%%%%%%%%%%%%%%%%%%%%%%%%%%%%%%%%%%%%%%%%%%%%%%%%%%%%%%%%%%%%%%%%%%%%%%%%
%%%%%%%%%%%%%%%%%%%%%%%%%%%%%%%%%%%%%%%%%%%%%%%%%%%%%%%%%%%%%%%%%%%%%%%%%%%%%%%%

\bigskip

%\newpage

\section{Acknowledgements}
\label{section:acknowledgements}

We are grateful to the LSC and the Virgo groups for the use of some of their
figures. We are indebted to STFC (U.K.), NSF (U.S.A.), the University of
Glasgow, the Royal Society of Edinburgh, and the Royal Society for support.

%%%%%%%%%%%%%%%%%%%%%%%%%%%%%%%%%%%%%%%%%%%%%%%%%%%%%%%%%%%%%%%%%%%%%%%%%%%%%%%%%%%
%%%%%%%%%%%%%%%%%%%%%%%%%%%%%%%%%%%%%%%%%%%%%%%%%%%%%%%%%%%%%%%%%%%%%%%%%%%%%%%%%%%
%%%%%%%%%%%%%%%%%%%%%%%%%%%%%%%%%%%%%%%%%%%%%%%%%%%%%%%%%%%%%%%%%%%%%%%%%%%%%%%%%%%

\newpage

\bibliography{refs}

\end{document}